\documentclass[twocolumn]{aastex7}

\usepackage{xspace}
\usepackage{fontawesome}

\definecolor{linkcolor}{rgb}{0.1216,0.4667,0.7059}
\newcommand{\codeicon}{{\color{linkcolor}\faFileCodeO}}
\newcommand{\pycodelink}[1]{\href{#1}{\codeicon}\@\xspace}

\newcommand{\sqp}{\texttt{squishyplanet}\@\xspace}

\newcommand{\kep}{Kepler-167\,e\@\xspace}
\newcommand{\jwst}{\textit{JWST}\@\xspace}
\newcommand{\kepler}{\textit{Kepler}\@\xspace}
\newcommand{\gaia}{{\it Gaia}}

\newcommand{\nautilus}{\texttt{nautilus}\@\xspace}

\newcommand{\eureka}{\texttt{Eureka!}\@\xspace}
\newcommand{\jedi}{\texttt{ExoTiC-JEDI}\@\xspace}
\newcommand{\kat}{\texttt{katahdin}\@\xspace}
\newcommand{\jwstpipe}{\texttt{jwst}\@\xspace}

\newcommand{\jax}{\texttt{JAX}\@\xspace}

\newcommand{\asterias}{\texttt{asterias}\@\xspace}

\newcommand{\mth}{Matérn 3/2\@\xspace}
\newcommand{\dchi}{$\Delta \chi^2$\@\xspace}

\received{October 29, 2025}
\accepted{January 15, 2026}
\submitjournal{AJ}
\shorttitle{Oblateness of a Jupiter Analog}
\shortauthors{Cassese et al.}
\graphicspath{{./}{figures/}}

\usepackage{amssymb}
\usepackage{lipsum}
\usepackage{bm}

\usepackage{amsmath}
\usepackage{wasysym}
\usepackage{cancel}

\begin{document}

\title{A JWST Transit of a Jupiter Analog \\I. Constraints on the Oblateness of Kepler-167\,e}

\author[0000-0002-9544-0118]{Ben Cassese}
\affiliation{Department of Astronomy, Columbia University, 550 W 120th Street, New York NY 10027, USA}
\email{b.c.cassese@columbia.edu}

\author[0000-0002-4365-7366]{David Kipping} 
\affiliation{Department of Astronomy, Columbia University, 550 W 120th Street, New York NY 10027, USA}
\email{dmk2184@columbia.edu}

\author[0000-0001-6516-4493]{Quentin Changeat}
\affiliation{Kapteyn Institute, University of Groningen, 9747 AD Groningen, NL}
\email{q.changeat@rug.nl}

\author[0000-0003-4755-584X]{Daniel A. Yahalomi}
\altaffiliation{Flatiron Research Fellow}
\affiliation{Center for Computational Astrophysics, Flatiron Institute, 162 5th Avenue, New York, NY 10010, USA}
\affiliation{Department of Astronomy, Columbia University, 550 W 120th Street, New York NY 10027, USA}
\email{daniel.yahalomi@columbia.edu}

\author[0000-0003-1481-8076]{Justin Vega}
\affiliation{Department of Astronomy, Columbia University, 550 W 120th Street, New York NY 10027, USA}
\email{j.vega@columbia.edu}

\author[0000-0003-1728-8269]{Yayaati Chachan}
\affiliation{Department of Astronomy and Astrophysics, University of California, Santa Cruz, CA 95064, USA}
\email{ychachan@ucsc.edu}

\author[0000-0002-5494-3237]{Billy Edwards}
\affiliation{SRON Netherlands Institute for Space Research, Niels Bohrweg 4, 2333 CA Leiden, Netherlands}
\email{B.Edwards@sron.nl}

\author[0000-0003-2331-5606]{Alex Teachey}
\affiliation{Academia Sinica Institute of Astronomy and Astrophysics, 11F of AS/NTU Astronomy-Mathematics Building, No. 1, Section 4, Roosevelt Rd, Taipei 10617, Taiwan, R.O.C.}
\email{amteachey@asiaa.sinica.edu.tw}

\begin{abstract}
In October 2024 JWST observed a transit of \kep, a Jupiter-analog planet on a 1000+ day orbit. These observations, recorded over a long baseline of nearly 60 hours, were designed to search for signatures of planetary oblateness and/or exomoons comparable to Ganymede. In this first in a series of studies analyzing these data we report on constraints on \kep's oblateness. We explored a large grid of data reduction pipelines and modeling choices, including a new entirely independent reduction pipeline (``\texttt{katahdin}'') and two new treatments for limb darkening. We find that under a Bayesian model comparison framework the data are fit equally well by both spherical and oblate planet models, and that our ability to constrain the oblateness is negatively impacted by the influence of exposure-long trends. Using the most conservative of our posteriors, we place a 95\% upper bound on the projected oblateness of $f<0.097$, which corresponds to a rotation period of $P\geq7.11$\,hours if the planet's spin axis is aligned with the sky plane. We note, however, that the final bound depends on the choice of reduction pipeline and systematics model, and that our suite of end-to-end analyses produced bounds as low as $f<0.065$ at 95\%. We conclude that leveraging JWST to make tighter constraints on planetary oblateness will require further investigation into mitigating exposure-long trends and correlated noise.
\end{abstract}


\section{Introduction} \label{sec:intro}

Two decades ago, \citet{hui_seager_2002} concluded an early study on exoplanet transit modeling with the following passage: ``...it is clear that future high-precision measurements of extrasolar planetary transits will present a very interesting challenge—there are several sources of small secondary fluctuations... that are within observational reach and that would require some effort to disentangle. The rewards of such efforts will be substantial—from detection of moons and rings to measuring the (projected) planet oblateness...''

However, in the years since these bold predictions these exact ``substantial rewards'' have yet to materialize. Though the field has grown wildly in that span and made innumerable discoveries concerning the architectures, atmospheres, and alignments of exoplanetary systems, there has yet to be a firm detection of a surefire exomoon or of non-zero planetary oblateness. The James Webb Space Telescope \citep{gardner_jwst_2006}, however, offers an opportunity to revisit attempts to extract these ``secondary fluctuations'' from an extremely precise, high-cadence observation of an exoplanet transit.

This article is the first in a series focused on the analysis of JWST observations collected in October of 2024 as a part of program GO:6491, ``Revealing the Oblateness and Satellite System of an Extrasolar Jupiter Analog'' (PI: Cassese). As described in Section \ref{sec:observations}, this remarkably-Jupiter like planet is one of the coldest, longest-period known transiting exoplanets, and also happens, fortunately, to transit a (relative for \kepler stars) bright, quiet, sub-Solar radius star. In other words, it is an ideal candidate to search for features that are common among the cold gas giants of our solar system but have yet to be observed beyond it, including Galilean-like moons or rotation-induced flattening. This article details our analysis concerning the planet's oblateness, while Kipping et al. 2025 (hereafter Paper II) details our search for a Ganymede-like satellite.

To begin, we define planetary oblateness. In this context we refer to oblateness as the relative expansion of a planet's equatorial radius relative to its polar radius caused by the centrifugal forces of the planet's rotation \citep{barnes_fortney}. In this conceptualization, the planet's equilibrium shape is an oblate spheroid, with two equal length axes in the equatorial planet $R_{\text{eq}}$ and a shorter polar radius $R_\text{p}$. We define the planet's true oblateness $f^*$ as:

\begin{equation}
    f^* = \frac{R_\text{eq} - R_\text{p}}{R_\text{eq}}.
\end{equation}

Note that this effect and equilibrium shape is different from the distortions caused by tidal affects felt by planets residing very near their host stars \citep{leconte_2011}. Attempts to detect this tidal deformation is another active area of research \citep{akinsanmi_2019, barros_2022, akinsanmi_2024} but are distinct from this effort.

In our solar system, Saturn is the most oblate of the gas giants with $f^*$ near 0.1, followed by Jupiter with $f^*$ near 0.05. These gas giants have rotation periods of 10.6 and 9.9 hours, respectively, which illustrates that rotation rate alone does not set a planet's oblateness: instead, it instead also influenced by additional properties such as mass and internal structure. \citet{barnes_fortney} show that by using the Darwin-Radau approximation to account for the redistribution of mass caused by the equatorial swelling, $f^*$ is related to the planet's rotation by the following:

\begin{equation} \label{eq:f_to_spin}
    f^* = \Omega^2\frac{R_\text{eq}^3}{G M_p}\left[\frac{5}{2} \left(1 - \frac{3}{2} \mathbb{C}\right)^2 + \frac{2}{5}\right]^{-1},
\end{equation}

where $G$ is the gravitational constant, $M_p$ the mass of the planet, $\mathbb{C}$ is the moment of inertia of the planet multiplied by $M_p^{-1} R_\text{eq}^{-2}$, and $\Omega$ the rotation rate of the planet. For uniform density planets, this relation is exact and $\mathbb{C}$ is known precisely ($\mathbb{C}=0.4$). For more realistic planets however, $\mathbb{C}$ can be approximated from interior structure models and falls near $\mathbb{C}=0.25$ \citep{hubbard_1984}.

Since we can generate approximations for $\mathbb{C}$ and constrain $R_\text{eq}$ and $M_p$ from measurements, Eq. \ref{eq:f_to_spin} implies that we can map directly from $f^*$ to a planet's rotation rate. This is a fundamental quantity capable of providing insights into giant planet formation and atmospheric dynamics that cannot be probed in any other way for planets not amenable to direct imaging \citep{batygin_2018, bryan_2018}. In addition to these rewards of measuring $f^*$, there are risks to \textit{not} doing so as well. Namely, using a spherical planet model to infer the radius of an oblate planet will lead to a biased estimate of the planet's radius, which in turn will bias inferences on the planet's density and composition \citep{burton_2014, berardo_bottleneck_2022}.

Several techniques have been proposed to probe $f^*$, including by comparing variations in transit depth between different epochs that may be caused by the precession of an oblate planet's spin vector, which in turn would alter its projected area and resulting transit depth \citep{carter_winn_depth_2010, biersteker_2017}. Similarly, precession could also shift the contact times of a transit, leading to a potentially detectably transit timing variation \citep{berardo_ob_effects_2022}. Alternatively, an oblate planet may imprint a detectable signal on a spectroscopic Rossiter–McLaughlin measurement \citep{akinsanmi_2020}.

However, arguably the most straightforward way to probe $f^*$ is to search for deviations from spherical transit models on the ingress and egress of a transit. Since the moment of first contact between an oblate planet and the stellar limb will appear early or late relative to a spherical planet with the same projected radius, spherical planet models are unable to fully capture the complex shape of the ingress and egress \citep{barnes_fortney}\footnote{Note that once fully in transit, an oblate planet will appear nearly indistinguishable from a spherical planet of the same projected area. Nearly, but not exactly, since they'll subtend slightly different portions of the star. Since the star's intensity varies with projected radius (limb darkening), the total flux blocked by these different shapes will vary slightly. However, this effect is $<1$ ppm \citep{squishyplanet}.} Therefore, the best-fitting spherical model will contain small \citep[10s-100s of ppm, ][]{squishyplanet} ``wiggles'' in the residuals during ingress and egress. By forward modeling a more complex simulation of an oblate transistor, we can fit these wiggles, place constraints on the oblateness, and perform model comparison with a spherical planet model.

There is an important caveat to this technique, however: it is only sensitive to $f$, oblateness of the on-sky projected ellipse, not $f^*$. This projected oblateness is set by the combination of the true oblateness and two rotation angles (the planet's obliquity and rotation about the $\hat{z}$ axis of its orbital plane) and must necessarily be a lower bound on the true oblateness \citep{berardo_ob_effects_2022}. The mapping between true and projected oblateness is given in \citet{carter_winn_empirical_2010}, among others.

Several studies have attempted to measure $f$ in the past, including \citet{carter_winn_empirical_2010}, who analyzed \textit{Spitzer} data of HD 189733\,b, \citet{zhu_2014}, who analyzed a selection of \kepler planets, and \citep{price_2025}, who developed a new forward modeling package and applied it to \textit{K2} data of HIP-41378\,f. \citet{dholakia_2025} recently created their own forward modeling package and applied it to JWST data of WASP-107\,b and were able to place an upper bound of $f<0.23$. Finally \citet{lammers_2024} used \sqp \citep{squishyplanet}, another open-source package for modeling the transits of non-spherical planets, to place an upper limit of $f<0.16$ for the super puff Kepler-51\,d using JWST data as well. These latter three constraints were used to place limits on the rotation rate of $P\geq15.3$, $P\geq13$, and $P\geq40$ hours for HIP-41378\,f, WASP-107\,b and Kepler-51\,d, respectively. Note, however, that these limits are conditioned upon different assumptions about the true alignment of the planet relative to the sky frame. \citet{dholakia_2025} assume the projected and true oblateness are equal, while \citet{lammers_2024} assumes that the planet's true obliquity is $<45^\circ$. Note also that \citet{liu_kep_51} analyzed the same \jwst data underlying \citet{lammers_2024} using their own forward model \citep{jojo} and arrived at a more conservative bound of $f<0.2$.

We describe \kep and our observations in Sec. \ref{sec:observations}. We detail the grid of reductions and trend models explored in Sec. \ref{sec:mle_grid}. We then describe how we culled this extensive grid via maximum likelihood fits in Sec. \ref{sec:mle_results}. We describe our posterior inference via nested sampling across a smaller grid of models in Sec. \ref{sec:nested_sampling}. Finally, we offer our conclusions in Sec. \ref{sec:conclusion}.

\section{Target and Observations} \label{sec:observations}

This study focuses on \kep, which by several metrics is the most Jupiter-like known transiting exoplanet. For one, weighing in at $1.01\,^{+0.16}_{-0.15}$\,$M_{\mathrm{J}}$ and with a radius of $0.91\pm0.037$\,$R_{\mathrm{J}}$, its physical size closely resembles Jupiter's \citep{chachan_rvs_2022}. More intriguingly, however, with an orbital period of 1071.23\,days, an equilibrium temperature of 134\,K, and an eccentricity of $<0.29$ ($3\sigma$-bound), \kep exists today and likely formed well beyond its host star's snowline. These properties place \kep in a rarefied  group of only 3 transiting planets with well-constrained periods beyond 1,000\,days, and among this list, \kep is the only world that transits a star with a radius $<1$\,R$_\odot$\footnote{Retrieved from the NASA Exoplanet Archive \citep{nasa_exoplanet_archive} Sept. 2025.}.

\kep which was first discovered by \citet{kipping_discovery_2016} through analysis of two transits observed by the \textit{Kepler} mission. The planet has since been the target of three other observing campaigns prior to this study: \citet{dalba_spitzer_2019} recorded a partial transit with the \textit{Spitzer Space Telescope} and managed to establish a workable ephemeris by ruling out large TTVs; next, a group of citizen scientists from nine different countries operating through the Unistellar Telescope Network recorded an additional transit \citep{perrocheau_citizen_science_2022}; and finally, \citet{chachan_rvs_2022} analyzed radial velocity measurements taken with Keck-HIRES from 2017-2020 to constrain the planet's mass and eccentricity.

\kep is the fourth and most massive planet known to orbit Kepler-167 A, a K dwarf star whose parameters are summarized in Table \ref{tab:stellar_props}. The other component of the Kepler-167 binary, Kepler-167 B, is a main sequence M4V dwarf approximately 100x fainter than Kepler-167\,A in the \textit{Kepler} bandpass and has other magnitudes of $J$=16.32, $K$=15.38 \citep{kipping_discovery_2016} and $G$=19.23 \citep{gaia_dr3}. With a projected semimajor axis of $\sim700$\,au and a period of $P_\text{orb} > 15,000$\,yrs, this companion has no meaningful impact on radial velocity measurements. Similarly, though its position of approximately 2.1\arcsec\, NE of the brighter companion places it within \textit{Kepler}'s extraction aperture, its relative faintness guaranteed that any contamination to \textit{Kepler}-derived planetary radii fell below measurement uncertainties \citep{chachan_rvs_2022}. For the purposes of this study, we note that Kepler-167\,B fell outside the 1.6\arcsec square slit (S1600A1) used in the JWST observations described below and that we did not investigate the possibility of contamination from the edges of its PSF.

Our data consist of a 59.78\,hour time series of observations with JWST's NIRSpec instrument centered on \kep's predicted transit on October 25th, 2024. This long baseline was selected to capture the transit of any moons in Ganymede-like orbits around the planet, which could lead or lag the planet by many hours. We used the Bright Object Time Series mode and the Prism/Clear grating/filter combination to collect spectral observations between 0.6-5.3\,$\mu$m at a typical resolution of $R=100$. Groups were read out using the NRSRAPID pattern, meaning no frames were averaged onboard the spacecraft, as is typical for time series observations. We used the standard SUB512 subarray used for Prism observations (a 32\,x\,512 pixel cutout of the NRS1 detector that contains no reference pixels) and 6 groups per integration for each of our 134,208 total integrations. All told, the dataset contains $n_{\mathrm{pix}}$ x $n_{\mathrm{groups}}$ x $n_{\mathrm{integrations}}$ = 13.19 billion individual flux measurements. To our knowledge, as of writing these data represent JWST's longest continuous time series observation.

Due to the extreme length of the time series and onboard processing/storage requirements of the observatory, our observations were split into 6 distinct exposures with a small pause between each. During this time, the detector was reset and briefly entered its passive continuous-readout mode. The JWST Engineering Mnemonic Database also indicates that the High Gain Antenna moved during these pauses, as well as several times within the science exposures. Exposures 1-2 contain no planetary transits; exposures 3-5 contain the transit of \kep; exposure 6 contains a transit of Kepler-167\,c, a super-Earth planet on a 7.4\,day orbit \citep{rowe_2014, kipping_discovery_2016}. As seen in Fig. \ref{fig:wlcs} and discussed in Sec. \ref{sub:trend_models}, each of these exposures were influenced by their own decreasing and nonlinear background trend. We note that no pixels are flagged as saturated by the \jwstpipe pipeline \citep{jwst_pipeline} in the median image.

With the exception of Fig. \ref{fig:wlcs}, the remainder of this manuscript deals only with the first 5 exposures, as we left out exposure 6 to avoid the computational complexity of modeling Kepler-167\,c and an additional set of trend parameters. This is justified in this study since exposure 6 would only affect our inference by indirectly informing shared parameters such as Gaussian Process hyperparameters or the length scale of an exponential trend; exposures 3 and 4, which contain the ingress and egress respectively, contain most of the constraining power for all planet-related parameters. Since Paper II deals with a search for moons which in principle could appear anywhere in the observation, it retains some of exposure 6 but manually masks just the transit of Kepler-167\,c again to avoid computational complexity.

\begin{deluxetable}{lr}
\tablecaption{Stellar Properties of Kepler-167\,A\label{tab:stellar_props}}
\tablewidth{0pt}
\tablehead{
\colhead{Property} & \colhead{Value}
}
\startdata
\textbf{Identification} & \\
\hline
Target Name & Kepler-167 A\\
Gaia ID\tablenotemark{a} & Gaia DR3 2051933102555095936 \\
Coordinates (J2000)\tablenotemark{a} & $19^{\mathrm{h}}30^{\mathrm{m}}38.05^{\mathrm{s}}$ \\
& $+38\degr20\arcmin44.02\arcsec$ \\
\hline
\textbf{Photometry} & \\
\hline
$V$ mag\tablenotemark{b} & $14.28 \pm 0.01$ \\
$J$ mag\tablenotemark{c} & $12.446 \pm 0.022$ \\
$K_s$ mag\tablenotemark{c} & $11.832 \pm 0.022$ \\
$G$ mag (\gaia)\tablenotemark{a} & $13.99$ \\
\hline
\textbf{Astrometry\tablenotemark{a}} & \\
\hline
Parallax (mas) & $2.92 \pm 0.013$ \\
Distance (pc)\tablenotemark{d} & $339.6 \pm 2.1$ \\
PM RA (mas yr$^{-1}$) & $15.01 \pm 0.014$ \\
PM cos Dec (mas yr$^{-1}$) & $36.35 \pm 0.015$ \\
RV (km s$^{-1}$) & $-26.79 \pm 2.20$ \\
\hline
\textbf{Physical Properties\tablenotemark{d}} & \\
\hline
Mass ($M_\odot$) & $0.777\,^{+0.034}_{-0.031}$ \\
Radius ($R_\odot$) & $0.749\,\pm 0.020$ \\
$T_{\mathrm{eff}}$ (K) & $4884\,^{+69}_{-75}$\\
log $g$ &  $4.579\,^{+0.027}_{-0.025}$ \\
$\mathrm{[Fe/H]}$ (dex) &  $0.020\,\pm0.067$\\
Age (Gyr) & $7.1\,^{+4.4}_{-4.6}$\\
\enddata
\tablenotetext{a}{\citet{gaia_dr3}}
\tablenotetext{b}{\citet{simbad_vmag}}
\tablenotetext{c}{\citet{simbad_2mass_phot}}
\tablenotetext{d}{\citet{chachan_rvs_2022}}
\end{deluxetable}

\section{Initial Model Exploration} \label{sec:mle_grid}

\subsection{Summary of Approach} \label{sub:oblate_summary}

A planet's oblateness, as discussed in Sec. \ref{sec:intro}, induces 10s-100s of ppm ``wiggles'' during a planet's transit ingress and egress that cannot be captured by a spherical planet model \citep{barnes_fortney}. While at first brush the scale of this signal appears comparable with atmospheric features that JWST regularly targets \citep[e.g.,][]{alam_compass_2025}, it is important to note a key difference between measuring oblateness and measuring variations in transit depth. While those chasing a planet's spectrum can use the entire time series to inform their depth measurements, an oblateness signal presents itself on a much shorter timescale. The signatures of oblateness only imprint themselves on the light curve during ingress and egress, and even worse, are (generally) not symmetric across both. In the case of \kep, the relevant timescale is $\sim15-30$\,min.

This requirement of a time-resolved signal introduces a number of potential pitfalls. To name a few, a noisy reduction that allows excessive outliers or errant pixels can ruin an oblateness measurement with just a handful of bad points; a slight mis-parameterization of the exposure-long trend could influence the slope of ingress/egress and either mask or inject a false oblateness signal; similarly, an inaccurate or insufficiently flexible limb darkening model can also pollute an oblateness measurement. This latter concern is especially relevant since limb darkening changes most rapidly near the edge of the star, exactly where quadratic models tend to break down and where our oblateness signal is strongest \citep{claret_2000}.

To bluntly summarize: measuring planetary oblateness is challenging, and one's modeling choices can easily either mask the signal of interest or create the deception of a false signal. With this as motivation, we chose to investigate a large grid of analyses spanning multiple reductions, trend models, initial binning cadence, and limb darkening models.

Instead of jumping straight to MCMC or nested sampling explorations of the posterior distribution, we instead begin by optimizing the likelihood function to arrive at a single point estimate of best-fitting parameters, which we could then use to compute metrics such as the Bayesian Information Criterion (BIC), Akaike Information Criterion (AIC), the root mean squared (RMS) of the residuals, and the median absolute deviation (MAD) of the residuals. Additionally, we also performed the same suite of tests several times by holding out random subsets of data and examining the best-fitting model's predictions on the held out points after fitting.

To be clear, at this step our goal was not to actually measure oblateness or to choose between oblate or spherical models using these fits: instead, our goal was simply to narrow down the search space of viable models ahead of more thorough but computationally expensive fits that probe the joint posteriors.

Our grid consisted of all combinations of the following: 3 different reduction pipelines, each ran with 6 different settings, for a total of 18 unique reductions; 6 unique binning cadences per reduction, producing a total of 108 unique light curves; 3 different heuristic trend models to capture exposure-long drifts; 3 different Gaussian Process additions to the trend model\footnote{One of the trend options was simply a constant trend with the intention of letting the GP handle all drifts, and one of the GP options was to leave off the GP. Consequently, we did not test models that used a constant trend but no GP.}; and, finally, 5 different limb darkening models, including two novel parameterizations. All told we attempted to fit 4320 unique light curve/forward model combinations. To ensure our final likelihood values robustly reflect the true global minimum, we repeated each fit 3 times with 3 different initial random seeds. All fits were run on Yale University's Bouchet high-performance computing cluster.

The remainder of this section is dedicated to a description of each of these models, while the results of the point-estimate fits are presented in Sec. \ref{sec:mle_results}.

\subsection{Reduction Pipelines and Light Curve Generation} \label{sub:oblate_reduction_pipelines}

\begin{figure*}
    \centering
    \includegraphics[width=0.9\linewidth]{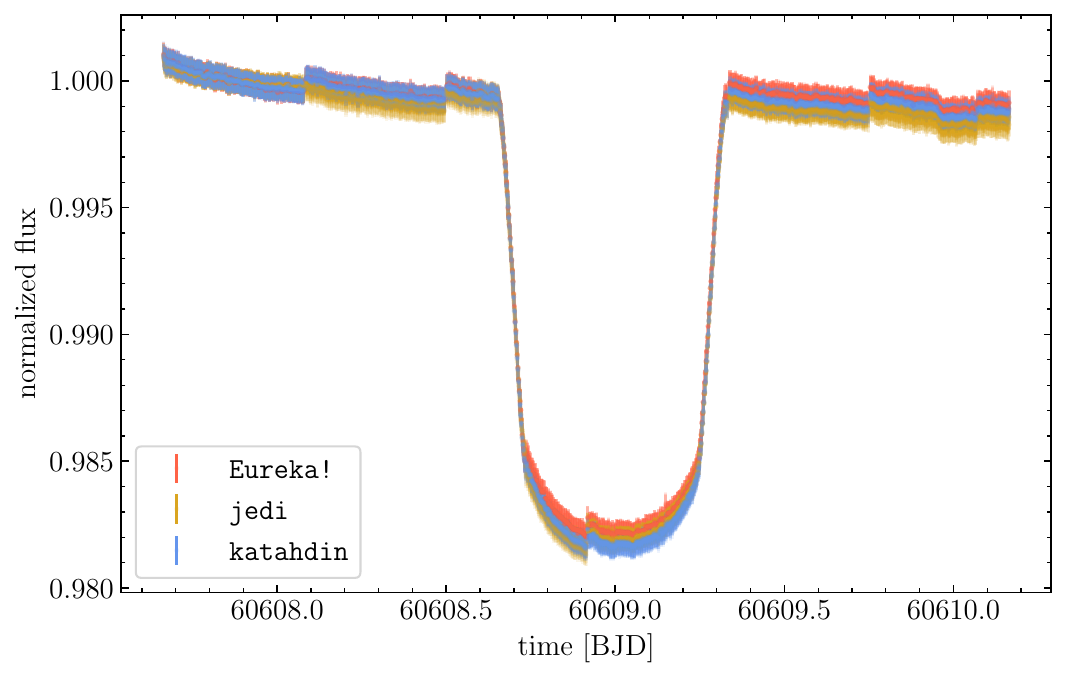}
    \caption{All 18 white light curves, each normalized by the median of flux in exposure 1. The discontinuities occur at exposure boundaries; the large dip in the middle corresponds to the transit of \kep, while the smaller dip in exposure 6 corresponds to a transit of Kepler-167\,c. Note that moving forwards this study only considers the first 5 exposures.}
    \label{fig:wlcs}
\end{figure*}

\subsubsection{Eureka!} \label{subsub:eureka}

Our grid includes 6 light curves derived from the \eureka pipeline \citep{eureka}. \eureka is among the most commonly used pipelines with the transiting exoplanet community and has heritage from the Cycle 1 The Transiting Exoplanet Community Early Release Science Program \citep{transit_ers_2018}. It is a sprawling and ambitious package that aims to be an ``end-to-end'' tool capable of performing both data reduction and model fitting for every instrument and mode available for time series observations aboard JWST. To this end, while \eureka includes some functionality unique to itself, in many cases it simply acts as a wrapper for other more specialized packages.

For instance, \eureka's initial two stages of data reduction (i.e. the steps run to convert uncalibrated data into per-pixel, per-integration rates) directly call the STSci-supported \jwstpipe pipeline \citep{jwst_pipeline}. Note however that as of writing, \eureka is incompatible with the most recent versions of \jwstpipe out of the box, and though it can accept products generated externally using these versions, it and must be cloned/tweaked to run Stage 1 itself. \eureka has rigid requirements for directory structures and does not easily support incremental steps, and instead relies on ``eureka control file'' text documents to configure large, contiguous sequences of pipeline operations. This has the benefit of making reductions easier to reproduce, but makes it challenging to adopt workflows other than the one described in the documentation (e.g. parallelization on a computing cluster).

Like all reduction pipelines, \eureka contains dozens of knobs one could turn when creating a spectral light curve. We chose to vary two of these across a mini-grid of 6 reductions: the method for subtracting the $1/f$ noise and the width of the extraction aperture.

$1/f$ noise, as its commonly referred to in the exoplanet literature thanks to its non-white power spectrum, refers to correlated noise imprinted on each frame by NIRSpec's readout electronics \citep{rauscher_irs2_2017}. Since NIRSpec reads out one pixel at a time and proceeds down each column before pausing and jumping to the top of the next column, this correlated noise produces apparent vertical ``banding'' in all NIRSpec images. Note that this is a per-group effect, and that to correct for it requires operating on individual group images \citep{rustamkulov_lab_timeseries_2022}. While detector-induced correlations certainly exist on longer timescales, discussions of $1/f$ noise in \textit{rate} data, not \textit{group} data, do not have a known physical basis. Note also that treatment of $1/f$ noise diffs between NIRSpec and NIRCam: although both use similar detectors/electronics, NIRCam is oriented such that readout occurs along the spectral, rather than spatial, dispersion axis.

To begin, we ran \jwstpipe v1.19.0 with three different \texttt{correct\_flicker\_noise} algorithms. The first of these simply skipped the step, which we note is the default \jwstpipe behavior. The next, \texttt{median}, takes a column-by-column median of the background pixels in each group and subtracts it. This does not remove correlations between columns, but since the dominant timescale of $1/f$ noise is comparable to the height of the subarray \citep{moseley_read_noise_2010, espinoza_commissioning_2023}, it is effective in removing the vertical banding. Finally, we used \texttt{fft}, which runs the \texttt{nsclean} algorithm \citep{rauscher_nsclean_2024}. This routine masks the science portion of an image, Fourier transforms the remainder, then performs its corrections in the Fourier domain before converting back into a sequence of images.

We then fed each of these three intermediate products into \eureka's Stage 3 reduction to create a spectral light curve. Here we varied only the extraction aperture, using either 3 or 6 pixels. In all cases we relied on \eureka's implementation of optimal spectral extraction from \citet{horne_optimal_extraction_1986}.

\subsubsection{ExoTiC-JEDI} \label{subsub:exotic_jedi}

Our next pipeline was \jedi, first used in \citet{alderson_wasp_ers_2023}. Like \eureka, \jedi relies heavily on the \jwstpipe pipeline for Stages 1 and 2 (uncalibrated data to rate images), then diverges to use its own routines for Stage 3 (images to spectral light curves). Unlike \eureka, however, by default it runs its own background subtraction in Stage 1. This routine requires users to input a ``width'' in units of FWHM to mask the science data when performing background subtraction: we ran both 5 and 15.

We then passed these intermediate products through \jedi's Stage 3 routines. Here we chose to vary the extraction aperture width again, and used 3, 6, and 8 pixels. We also experimented with turning on and off the $1/f$ correction and with different orders for the polynomials that model the trace center and width but found that these alterations did not affect the scatter in the final reductions as aggressively as the background mask width and spectral extraction aperture.

\subsubsection{katahdin} \label{subsub:katahdin}

Our final pipeline, called \kat, is our own creation developed over the course of this analysis. A separate publication describing \kat is in preparation, but here we include a brief overview of its operations.

Compared to \eureka and \jedi, \kat takes an alternative approach to JWST data reduction and does not rely on \jwstpipe for anything other than bad pixel masks and when creating priors. Rather than perform a series of deterministic transformations based on static reference files, \kat uses a Bayesian framework and jointly fits for nuisance parameters alongside science parameters.

The version of \kat used in this study fit a Gaussian Process + pedestal drift model to each frame to correct for $1/f$ noise. The priors for the GP were created via fits to dark calibration frames collected with the same subarray. After $1/f$ correction, each exposure was treated separately, and each pixel was then fit independently with a model that jointly considered per-integration rates, per-integration biases, and shared nonlinearity coefficients and gains.

\kat is written in JAX \citep{jax}, an autodifferentiable framework underpinning several large-scale machine learning efforts from Google. \kat leverages this differentiability to both optimize its likelihood model (each pixel requires fitting 10s of thousands of parameters simultaneously) and to estimate uncertainties on each parameter via the Laplace Approximation.

\kat only includes functionality to process the equivalent of Stage 1 of \jwstpipe, meaning it includes no new algorithms for spectral extraction. In an effort to maintain true independence from the other pipelines, however, we opted not to use either \jwstpipe, \eureka, or \jedi for the equivalent of Stage 3 processing. Instead, we used the following simple routine: looking at just exposure 1, which does not contain any of \kep's transit, we flagged all pixels that fell above various signal-to-noise thresholds and summed them column-wise. This was justified by the minimal background signal left following the GP $1/f$ correction. Our 6 \kat light curves correspond to using SNR thresholds 3, 5, 8, 10, 15, and 25, where as ordered each light curve relied on fewer and fewer pixels. As discussed in Sec. \ref{sub:mle_takeaways}, however, we only used the latter four of these in our final nested sampling fits.

\kat requires access to a modern GPU cluster to perform its reduction and is not currently designed to be run out-of-the-box. However, it is under active development, and our intention is for it to serve as a truly independent comparison to other pipelines in the literature. While it is currently accepted best practice to use more than one reduction in atmospheric analyses (e.g. the COMPASS survey \citet{compass}, and the Hot Rocks survey \citet{hot_rocks}), including an orthogonal pipeline that does not share source code with the others and operates under a different statistical philosophy would greatly bolster assertions of pipeline-independent detections.

\subsubsection{Binning/Cleaning Procedure} \label{subsub:oblate_bin_and_clean}

Each of the three pipelines described above created spectral light curves with 134,208 time stamps and varying numbers of wavelength channels. To convert each of these into white light curves, we used the following approach across all 18 reductions. We first queried the JWST Engineering Mnemonic Database for a time series of all moments when the observatories High Gain Antenna (HGA) was moving. These moves are prohibited during non-time series observations as they can cause significant transient pointing errors, but allowed during long stares. After visual inspection we mask out any time stamp that occurred within 20 seconds following any HGA move: this resulted in losing 0.07\% of the data. Note that the HGA moved several times within the exposures, but also during the pauses between exposures.

Next, we reshaped the data into even bins of [32, 48, 96, 233, 466] integrations, which correspond to [0.86, 1.28, 2.57, 6.23, 12.46] minutes, respectively. We also created a 5 minute binning by discarding the first 115 integrations from each exposure and reshaping the remaining points into 187 integrations/bin. For each of these reshaped light curves, we computed a robust standard deviation within each time and spectral bin by calculating the median absolute deviation and multiplying by 1.4826, then rejected any points that fell beyond 5$\sigma$ of the mean. We then summed along the spectral axis, computed the mean of the white light flux within the time bin, and adopted the standard error of the mean as its uncertainty. This procedure therefore disregards the pipeline-reported uncertainties on each data point.

\subsubsection{Pipeline Comparisons} \label{subsub:pipeline_compare}

\begin{figure*}
    \centering
    \includegraphics[width=0.9\linewidth]{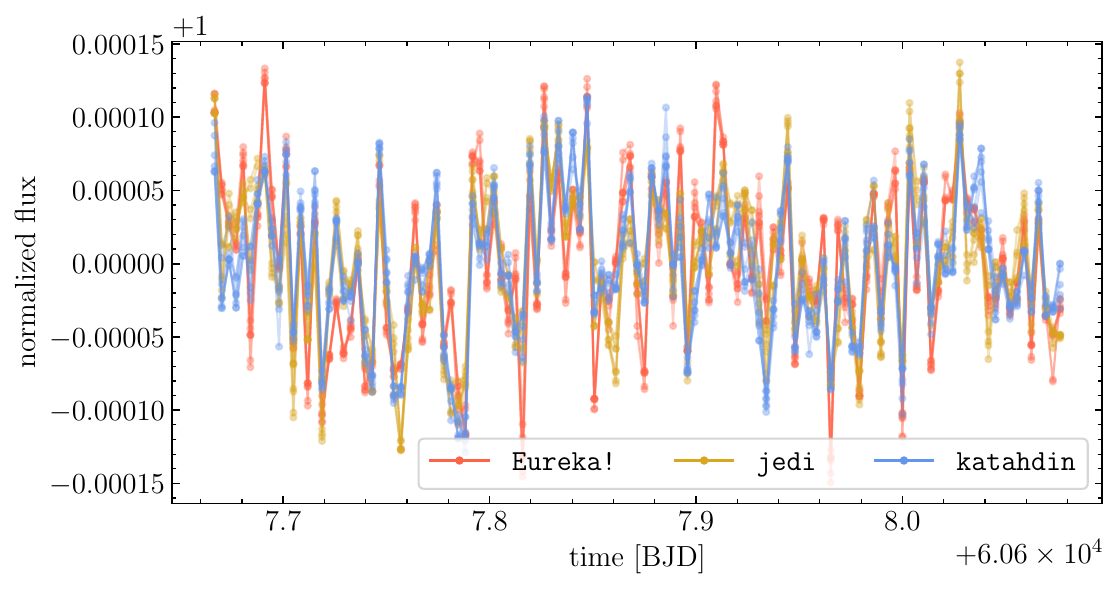}
    \caption{A zoom-in on exposure 1 of all 18 5-minute light curves after dividing out the best-fit quadratic trend from each. Note that all reductions tend to agree on the overall structure of the time series, and that reductions derived from the same pipeline tend to agree with each other.}
    \label{fig:exp_1_zoom}
\end{figure*}

\begin{figure*}
    \centering
    \includegraphics[width=0.9\linewidth]{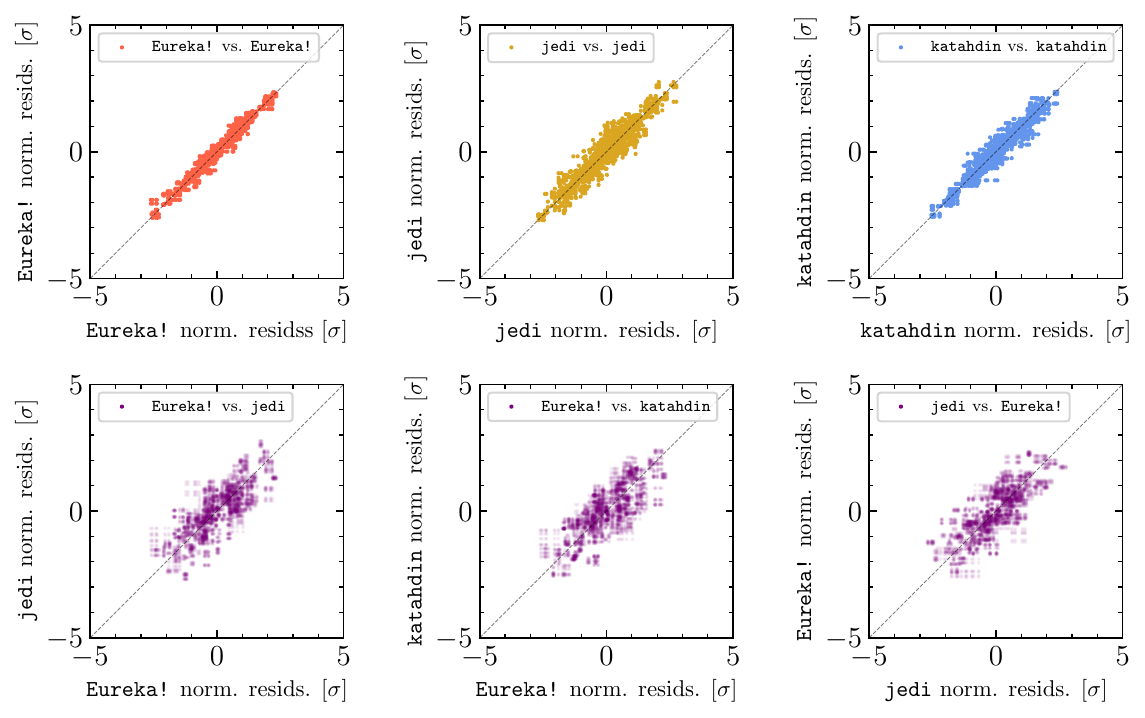}
    \caption{Correlations between the different light curves plotted in Fig. \ref{fig:exp_1_zoom}. Each point represents a comparison between two different reductions at a single data point; the panels in the top row are limited to comparisons between reductions that used different settings of the same underlying pipeline, while the panels in the bottom row show comparisons between reductions created with different pipelines. As noted, the three pipelines tend to agree, and light curves derived from the same pipeline using different settings are even more strongly correlated.}
    \label{fig:wlc_correlations}
\end{figure*}

Here we take the opportunity to discuss the white light curves themselves absent any reference to their influence on our fit results. Fig. \ref{fig:wlcs} shows all 18 reductions plotted in full in their 5-minute cadence forms, each normalized by the median of flux in exposure 1. Fig. \ref{fig:exp_1_zoom} shows a zoomed-in view of just exposure 1 after dividing out the best-fit quadratic trend from each time series. Two immediate features jump out here: first, light curves that were produced by the same pipeline tend to group together, regardless of their individual pipeline settings. Secondly, all three pipelines tend to agree with one another. Fig. \ref{fig:wlc_correlations} validates both of these conclusions by plotting the correlations between white light curves derived from the same pipeline vs. those from two different pipelines: both cases are positively correlated, but the intra-pipeline distributions are tighter.

\subsection{Limb Darkening Models} \label{sub:LDmodels}

\begin{figure}
    \centering
    \includegraphics[width=0.9\linewidth]{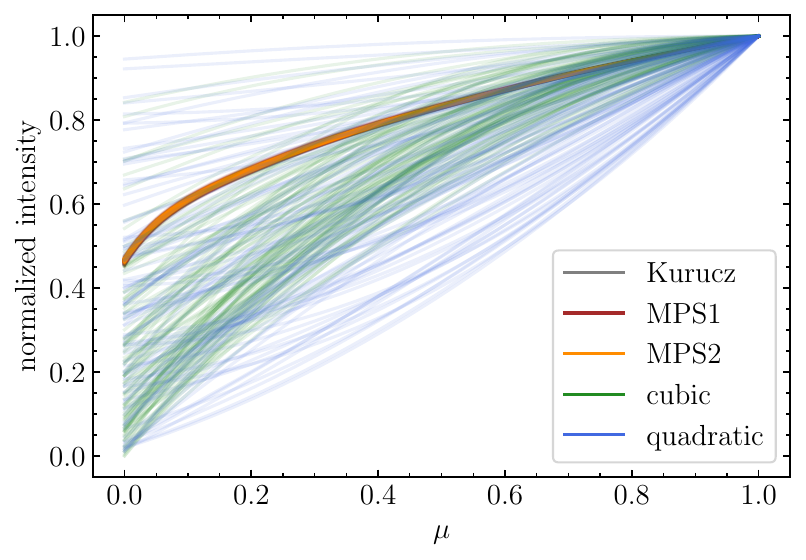}
    \caption{Draws from the priors of the five different limb darkening models. Note that all three of the physical models land essentially on top of one another, but there is still a finite width to their distribution.}
    \label{fig:ld_prior_draws}
\end{figure}

Here we describe our three different limb darkening frameworks and resulting 5 different limb darkening models. Draws from each of the 5 priors are illustrated shown in Fig. \ref{fig:ld_prior_draws}.

\subsubsection{Uninformative Quadratic} \label{subsub:quad_ld}

Our first limb darkening model is the most commonly used in Bayesian analyses of exoplanet transits: ``uninformative'' quadratic limb darkening. This is a favorite of the community due to a) its admission of analytic solutions to the transit problem and b) the ease of placing priors on the coefficients for Bayesian analysis. By re-parameterizing the $u$-coefficients via the transformation in \citet{kipping_quad_ld_2013}, one can sample all positive and monotonic quadratic profiles simply by sampling uniform distributions bounded between 0 and 1.

Quadratic limb darkening does come with drawbacks, however. For one, it struggles to capture theoretical limb darkening profiles generated from stellar atmosphere models; these tend to be much better expressed with nonlinear models such as the 4-parameter nonlinear function from \citet{claret_2000}. For another, even if one does not want to base their limb darkening profile on theoretical models, quadratic profiles are relatively inflexible at the limb of the star and are incapable of expressing steep curvature. For an example of this, see Fig. \ref{fig:asterias_varied_poly_order}, which demonstrates that the quadratic approximation to a stellar atmosphere model generated for our target/bandpass can produce spurious spikes at ingress/egress with magnitudes comparable to our oblateness signal of interest. Since our goal is a confident detection of a signal that presents itself as the planet crosses this region, we explored alternative parameterizations in addition to this standard model.

\subsubsection{Uninformative Cubic}
\label{subsub:cubic_ld}

Motivated by the shortcomings of quadratic limb darkening, we also included a more flexible model in our grid: cubic polynomials. Cubics present several practical challenges: for one, one of the most commonly used transit models, \citet{mandel_agol_2002}, is explicitly built around quadratics. This is thankfully circumvented by using a modeling package such as \sqp that implements the arbitrary-order limb darkening framework presented in \citet{agol_luger_dfm_2020}.

Even with the ability to forward model them, however, it is not immediately clear how to impose priors on cubic coefficients. \citet{kipping_cube_ld_2016} showed that when assuming 1) the profile is positive everywhere, 2) the profile is monotonically decreasing from the center to the limb, and 3) the curl is negative at the limb, it is possible to re-parameterize the coefficients such that when sampling from three uniform distributions, 94.4\% of the allowable parameters are accessible, though 2.6\% of all draws will violate one of the three criteria.

Rather than implement this approach, we chose instead to use a novel technique that sacrifices fitting one extra parameter in order to avoid rejection sampling and access the full parameter space. More details can be found in \citet{hattori_2025} (in prep) which explores techniques for sampling higher-order polynomials. To briefly summarize, we sample from a 4D Dirichlet distribution (practically, during sampling, we independently sample 4 gamma distributions with $\alpha=1$, then normalize). We then multiply each of these vectors with the following matrix:  [[0, 0, 0], [1, 0, 0], [0, 0, 1], [0, 1.5, -0.5]]. The resulting 3D vectors are polynomial coefficients that create profiles guaranteed to be positive everywhere, monotonic, and concave.

\subsubsection{Physically Informed Limb Darkening} \label{subsub:asterias}

While above we described an alternative to quadratic limb darkening that is comparatively more flexible, here we we present another alternative that takes a different approach. Instead of maximizing a model's flexibility, we now instead prioritize the model's interpretability by fitting for physical characteristics of the host star rather than coefficients in an abstract model.

To do this, we start by downloading a subset of the pre-computed stellar atmosphere models that underpin \texttt{exotic-ld} \citep{exotic_ld}, an open-source package for computing theoretical limb darkening profiles. Each file corresponds to a star with a certain $T_{\mathrm{eff}}$, log\,$g$, and metallicity; each file contains several spectra of that star corresponding to different distances from the projected center. We also download the appropriate instrument response curve for our specific filter/grating combination from \texttt{exotic-ld}.

We then diverge from \texttt{exotic-ld}, which expects users to supply a single value for $T_{\mathrm{eff}}$, log\,$g$, and metallicity. That package then finds either the closest-matching star in a given model grid or the weights of the linear combination that best interpolates to that star. \texttt{exotic-ld} then integrates the (weighted combination of) stellar spectra convolved with the instrument sensitivity to arrive at an empirical intensity profile, or how the brightness of the star changes as a function of distance to the projected center. The code then approximates this empirical profile with a user-supplied functional form, which can be quadratic, nonlinear-4 parameter, square root, or any other function to land on the final parametric limb darkening model.

By contrast, we instead integrate \textit{every} star in a stellar model grid, or at least the stars that may plausibly resemble our target. We then store these empirical profiles as \jax arrays. Then, during sampling, we draw a vector of $T_{\mathrm{eff}}$, log\,$g$, and metallicity, interpolate between the grid of empirical profiles, then fit an 8th-order polynomial to the resulting points. To avoid overfitting, we first interpolate the empirical profile with a monotonic cubic spline and resample it on a dense grid.

We use a high-order polynomial instead of a more standard nonlinear model because \sqp is built on an expanded implementation of the transit algorithm presented in \citet{agol_luger_dfm_2020}. This presented an analytic solution to the amount of flux blocked by a spherical planet crossing in front of a star with arbitrary-order limb darkening. The driving motivation behind \sqp's development was to extend this algorithm for oblate planets via use of numerical integration where necessary precisely so that one could approximate many different limb darkening models while fitting oblate transits.

\begin{figure}
    \centering
    \includegraphics[width=0.75\linewidth]{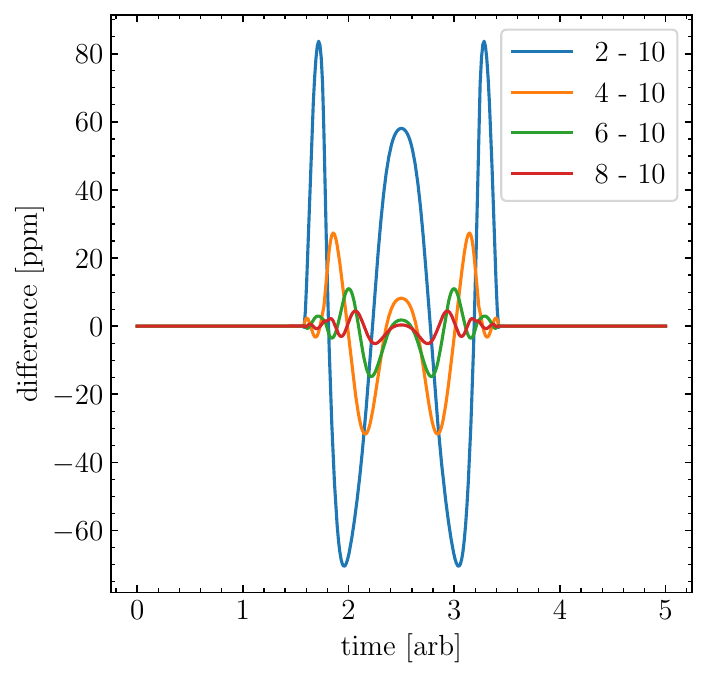}
    \caption{The impact of limb darkening polynomial order on transit light curves. We first simulated a transit of a \kep-sized planet across a Kepler-167-like star, using a 10th order polynomial approximation to the stellar intensity profile generated using the Kurucz stellar grid. We then simulated 4 other transits using successively lower polynomial orders for the stellar intensity approximation, and differenced the resulting transits with the fiducial one. The differenced curves, shown above for 2nd-10th order, 4th-10th order, etc, reveal that low-order approximations impart residuals of 10s of ppm over the course of the transit.}
    \label{fig:asterias_varied_poly_order}
\end{figure}

Fig. \ref{fig:asterias_varied_poly_order} demonstrates why such a high-order is necessary for the polynomial when dealing with physically-informed models that curve dramatically near the limb. We first generated a transit for a fiducial planet and a 10th order approximation to the interpolated profile. We then fit this fake, noiseless light curve with models that were limited to successively lower-order polynomial limb darkening models. Note that low polynomial orders, especially the quadratic approximation, impart residuals on ingress/egress that has an amplitude comparable to our signal of interest.

One may fairly ask in all of this why we do not simply fix the limb darkening coefficients before a fit as is occasionally done elsewhere in the literature \citep[e.g.][]{alam_hst_spectrum_2022}. Doing so, however, runs the risk that a slight misparameterization could irreparably inject or suppress an oblateness signal \citep[and/or alter a planet's spectrum,][]{mercier_2025}. Further, even if one were to fully trust a given stellar atmosphere model \citep[which would be unwise, given the evidence that they do not perfectly represent observed empirical profiles][]{espinoza_ld_2015, patel_ld_2022}, every star comes with uncertainties on its physical characteristics as well.

\begin{figure}
    \centering
    \includegraphics[width=0.75\linewidth]{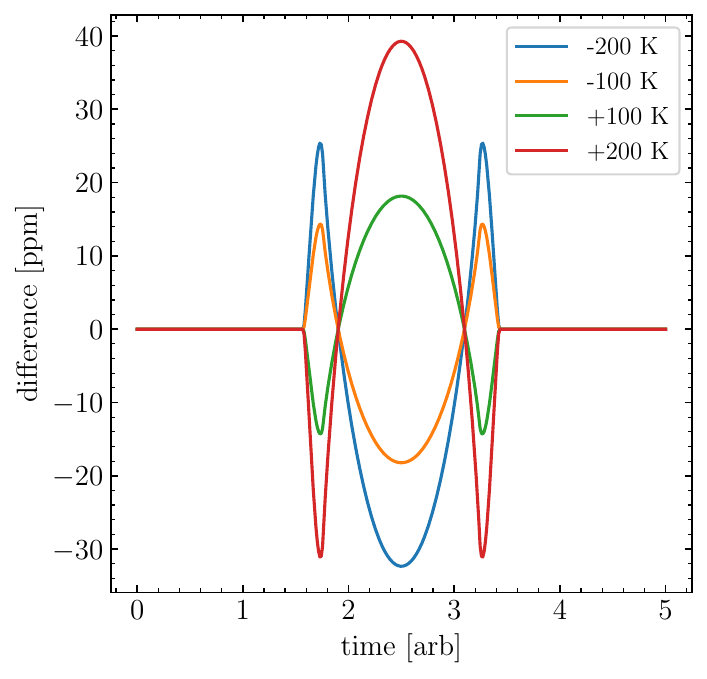}
    \caption{Similar to Fig. \ref{fig:asterias_varied_poly_order}, but now showing the difference between transits across stars with varying stellar temperature.}
    \label{fig:asterias_varied_temp}
\end{figure}

The implications of this are demonstrated in Fig. \ref{fig:asterias_varied_temp}. Here we created a fiducial transit using the mean parameters of \kep and the stellar model grid from \citet{kurucz_grid}. We then ran several additional models where we fixed the stellar parameters along a grid of varying temperatures and differenced the pairs of light curves. Note that that the discrepancies reach a maximum on the limbs, exactly where an oblateness signal will manifest.

The procedure laid out above where we fit stellar parameters rather than heuristic function coefficients offers several practical advantages. First, one can naturally assign target-specific priors to the coefficients that control limb darkening: these are simply the mean and uncertainties on $T_{\mathrm{eff}}$, log\,$g$, and metallicity. Secondly, since all of the grid profiles can be pre-computed, we can store these grids in memory and avoid any file I/O or numerical integrations during runtime. In fact, since we have implemented this framework in \jax, the whole algorithm is autodifferentiable: one can take the gradient of the limb darkening coefficients with respect to the stellar properties.

We have assembled this routine into an open-source python/\jax package we call \asterias, after the Latin genus of the common star fish, and hope that the community finds it useful for modeling the limbs of stars. We emphasize that \asterias is built on the work of \citet{exotic_ld} and their efforts to compile stellar atmosphere model outputs into standardized file types, and that any efforts that rely on \asterias should also cite \citet{exotic_ld}.

In this study we used the ``Kurucz'' grid  of stellar atmosphere models \citep{kurucz_grid} and the two MPS grids of stellar atmosphere models \citep{mps_grid_a, mps_grid_b}. We opted not to use the STAGGER 3D grid \citep{stagger_grid} since its non-uniform grid spacing interfered with our interpolation routines, or the Phoenix grid \citep{phoenix_grid} since its unique treatment of the very edge of the star did not mesh well with the others. All combinations of pipeline/trend models were passed through all three of these models.

\subsection{Trend Models} \label{sub:trend_models}
As seen in Figs. \ref{fig:wlcs}, our data display a general downwards trend over the course of each exposure, along with offsets between each exposure with amplitudes that vary across our reduction pipelines. This section describes the three parametric forms we tested to describe this trend. Note that unlike Paper II, all trend parameters were treated like all other parameters in the model and that we did not employ any profile likelihood.

\subsubsection{Linear} \label{subsub:lin_trend}

Our first trend model is a simple linear trend in time, where each of the five exposures get their own unique intercept and slope. Linear trends are common components of systematics models used in other studies \citep[e.g.][]{rustamkulov_ers_prism_2023, teske_compass_2025, alam_compass_2025}. Even visually in Fig. \ref{fig:wlcs}, however, it is clear that our data are not well described by a linear trend alone. We include it here to test pairings with different overlying Gaussian Processes.

Note that we do not include a quadratic trend as initial tests revealed that the best-fitting quadratics were often not monotonic and/or concave. As discussed in Paper II, this tendency for the trends of the in-transit exposures to reverse direction mid exposure led us to abandon fully unrestricted quadratics in this article. Experiments with restricted quadratics (e.g. enforcing monotonicity during fitting) produced similar results to the exponential model described below but at the cost of additional free parameters.

\subsubsection{Exponential} \label{subsub:exp_trend}
Our second trend model is an exponential decay. We parameterize the trend with the following form:

\begin{equation}
    f_i(t) = \left(\frac{d_i e^\tau}{e^\tau - 1} \right)e^{-\tau t} + b_i
\end{equation}

where $d_i$ parameterizes the total ``drop'' of each exposure $i$, $b_i$ the starting height of each exposure $i$, and $\tau$ the length scale of the exponential drop off in the trend. Time $t$ is normalized between 0 and 1, and the prefactor is derived by evaluating the exponential at $t=0$ and $t=1$, which produces the apparently suppressed units. For small $\tau$, this reduces to a linear model, while increasing $\tau$ forces a more aggressive exponential decay. We choose to fit for $\tau$ with large priors, but enforce that the same $\tau$ is shared across all 5 exposures. This choice was driven by initial testing that revealed letting $\tau$ vary with each exposure produced very similar values for exposures 1, 2, and 5, but markedly different values for exposures 3 and 4, where it is degenerate with the planetary parameters. By using only one $\tau$, we also add only one extra free parameter to the model compared to the linear trend. Note, however, that Sec. \ref{sec:nested_sampling} includes a discussion on leaving out exposures 1,2, and 5 to investigate whether their contributions to $\tau$ might drive the inference on $f$, but that this had little effect. Note also that this model is distinct from the exponential model used in Paper II, which fixes $\tau$ but includes an additional linear component in time.

\subsubsection{Constant (GP-only)} \label{subsub:const_trend}

Our final trend model is simply a flat constant for each of the 5 exposures. Looking to Figs. \ref{fig:wlcs}, it is clear that this is a fairly terrible choice in isolation. However, when paired with one of the Gaussian Processes described in Sec. \ref{sub:gps}, this essentially gives the GP full control over both short and long term trends. Since this trend model also has 5 fewer parameters than the linear model and 6 fewer than the exponential, it is in principle possible that it ends up the preferred choice by way of model simplicity.

\subsubsection{Roads Not Taken} \label{subsub:trends_not_used}

In addition to the trend in time, it is common for other studies to include various other regressors in their systematics models as well, such as the time series of trace centroid movements \citep[again, e.g.][]{rustamkulov_ers_prism_2023, teske_compass_2025, alam_compass_2025} or even simultaneously collected engineering telemetry data \citep{wallack_compass_2024}. However, initial experimentation with these quantities, and with guide star data, did not dramatically improve the quality of our fits. Consequently, we instead choose to use a Gaussian Process on top of our mean model to capture short-term correlations in time rather than rely on any specific external regressor.

Additionally, as discussed further in Paper II, we note that observations such as these that are composed of multiple exposures present a unique challenge compared to observations that fit entirely between two detector resets. Since the number of free parameters generally tracks the number of exposures, our choice of trend models was relatively constrained in order to keep the final dimensionality of our models tractable. To illustrate this, our linear trends required fitting an additional 10 free parameters (2 for each exposure), already more parameters than the astrophysical parameters driving the transit shape. Paper II offers a potential solution to this limitation in the form of profile likelihood, though as mentioned, that technique was not used in this study.

\subsection{Gaussian Processes} \label{sub:gps}

To augment the mean trend models described above, we also tested 3 overlying Gaussian Processes to complete our trend model. These were each simple, single-kernel GPs implemented via the \texttt{tinygp} package \citep{tinygp}.

\subsubsection{Exponential Squared}

The first GP is a simple exponential squared kernel, parameterized by an amplitude, length scale, and an additional ``jitter'' term add to the diagonal of the covariance matrix. Each free parameter was fit in log space to avoid nonphysical combinations.

\subsubsection{Matérn 3/2}

The second GP is a simple Matérn 3/2 kernel, again parameterized by an amplitude, length scale, and jitter term. This kernel decays faster with lag than the exponential squared kernel, meaning it has a stronger emphasis on local correlations. Again, we fit each parameter in log space.

\subsection{None}

The final GP isn't a GP at all: instead, we simply use the mean models from Sec. \ref{sub:trend_models} on their own as the entire trend. This allowed us to probe the GP's influence on the correlated residuals. We did not fit the constant mean model with no GP in light of the obvious downwards trend in each exposure.

\section{MLE Results} \label{sec:mle_results}

\subsection{MLE Fitting Procedure} \label{sub:mle_fitting}

By writing our forward model in an entirely autodifferentiable framework, we were able to leverage robust gradient descent algorithms for this optimization even in relatively high-dimensional ($n>20$) spaces. We implemented each forward model described above in \jax \citep{jax}, an autodifferentiable framework developed and maintained by Google. This allowed us to take the gradient of our likelihood function with respect to each of our free parameters, which in turn allowed us to use a gradient-informed optimization algorithm. We emphasize that all parts of the model, including \sqp's numerical integrals and \asterias' stellar model interpolation, were all included in this differentiable toolchain.

Practically, though, we were interested in maximum \textit{likelihood} solutions, not maximum \textit{a-posteriori} solutions, we still involved a prior in the fit to simply help keep all parameters near unity until a final transformation back to physical quantities. More specifically, we fit each parameter in an unconstrained space, then used a sigmoid to map these vectors into the unit cube, then used the transformation from these cube to physical values using the prior transformation described later in Sec. \ref{sub:priors}. By following this procedure we allowed the optimizer to explore all valid parameter combinations while ensuring the gradients remained numerically stable and the parameter values remained within physical limits. 

Each model consisted of at least the radius ratio $R_p/R_s$, impact parameter $b$, time of mid transit $T_0$, projected oblateness $f$, projected ellipse rotation $\theta$, and separation $a/R_s$. Note that for these MLE fits, we fixed the period to a fiducial value of 1071.23205 days from \citet{chachan_rvs_2022}, then fit in $a/R_s$, rather than in pseudo stellar density $\rho_*$ and period $P_\text{orb}$, as we do in Sec \ref{sec:nested_sampling}. In addition to these parameters, each model has 2-4 limb darkening parameters, 5-11 trend parameters, and 0 or 3 GP parameters.

We used several rounds of the ADAM optimizer \citep{adam} with a learning rate of 0.01, followed by a final round with a learning rate of 0.001. Optimization was terminated when the difference in likelihood values between 10 iterations fell below 0.01 or the cluster job ran out of time. As discussed earlier, we performed 3 fits for each light curve/model in the grid of 4320 combinations, then retained the maximum likelihood value of each trio. In total, 4293 of the models reached acceptable convergence.

Finally, we note that though our grid includes light curves with different cadences, when generating forward models, we did not directly use each light curve's underlying time stamps. Rather, since ``binning is sinning'' \citep{kipping_binning_sinning_2010}, we actually generated light curves that were over-sampled by factors ranging from 1 to 9 depending on the cadence. When comparing to the actual data, we then integrated over this denser time series using the correction from \citet{starry}. The over-sample factors were chosen through numerical experiments which drew planet parameters from the prior, generated model light curves from the prior that were over-sampled by a factor of 15, then generated additional light curves at various other over-sample factors. These additional curves were then differenced with the high-resolution curve, and we selected the smallest factor for which the difference between the two never exceeded 5\,ppm.

\subsection{Intra-light curve results}

To draw any guiding conclusions on what trend, limb darkening, or Gaussian Process we should use going forwards, we need to ensure that we are fairly comparing only legal pairs of fits from our grid. Model selection criteria such as the BIC and AIC assume that models are fit to the same underlying data, and this is not necessarily the case when it comes to our grid: each unique reduction/cadence combination needs to be treated as its own data set, meaning that for each trend, limb darkening, and GP combination, we can compute 108 different BICs, AICs, $\chi^2$, etc.

We note that BIC and AIC are imperfect model comparison frameworks as they assume the underlying likelihood surface is Gaussian. As shown at in Fig. \ref{fig:rep_corner}, which shows a representative posterior, and as discussed in \citet{dholakia_2025}, the posterior of an oblateness fit is not well described by a Gaussian. However, we accept this imperfection for the purposes of this initial model down-select. We performed the same optimization described above but held out random subsets of data to examine the out-of-sample RMS, and found that the resulting patterns were much noisier than the BIC and AIC patterns described below.

We also note that the point-estimates found by our MLE solver occasionally fell far from the credible ranges of the posteriors described in Sec. \ref{sec:nested_sampling}. This is in part because these fits optimized the likelihood, not the posterior, which allowed them to explore regions of parameter space at the edges of the prior. The physical limb darkening fits in particular often landed on point-esimtates where the temperature and metallicity of the star fell $>5\sigma$ from the prior mean. This is also in part, however, because in these relatively high-dimensional spaces there is no guarantee that the maximum likelihood or posterior mode are within the typical set. By using a gradient-based optimizer capable of climbing narrow likelihood spikes, we can land in regions with high posterior density but negligible prior volume.

Under this framework, we draw the following conclusions.

\begin{figure*}
    \centering
    \includegraphics[width=0.83\linewidth]{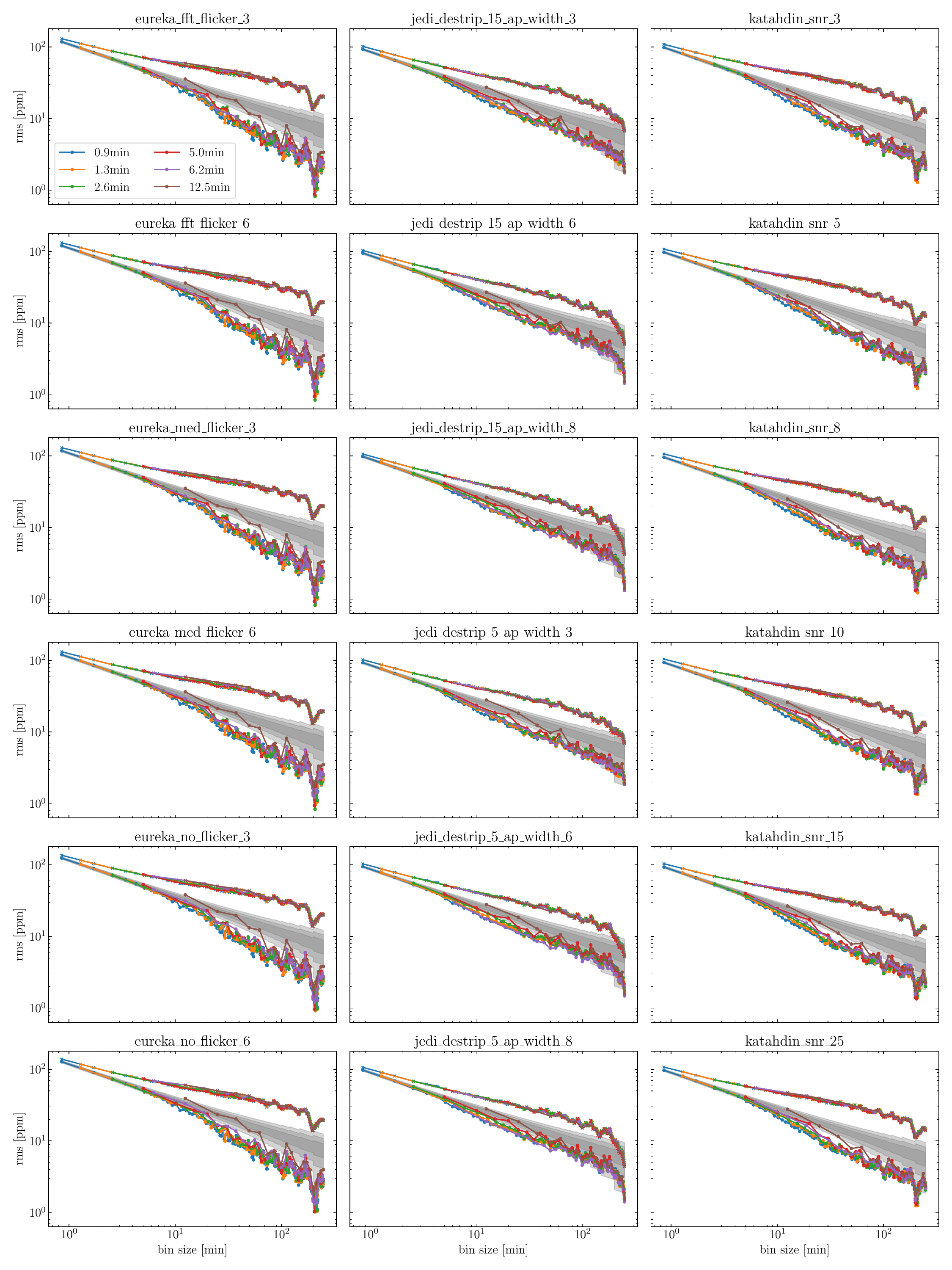}
    \caption{Bin size vs. rms plots for a subset of the best-fitting models from our grid. This displays one line for every fit that used an exponential trend and quadratic limb darkening, then either an \mth kernel (dots) or no overlying GP (crosses). The $1\sigma, 2\sigma,$ and $3\sigma$ contours for expected $\sqrt{N}$ binning are shown via grey contours. We see that the residuals of the GP fits consistently bin down faster than white noise after the bin size exceeds roughly 10 minutes, but that the no-GP fits immediately diverge from white noise behavior. We also note that the cadence the fit was performed on does not seem to influence residual binning behavior.} 
    \label{fig:mle_rms}
\end{figure*}

\textbf{A Gaussian Process is favored but may overfit.}
For each of the 108 unique light curve/cadence combinations, the models with no GP were consistently the last-ranked model no matter what standard model selection framework you choose. BIC, AIC, \dchi, MAD, and RMS all rank any model with “none” below any model choice that uses either a \mth or exponential squared kernel. However, as shown in Fig. \ref{fig:mle_rms}, the residuals of fits using a GP tend to bin down faster than expected for white noise once the bin size reaches approximately 10 minutes. When not using a GP, however, the residuals display correlated noise and bin down worse than white noise immediately. This indicates that our deterministic trend model is not sufficient for ``whitening'' the data on its own.

\textbf{Initial light curve binning has limited influence.}
As seen in Fig. \ref{fig:mle_rms}, fits to lower-cadence binned light curves (e.g. 5 minutes) end up with residual RMS that closely resembles $\sqrt{N}$ binning of the residuals for fits to higher-cadence light curves (e.g. 0.9 min). This is reassuring, since it was possible that different binning schemes could have exposed or suppressed correlated noise. Indeed, as shown in Fig. \ref{fig:gp_scales}, which shows the best-fitting length scale for a subset of fits, the different cadences do seem to prefer slightly different length scales. However, moving forwards we likely need to consider only one cadence.

\textbf{Exponential with shared $\tau$ is the best trend.}
For each of the 108 unique light curve/cadence combinations, the top 8 models all use an exponential trend when using either AIC or BIC as a model selection criteria. In fact, when using AIC, there is always a strict ranking among trend types when restricting ourselves to models that use a Gaussian Process: all exponential models outrank any model using a linear trend, and all linear trend models outrank any model using a constant trend where the GP is responsible for all short and long term correlations. There are a few exceptions to this strict pattern when using BIC, which applies a harsher penalty for the exponential trend's extra free parameter than the AIC does, but the general pattern is clear.

\textbf{\mth kernels are preferred over exponential squared kernels.}
When comparing like-for-like models that only differ according to their GP type, both AIC and BIC prefer the \mth kernel in 93\% of cases. Digging a little deeper, we note an interesting behavior: though models using exponential squared kernels occasionally outperform their \mth counterparts and can reach similar likelihood values, they can only do so by increasing the ``jitter'' term in the GP. In other words, models using exponential squared kernels consistently use larger implied error bars than models using \mth kernels, and as a consequence, have much worse $\chi^2$, RMS, and MAD behaviors. This is illustrated in Fig. \ref{fig:expsq_jitter}, which shows the distribution of $\Delta \chi^2$ between \mth and exponential squared models, and distribution of the ratio of the best-fit jitter terms between the two.

\begin{figure}
    \centering
    \includegraphics[width=\linewidth]{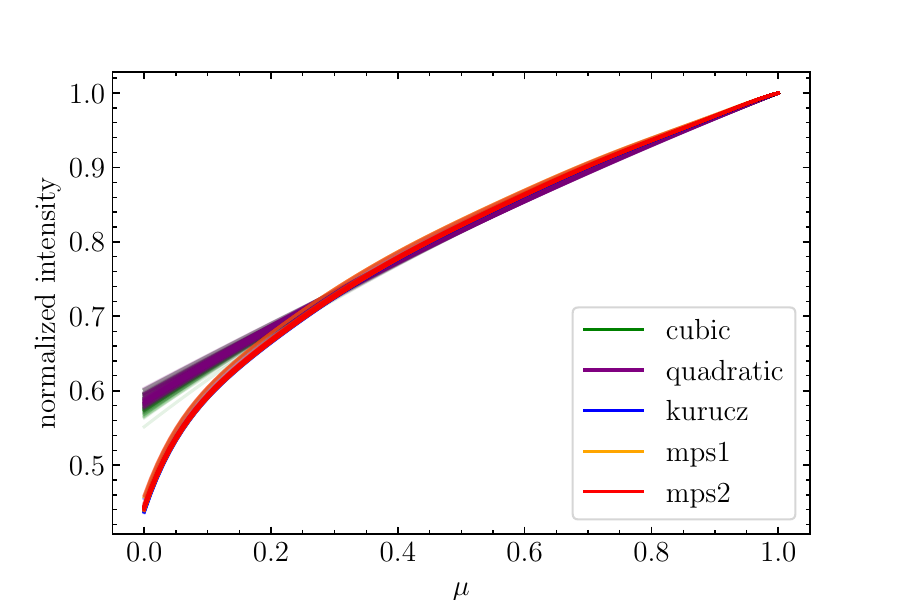}
    \caption{The MLE limb darkening profiles for all models using an exponential trend and a \mth GP kernel. Note that all three of the physical models stack on top of one another, while the quadratic and cubic models agree with each other and prefer a gentler curve on the limb.} 
    \label{fig:mle_ld_profiles}
\end{figure}

\textbf{Quadratic and cubic limb darkening agree well.}
Fig. \ref{fig:mle_ld_profiles} show the MLE limb darkening profiles for all fits that used an exponential trend and a \mth GP kernel. We see that despite the additional flexibility of the cubic parameterization, all best-fit cubics agree quite well with the best-fit quadratics. Note that not every quadratic profile can be represented by a cubic profile in our parameterization: we have imposed that the cubics must be monotonic and concave, while the quadratics need only be monotonic.

\subsection{Inter-light curve results}

\begin{figure}
    \centering
    \includegraphics[width=0.9\linewidth]{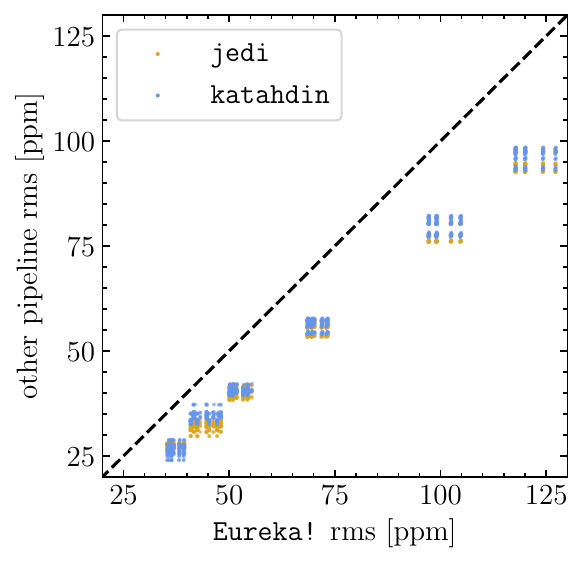}
    \caption{Comparison between the RMS of the residuals for fits that used the \eureka pipeline and fits that used either \jedi or \kat. Each point represents a like-for-like comparison where every modeling choice is the same (cadence, trend type, GP kernel, etc) except for the reduction used.} 
    \label{fig:eureka_underperform}
\end{figure}

While we can't use model comparison metrics like BIC or AIC to compare across pipelines/cadences, we did note one consistent trend when examining the residuals of the best-fitting light curves: \textbf{\texttt{Eureka!} underperforms \texttt{ExoTiC-JEDI} and \texttt{katahdin}}. Fig. \ref{fig:eureka_underperform} shows the RMS residuals for every like-for-like comparison between each \eureka model and every \jedi and \kat model. We see that while \jedi and \kat tend to have a similar scatter to their residuals, the RMS of the residuals for \eureka fits always fall below the 1-1 line.

\begin{figure}
    \centering
    \includegraphics[width=0.9\linewidth]{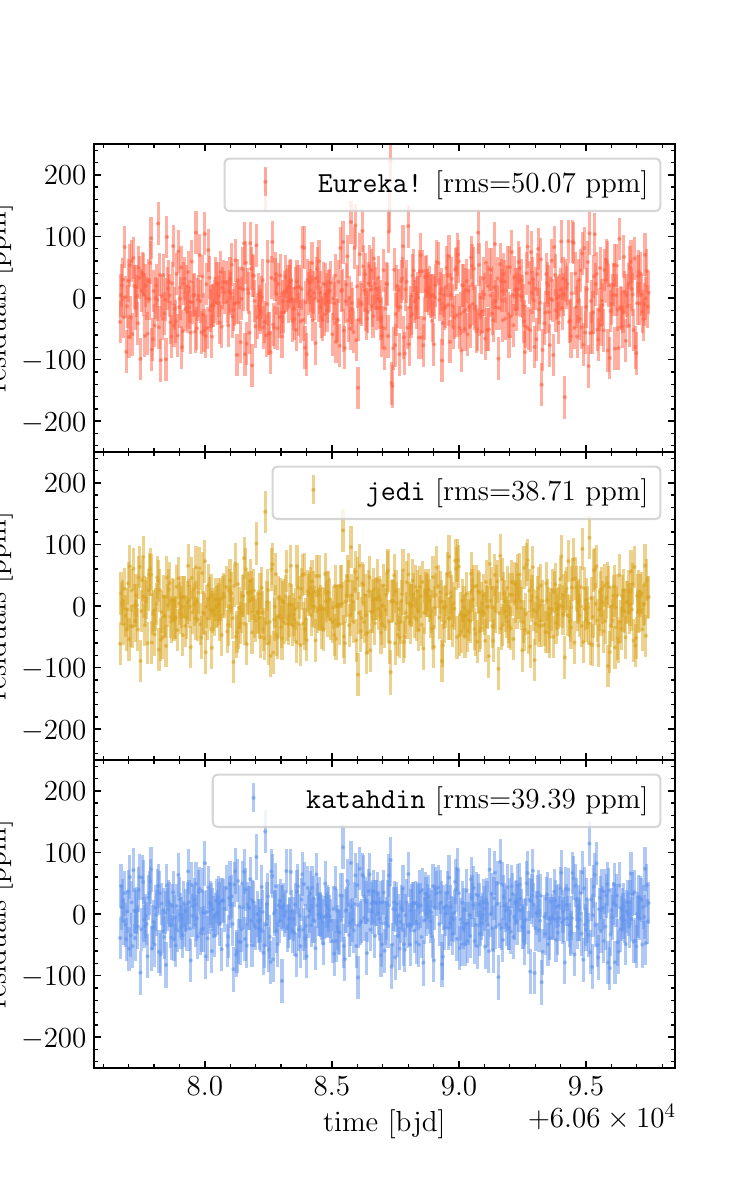}
    \caption{The time series of residuals for three fits from the MLE grid that differ only by which reduction was used. All fits here used an exponential trend, quadratic limb darkening, 5-minute cadence, and an \mth kernel, though we find that this pattern of \jedi and \kat achieving comparable residual RMS values while \eureka arrives at a solutions with higher scatter holds for all modeling combinations.} 
    \label{fig:example_mle_resids}
\end{figure}

This is also illustrated in Fig. \ref{fig:example_mle_resids}, which shows the time series of residuals for three fits that differ only by what reduction was used: the top panel used eureka\_fft\_flicker\_3, the middle jedi\_15\_3, and the bottom kathadin\_snr\_10. We see that the residuals to the \eureka fit have both a larger scatter and systematic excursions within the transit that are absent from both \jedi and \kat. It is for this reason that we do not consider \eureka in Paper II.

\subsection{MLE Takeaways} \label{sub:mle_takeaways}

The purpose of this grid of MLE fits was to inform what models and reductions to use for nested sampling fits. To that end, we make the following cuts from our grid of 4320 possibilities:

\begin{itemize}
    \item We will only consider \kat and \jedi reductions. This is motivated by the consistently higher scatter in residuals for fits to any reductions that use the \eureka pipeline.
    \item Since the bin size vs. RMS behavior appears to be similar for all tested cadences, we will consider only one cadence moving forwards. We choose to this to be 5 minutes to match Paper II.
    \item We will only use \mth kernels, not exponential squared kernels, as the latter produce significantly higher RMS residuals and require relatively inflated jitter values.
    \item Instead of considering all 3 physical limb darkening models, we will only retain the Kurucz grid. This is motivated by the close agreement between all physical models near the limb, and the fact that the Kurucz grid is the sparsest of the 3 tested and therefore the fastest to interpolate.
\end{itemize}

These cuts leave us with 72 unique combinations: 12 light curves (6 each for \jedi and \kat), 3 limb darkening treatments, and 2 planet shapes (oblate or spherical).

Following this MLE grid of fits, we experimented with a more aggressive outlier rejection scheme when extracting \kat light curves: instead of simply summing all pixels that fell above a certain SNR threshold, we first rejected all pixels whose variances or median signal value was more than a factor of 2 higher or lower than both of its row neighbors. We applied this filtering to SNR thresholds 8, 10, 15, and 25, leaving us with actually 60 unique combinations instead of the full 72.

\section{Nested Sampling} \label{sec:nested_sampling}

\subsection{Priors} \label{sub:priors}

Our Bayesian analysis requires placing priors on each of our parameters of interest. In general, we opted for wide, flat priors that encompassed reasonable posterior ranges. We made four exceptions to this practice. Firstly, for the physical limb-darkening model, we used the distributions for $T_{\mathrm{eff}}$, log\,$g$, and metallicity presented in \citet{chachan_rvs_2022}'s analysis of a high-resolution spectrum of Kepler-167 A. 

Secondly, when modeling $f$, the oblateness of the planet, we assumed that the distribution of \textit{true} oblateness was uniform between 0 and 0.5. Above this upper bound, the centrifugal forces at the planet's equator would overpower the gravitational attraction and the planet would come unbound \citep{berardo_ob_effects_2022}. However, when accounting for a planet's precession/obliquity, a planet's \textit{projected} oblateness can only by equal to \textit{or less than} its true oblateness. \citet{berardo_ob_effects_2022} show that if one assumes that planets' rotation axes are uniformly distributed on the sphere, the median projected $f$ is 70-75\% of the true $f$. 

This choice to focus on true rather than projected, oblateness may or may not be a fair assumption. On the one hand, the planets within our solar system exhibit a wide range of obliquities (from Mercury at nearly 0.0\degr to Uranus at nearly 90\degr), but on the other, one could conjure dynamical arguments that planets may be preferentially spin aligned with the their orbits \citep[such as Mercury's spin-orbit resonance,][ though this particular mechanism would not be at play for this planet]{murray_dermott}. Assuming a random distribution of orientations, however, is a more conservative choice than assuming perfect orbital alignment.

\begin{figure}
    \centering
    \includegraphics[width=0.9\linewidth]{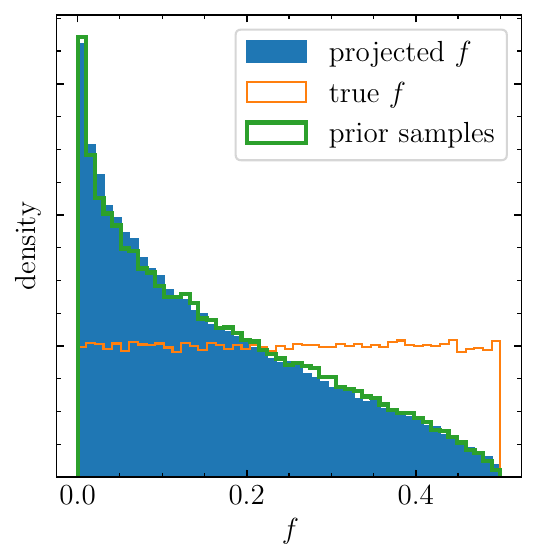}
    \caption{Illustration of the prior for the projected flattening used in the nested sampling fits.}
    \label{fig:f_prior}
\end{figure}

To account for this suppression of large projected $f$s, we used \sqp to generate 100,000 planets with random pole orientations and true $f$'s uniformly distributed between 0 and 0.5. The resulting distribution was well-fit by a Beta distribution with $\alpha=0.8$, $\beta=1.66$, which we adopted at our prior for projected $f$ in all following fits. This procedure is illustrated in Fig. \ref{fig:f_prior}.

Thirdly, for pseudo stellar density $\rho_*$, we use the same prior as in Paper II that was derived via fits to \kepler and Spitzer transits: $\mathcal{N}(2819.5, 114.6)$\,kg/m$^3$. Fourth and finally, we used the same distribution for the period $P_\text{orb}$ as in Paper II: $\mathcal{N}(1071.23325, 0.00055)$\,days.

The remaining free parameters are $R_p/R_s$, $b$, and $T_0$ for the planet, either $\{q_1, q_2\}$, $\{T_{eff}, \log(g)$, metallicity\}, or $\{u_1, u_2, u_3, u_4\}$ for limb darkening, \{5 constants, 5 slopes, and one $\tau$\} for the trend, and a length scale, amplitude, and jitter for the GP. Note that when fitting oblate planets, $R_p/R_s$ is the radius of a spherical planet with the same projected area as the oblate planet in question. This avoids the extreme correlation between the projected major axis radius and $f$, which when combined set the transit depth.

\subsection{Fitting Algorithm} \label{sub:nautilus}

In principle, we could have opted to use a Hamiltonian MCMC routine such as the NUTS algorithm in a package like \texttt{numpyro} \citep{NUTS, numpyro}, since our entire model (including the numerical integrations behind \sqp and the stellar atmosphere interpolation within \asterias) is autodifferentiable. However, we opted instead to use a nested sampling routine instead, as this allowed us to compute Bayesian evidences alongside our joint posteriors. See \citet{skilling_ns_2004} for the inception of nested sampling, \citet{buchner_nested_review_2023} for a recent review, and \citet{aston_ns_primer_2022} for an accessible primer.

The astronomical community has access to numerous nested sampling packages in 2025, including but not limited to \texttt{MultiNest} \citep{multinest_1, multinest_2, multinest_3}, \texttt{dynesty} \citep{dynesty}, and \texttt{UltraNest} \citep{ultranest}. Though in principle any of these options could produce our desired outputs, we opted to use \nautilus, a python-based nested sampling package \citep{lange_nautilus_2023}. During sampling, \nautilus trains a simple neural network to help determine regions of improved likelihood without having to call the expensive likelihood function as often; however, when computing the evidence, it always uses the true forward model, and the bias associated with constructing bounds in this way can be removed by re-running the sampler following initial exploration. We found that \nautilus converged much faster than alternative algorithms, and encourage those interested in the details of the algorithm to see \citet{lange_nautilus_2023}. In contrast, we opt to use \texttt{MultiNest} in Paper II, as there is significant literature precedent for this algorithm in the context of exomoon transit fitting \citep{kipping_2025_response}.

We first ran each of our 60 fits using 5,000 live points, a very conservative expansion above the default of 3,000. Sampling was terminated when the live points contained an estimated $<1\%$ of the evidence, at which time the contributions from the remaining live points were added without replacement. We used the default value of 1.1 for the \texttt{enlarge\_per\_dim} setting, which sets the expansion beyond the outer ellipsoidal bounds during sampling. We set the \texttt{discard\_exploration} flag to True, and on the second sampling pass continued until we accumulated at least 50,000 effective samples.\footnote{As shown in Tables \ref{tab:jedi_fits} and \ref{tab:kat_fits}, 4/60 fits did not reach this sample size due to termination via timeout. However, all jobs finished their initial exploration phase and reached at least 20,000 effective samples, still well above the \nautilus recommendation of 10,000.}

While one can run tests such as Gelman-Rubin statistic to check the convergence of MCMC runs, running diagnostics on nested sampling runs is less straightforward. \citet{fowlie_2020} suggest one such test, which in short checks that, when considering an ordered list of the likelihoods of all live points, the likelihood of new live points are consistently inserted at random indices. \citet{buchner_nested_review_2023} proposed an update to this test to correctly account for the discretized nature of the problem. However, these ad hoc tests still do not guarantee convergence. Instead, a more reassuring test one can run to validate a nested sampling run is simply to run it again with a different number of live points \citep{buchner_nested_review_2023, lange_nautilus_2023}. 

While we did not do this for every fit in our grid, we did perform one additional fit a that involved the most parameters (+2 for oblateness and shape, +4 for cubic limb darkening) using 10,000 live points. We found that the evidences computed between the 5,000 and 10,000 point runs differed by only 0.03, or 0.0002\%, which indicates our 5,000 point runs used a sufficiently high resolution.

\subsection{Nested Sampling Results} \label{sec:results}

Our highest-level takeaway from the nested sampling fits is this: when considering one light curve and using Bayes factor to perform model selection, the data are fit equally well by both spherical and oblate models. For all reductions, the Bayes factor for the best-fitting oblate and spherical models always falls below 10 \citep[the threshold for ``strong evidence'',][]{kass_1995}, indicating that the additional flexibility from adding extra parameters to the model neither helps achieve markedly better fits nor dramatically hurts the model via additional complexity. 

Furthermore, for most reductions, the data are fit equally well by all three of the tested limb darkening models. This is not the case for three \kat-based reductions (SNR thresholds 10, 15, and 25), where quadratic limb darkening is preferred over physical/Kurucz by a Bayes factor just over this threshold.

\begin{figure*}
    \centering
    \includegraphics[width=0.85\linewidth]{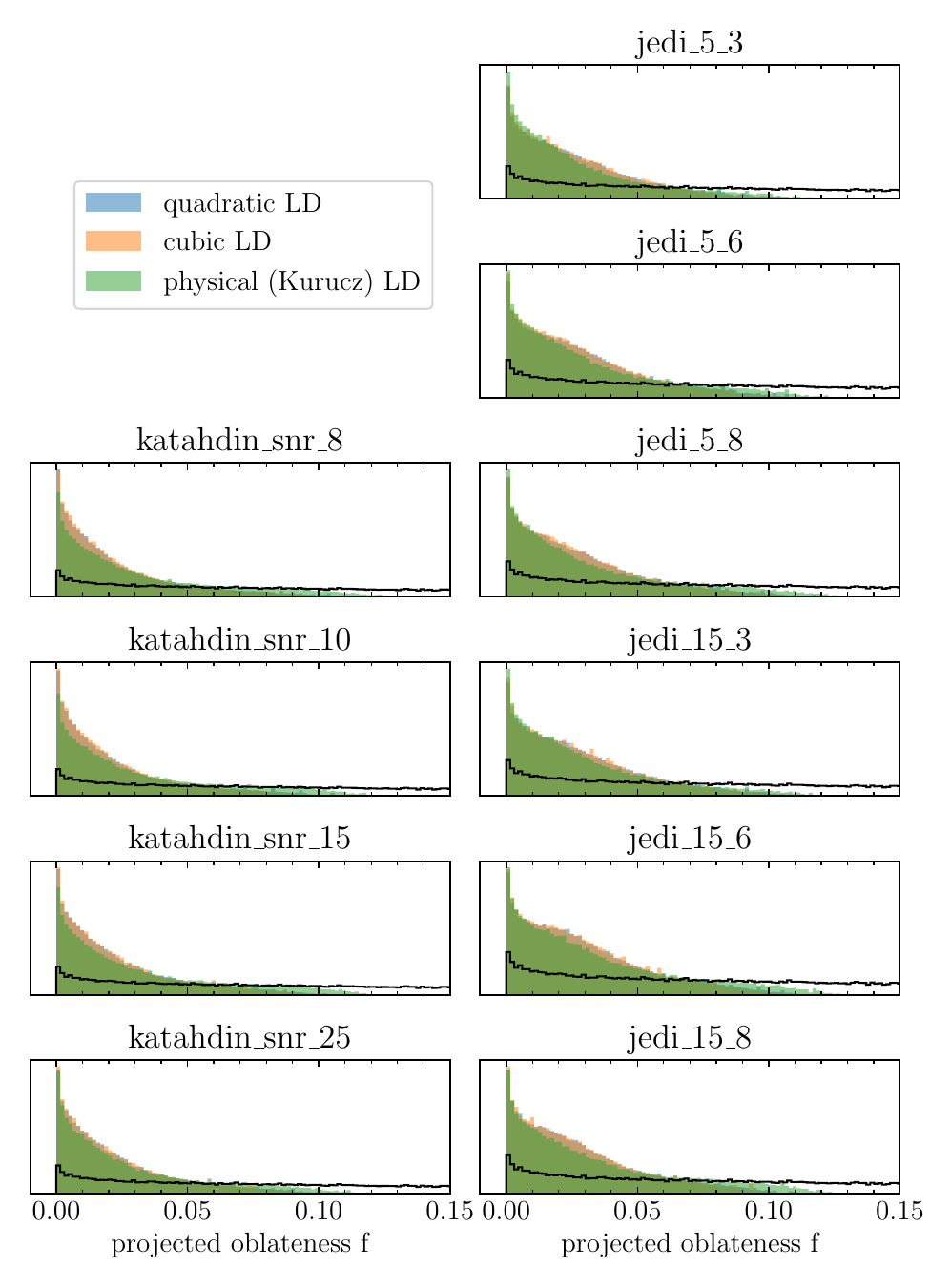}
    \caption{All of the marginalized $f$ posteriors from our grid of nested sampling fits. All panels use the same colors for quadratic, cubic, and physical limb darkening, and in all cases, the quadratic and cubic posteriors are nearly indistinguishable. The beta distribution used as a prior is overplotted in black in all panels.}
    \label{fig:nested_f_posts}
\end{figure*}

Summaries of each of the runs along with a representative full corner plot (Fig. \ref{fig:rep_corner}) can be found in Appendix. \ref{ap:nested}. The marginalized distributions for projected oblateness $f$ are shown in Fig. \ref{fig:nested_f_posts}, which breaks them down by limb darkening model and reduction. In general, we see that a) using the Kurucz physical limb darkening always allows for tails towards higher $f$ value, b) the quadratic and cubic posteriors are nearly indistinguishable, and c) all marginalized posteriors peak at $f=0$.

To place final upper bound constraints on $f$, we focus just on our oblate model fits. We first reject reductions are overruled via a Bayes factor of 10 or more by any other model fit to that reduction. This eliminated all physical/Kurucz oblate models that used the \kat pipeline, as each of these were overruled by spherical models with quadratic limb darkening. Next, we computed the 95th percentile of the marginalized $f$ posteriors of all remaining reductions, and selected the one with the largest value. This is came from jedi\_15\_6 using physical/Kurucz limb darkening, and leaves us with $f<0.097$ at 95\%.

We note, however, that by construction all other posteriors report smaller upper limits than the one we have settled on as our final bound. The most extreme reduction/model combination is katahdin\_snr\_8 using cubic limb darkening, whose posterior implies that $f<0.065$ at 95\%. This $\sim33\%$ difference in possible upper bounds reflects the sensitivity of our inference on $f$ has on our reduction and modeling choices.

In addition to the fits shown in Fig. \ref{fig:nested_f_posts}, we also repeated all 60 fits when leaving out exposure 1, exposures 1 and 2, and exposures 1, 2, and 5 to test whether the out of transit baselines were affecting the inferred trend shapes. However, there was little qualitative difference between these fits and the ones presented in Fig. \ref{fig:nested_f_posts} and Tables \ref{tab:jedi_fits} and \ref{tab:kat_fits}: all reductions peaked at $f=0$, and Bayesian evidence offered no preference between spherical and oblate models.

\section{Conclusions} \label{sec:conclusion}

\subsection{Implications for Rotation Period}
Having placed a bound on $f$, we can use Eq. \ref{eq:f_to_spin} to in turn place a bound on the planet's rotation rate $P$. As discussed in Secs. \ref{sec:intro} and \ref{sub:priors}, this equation references the planet's \textit{true} oblateness, $f^*$, while what we have here is a bound on the \textit{projected} oblateness $f$. If we temporarily assume that the true oblateness is well-matched by the projected oblateness, then we have $P\geq7.18$\,hours using Eq. \ref{eq:f_to_spin}, the posterior mode of $R_p/R_s$ from the jedi\_15\_6 + physical LD reduction, and the fiducial value for the star's physical size from \citet{chachan_rvs_2022}. Using an alternative formulation from \citet{de_Pater_Lissauer_2015}, as is done in \citet{lammers_2024}, gives $P\geq7.11$\,hours when using the same $\Lambda_2$ as Jupiter (0.165).

Marginalizing over the uncertainty in the planet's true orientation is challenging in this case since posteriors from different reductions have different marginal distributions for $f$. However, we can get a sense of its effects using the random samples used to generate the prior for $f$ in Fig. \ref{fig:f_prior}. If we take all of the randomly-generated $f$'s that fell below $f<0.097$ and investigate their corresponding $f^*$'s, we find the 95th percentile is $f^*<0.431$, corresponding to $P\geq3.35$ hours. If we additionally enforce that the true obliquity is $<45^\circ$, as in \citet{lammers_2024}, we find $f^*<0.125$ and $P\geq6.30$\,hours. These are crude bounds included for illustrative purpose more than anything, as they do not actually marginalize over any of our posteriors and are derived only by considering our 95\% upper limit on $f$.  However, we believe it worth restating that every reported period bound is conditioned on an assumption about the planet's true orientation.

\subsection{Challenges and Promise of Measuring Oblateness with JWST}

JWST is an amazingly capable facility, and these data are by far the most precise observations of this planetary system recorded to date. However, while we were still able to place a constraint on \kep's projected oblateness and its rotation rate, it is worth reflecting on the limitations of these data in light of how we arrived at these upper bounds.

As noted in Sec. \ref{sec:nested_sampling}, different reduction/model combinations lead to a wide range of bounds on $f$: our smallest upper bound is nearly 33\% smaller than our highest upper bound. Further, examination of Tables \ref{tab:jedi_fits} and \ref{tab:kat_fits} reveals that lowest reduced $\chi^2$ of our fits, that achieved for katahdin\_snr\_10, a spherical planet, and quadratic limb darkening, is 1.19. This indicates that even with our flexible GPs the models tested are still not fitting these data as well as we might expect. However, with multiple exposures to contend with, it is challenging to experiment with and fit more complex trend models in a typical Bayesian workflow. The models presented above already required fitting 21-25 free parameters and push the bounds of nested sampling routines when using a relatively expensive forward model like \sqp. While it may be possible to explore higher-dimensional spaces using Hamiltonian Monte Carlo routines, this would require abandoning Bayesian evidence as a model comparison tool. Paper II, which deals with even higher dimensional models involving a planet and a moon, explores using profile likelihood as an alternative technique that allows for computation of Bayesian evidence in higher dimensional spaces.

Though all final fits achieve an RMS scatter comparable to our proposal prediction of 39\,ppm per 5-minute bin, they only do so thanks to the inclusion of the GPs: as seen in Fig. \ref{fig:mle_rms}, leaving the GP out left substantial correlated noise in the residuals. As seen in Fig. \ref{fig:gp_scales}, the GPs often preferred length scales comparable to half the ingress/egress duration, which complicates an oblateness inference. Future efforts that suppress this correlated noise while not admitting such flexible trends would greatly benefit future observations to measure planetary oblateness.

For future oblateness studies with JWST, we the recommend following:

\begin{enumerate}
    \item If possible, observations should be designed to fit into one exposure. These are limited to 196,608 frames (\# integrations x \# groups) and 65,536 integrations. Multi-exposure observations such as these require more complex modeling to account for exposure-specific trends, and handling exposure resets mid-transit (between exposures 3-4) and right at the end of egress (between exposures 4-5) required more complex models with more free parameters than would have been necessary under an alternative observation setup. Note that the limitation is on frames/integrations, not time, and that it is possible to fit long transit observations into single exposures by selecting instrument modes with longer readout times. Balancing the tradeoff between expected SNR and number of exposures will require careful target-specific planning.
    \item Modelers should complete a full end-to-end analysis using multiple pipelines to ensure that constraints are robust to pipeline choice. This practice is helpful for atmospheric studies but crucial for those focused on oblateness as the relevant time scale is much shorter. Though both pipelines reduce the data equally well in the sense that the RMS of their MAPs are similar, the final bounds on $f$ are different. Since all reductions were fit equally well by spherical planet models, we did not have to address the issue of distinguishing between these reductions and only needed to report the most conservative upper bound. The first reported non-zero detection of oblateness, should it be made, should take care to ensure its conclusions are not driven by pipeline or modeling choices.
\end{enumerate}

\kep is now the first Jupiter-like planet beyond the solar system with a constraint on its oblateness and rotation rate. We encourage the community to carry out similar measurements for other targets, and eventually look forward to the insights obtained from a population-level study of planetary oblatenesses.

{\begin{acknowledgments}
This research has made use of the Astrophysics Data System, funded by NASA under Cooperative Agreement 80NSSC21M00561.

We thank the Yale Center for Research Computing for guidance and assistance in the computation run on the Bouchet Cluster.

Resources supporting this work were provided by the NASA High-End Computing (HEC) Program through the NASA Advanced Supercomputing (NAS) Division at Ames Research Center.

This work is based [in part] on observations made with the NASA/ESA/CSA James Webb Space Telescope. The data were obtained from the Mikulski Archive for Space Telescopes at the Space Telescope Science Institute, which is operated by the Association of Universities for Research in Astronomy, Inc., under NASA contract NAS 5-03127 for JWST. These observations are associated with program \#6491.

The JWST data presented in this article were obtained from the Mikulski Archive for Space Telescopes (MAST) at the Space Telescope Science Institute. The specific observations analyzed can be accessed via \dataset[doi: 10.17909/e50n-4y96]{https://doi.org/10.17909/e50n-4y96}.

Special thanks to donors to the Cool Worlds Lab, without whom this kind of research would not be possible: Douglas Daughaday,
Elena West,
Tristan Zajonc,
Alex de Vaal,
Mark Elliott,
Stephen Lee,
Zachary Danielson,
Chad Souter,
Marcus Gillette,
Jason Rockett,
Tom Donkin,
Andrew Schoen,
Mike Hedlund,
Ryan Provost,
Nicholas De Haan,
Emerson Garland,
Queen Rd Fnd Inc.,
Ieuan Williams,
Axel Nimmerjahn,
Brian Cartmell,
Guillaume Le Saint,
Robin Raszka,
Bas van Gaalen,
Josh Alley,
Drew Aron,
Warren Smith,
Brad Bueche,
Steve Larter,
Marisol Adler \&
Craig Frederick.

\end{acknowledgments}}

\vspace{5mm}
\facilities{\jwst (NIRSpec)}

\software{
    \texttt{astropy} \citep{astropy:2013, astropy:2018, astropy:2022}; \jax \citep{jax}; \sqp \citep{squishyplanet}; \texttt{tinygp} \citep{tinygp}
}

\appendix

\section{Additional MLE Diagnostics} \label{ap:mle}

\begin{figure}
    \centering
    \includegraphics[width=0.5\linewidth]{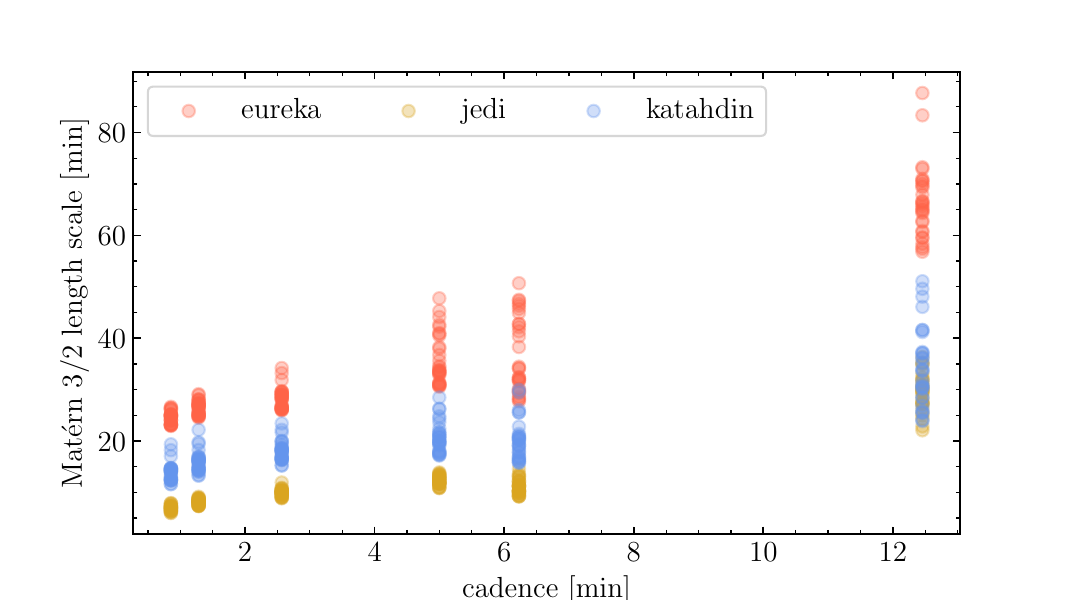}
    \caption{The best-fitting length scales for all fits that used an exponential trend and an \mth GP kernel.} 
    \label{fig:gp_scales}
\end{figure}

\begin{figure}
    \centering
    \includegraphics[width=0.5\linewidth]{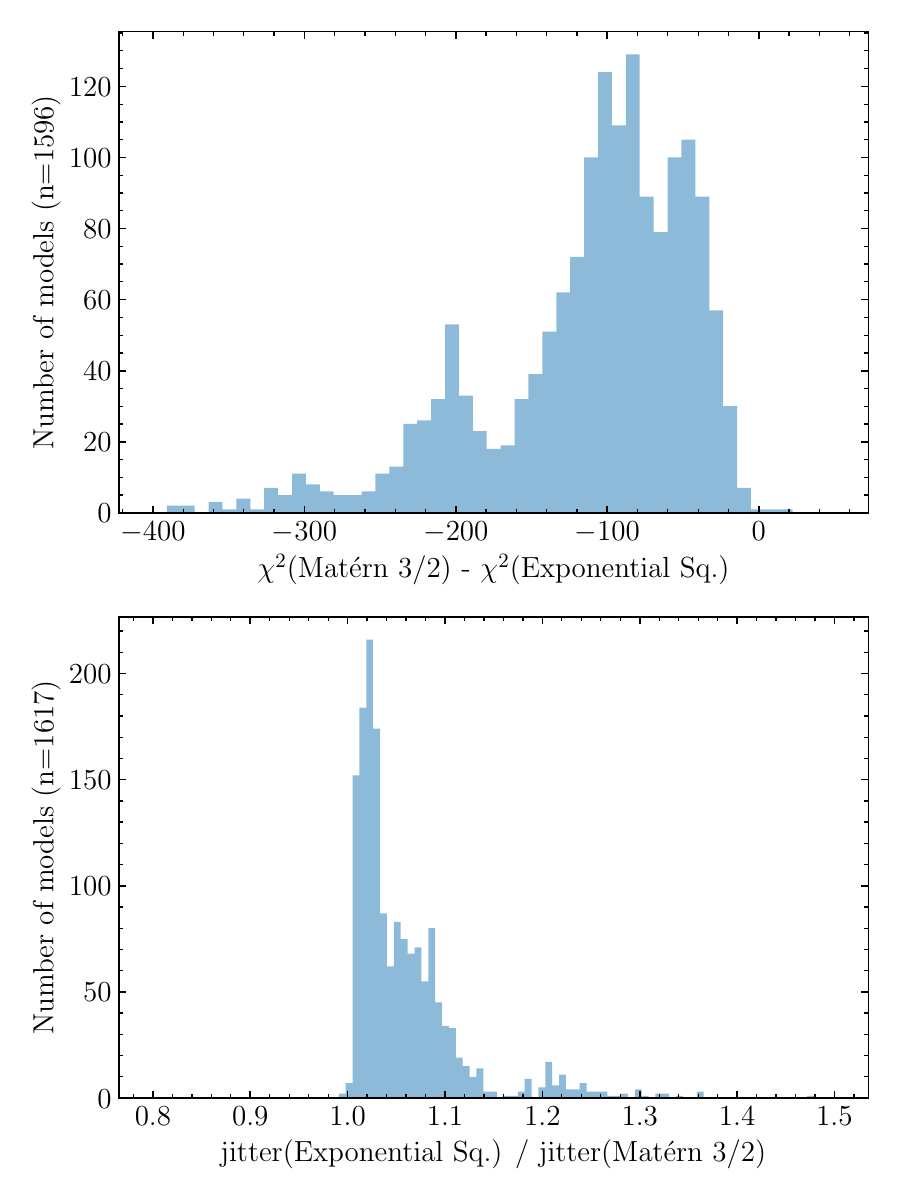}
    \caption{Illustration of how when comparing models that only differ by which GP kernel they used, the exponential squared models consistently prefer larger jitter and land on worse $\chi^2$.} 
    \label{fig:expsq_jitter}
\end{figure}

\section{Nested Sampling Details} \label{ap:nested}

\begin{deluxetable}{lccccccc}
\tablecaption{\jedi Nested Sampling Fits\label{tab:jedi_fits}}
\tablehead{
\colhead{Light Curve} &  \colhead{Oblate?} & \colhead{Limb Darkening} & \colhead{log $\mathcal{Z}$} & \colhead{ESS} & \colhead{MAP RMS [ppm]} & \colhead{MAP MAD [ppm]} & \colhead{MAP $\chi^2$}
}
\startdata
& True & quad & -4155.6 & 50009 & 38.32 & 25.00 & 730.71 \\
jedi\_5\_3 & True & cube & -4155.8 & 25008 & 38.33 & 24.93 & 730.98 \\
 & True & kurucz & -4154.8 & 50004 & 38.83 & 25.55 & 750.09 \\ \hline
 & False & quad & -4154.2 & 50027 & 38.31 & 24.98 & 730.38 \\
jedi\_5\_3 & False & cube & -4154.4 & 50019 & 38.40 & 25.08 & 733.58 \\
 & False & kurucz & -4153.3 & 50058 & 39.00 & 25.73 & 756.95 \\ \hline\hline
 & True & quad & -4198.3 & 50050 & 39.29 & 24.34 & 734.77 \\
jedi\_5\_6 & True & cube & -4198.5 & 50006 & 39.29 & 24.50 & 735.09 \\
 & True & kurucz & -4198.2 & 50025 & 40.44 & 25.86 & 778.78 \\ \hline
 & False & quad & -4197.0 & 50074 & 39.47 & 24.91 & 741.69 \\
jedi\_5\_6 & False & cube & -4197.1 & 50008 & 39.37 & 24.66 & 737.93 \\
 & False & kurucz & -4196.8 & 50040 & 40.37 & 25.77 & 775.98 \\ \hline\hline
 & True & quad & -4225.7 & 50039 & 41.13 & 26.67 & 752.95 \\
jedi\_5\_8 & True & cube & -4225.9 & 50001 & 41.49 & 26.78 & 766.34 \\
 & True & kurucz & -4226.4 & 50042 & 41.75 & 26.68 & 775.90 \\ \hline
 & False & quad & -4224.3 & 50083 & 41.33 & 26.52 & 760.40 \\
jedi\_5\_8 & False & cube & -4224.5 & 50004 & 41.22 & 26.33 & 756.48 \\
 & False & kurucz & -4225.0 & 50071 & 42.05 & 27.13 & 787.07 \\ \hline\hline
 & True & quad & -4153.7 & 50010 & 38.67 & 25.42 & 745.30 \\
jedi\_15\_3 & True & cube & -4153.7 & 25004 & 38.75 & 25.63 & 748.55 \\
 & True & kurucz & -4152.9 & 50019 & 39.08 & 26.14 & 761.42 \\ \hline
 & False & quad & -4152.4 & 50051 & 38.61 & 25.36 & 743.12 \\
jedi\_15\_3 & False & cube & -4152.4 & 50013 & 38.69 & 25.50 & 746.39 \\
 & False & kurucz & -4151.5 & 50066 & 39.36 & 26.49 & 772.45 \\ \hline\hline
 & True & quad & -4195.4 & 50002 & 39.65 & 25.86 & 747.78 \\
jedi\_15\_6 & True & cube & -4195.5 & 45147 & 39.71 & 25.93 & 750.22 \\
 & True & kurucz & -4195.5 & 50053 & 40.41 & 27.29 & 777.06 \\ \hline
 & False & quad & -4194.2 & 50036 & 39.63 & 25.98 & 747.08 \\
jedi\_15\_6 & False & cube & -4194.2 & 50013 & 39.78 & 25.90 & 752.82 \\
 & False & kurucz & -4194.2 & 50090 & 40.27 & 26.91 & 771.75 \\ \hline\hline
 & True & quad & -4222.0 & 50034 & 41.37 & 26.86 & 760.15 \\
jedi\_15\_8 & True & cube & -4222.1 & 26595 & 41.62 & 27.00 & 769.66 \\
 & True & kurucz & -4222.6 & 50005 & 42.10 & 27.59 & 787.43 \\ \hline
 & False & quad & -4220.6 & 50010 & 41.48 & 26.84 & 764.19 \\
jedi\_15\_8 & False & cube & -4220.7 & 50027 & 41.46 & 27.06 & 763.65 \\
 & False & kurucz & -4221.3 & 50030 & 42.09 & 27.77 & 786.90 \\ \hline\hline
\enddata
\tablecomments{At 5 minute binning, our data consisted of 595 flux measurements. To achieve a Bayes factor of 10, two models must be separated by a $\Delta\log\mathcal{Z} > 2.303$.}
\end{deluxetable}

\begin{deluxetable}{lccccccc}
\tablecaption{\kat Nested Sampling Fits\label{tab:kat_fits}}
\tablehead{
\colhead{Light Curve} &  \colhead{Oblate?} & \colhead{Limb Darkening} & \colhead{log $\mathcal{Z}$} & \colhead{ESS} & \colhead{MAP RMS [ppm]} & \colhead{MAP MAD [ppm]} & \colhead{MAP $\chi^2$}
}
\startdata
 & True & quad & -3346.1 & 50032 & 39.09 & 26.32 & 709.77 \\
katahdin\_snr\_8 & True & cube & -3346.7 & 50020 & 39.09 & 26.28 & 709.90 \\
 & True & kurucz & -3347.2 & 50015 & 39.17 & 26.79 & 712.80 \\ \hline
 & False & quad & -3344.4 & 50005 & 39.08 & 26.34 & 709.54 \\
katahdin\_snr\_8 & False & cube & -3345.0 & 50004 & 39.00 & 26.26 & 706.44 \\
 & False & kurucz & -3345.7 & 50069 & 39.16 & 26.74 & 712.69 \\ \hline\hline
 & True & quad & -3328.4 & 50002 & 38.55 & 24.67 & 686.66 \\
katahdin\_snr\_10 & True & cube & -3329.0 & 50006 & 38.41 & 24.79 & 681.54 \\
 & True & kurucz & -3330.7 & 50001 & 38.81 & 25.33 & 695.80 \\ \hline
 & False & quad & -3326.7 & 50051 & 38.51 & 24.67 & 685.20 \\
katahdin\_snr\_10 & False & cube & -3327.3 & 50006 & 38.67 & 24.71 & 690.77 \\
 & False & kurucz & -3329.2 & 50082 & 38.73 & 25.28 & 693.17 \\ \hline\hline
 & True & quad & -3307.6 & 50074 & 39.14 & 25.61 & 692.21 \\
katahdin\_snr\_15 & True & cube & -3308.2 & 31543 & 38.95 & 25.73 & 685.61 \\
 & True & kurucz & -3310.6 & 50012 & 39.20 & 25.53 & 694.42 \\ \hline
 & False & quad & -3305.9 & 50023 & 38.95 & 25.62 & 685.49 \\
katahdin\_snr\_15 & False & cube & -3306.6 & 50001 & 39.04 & 25.78 & 688.49 \\
 & False & kurucz & -3309.1 & 50007 & 39.21 & 25.35 & 694.96 \\ \hline\hline
 & True & quad & -3287.7 & 50005 & 40.54 & 26.78 & 695.96 \\
katahdin\_snr\_25 & True & cube & -3288.3 & 50009 & 40.46 & 26.62 & 693.21 \\
 & True & kurucz & -3291.3 & 50023 & 40.74 & 26.98 & 703.07 \\ \hline
 & False & quad & -3286.1 & 50011 & 40.32 & 26.56 & 688.34 \\
katahdin\_snr\_25 & False & cube & -3286.7 & 50002 & 40.28 & 26.52 & 687.12 \\
 & False & kurucz & -3289.7 & 50002 & 40.78 & 27.05 & 704.49 \\ \hline\hline
\enddata
\tablecomments{Notes: We reject the first two katahdin reductions, (SNR thresholds 3 and 5), as we see their best-fitting residuals are far worse than all other reductions. At 5 minute binning, our data consisted of 595 flux measurements. These models had been 21-25 free parameters. To achieve a Bayes factor of 10, two models must be separated by a $\Delta\log\mathcal{Z} > 2.303$.}
\end{deluxetable}

\begin{figure*}
    \centering
    \includegraphics[width=\linewidth]{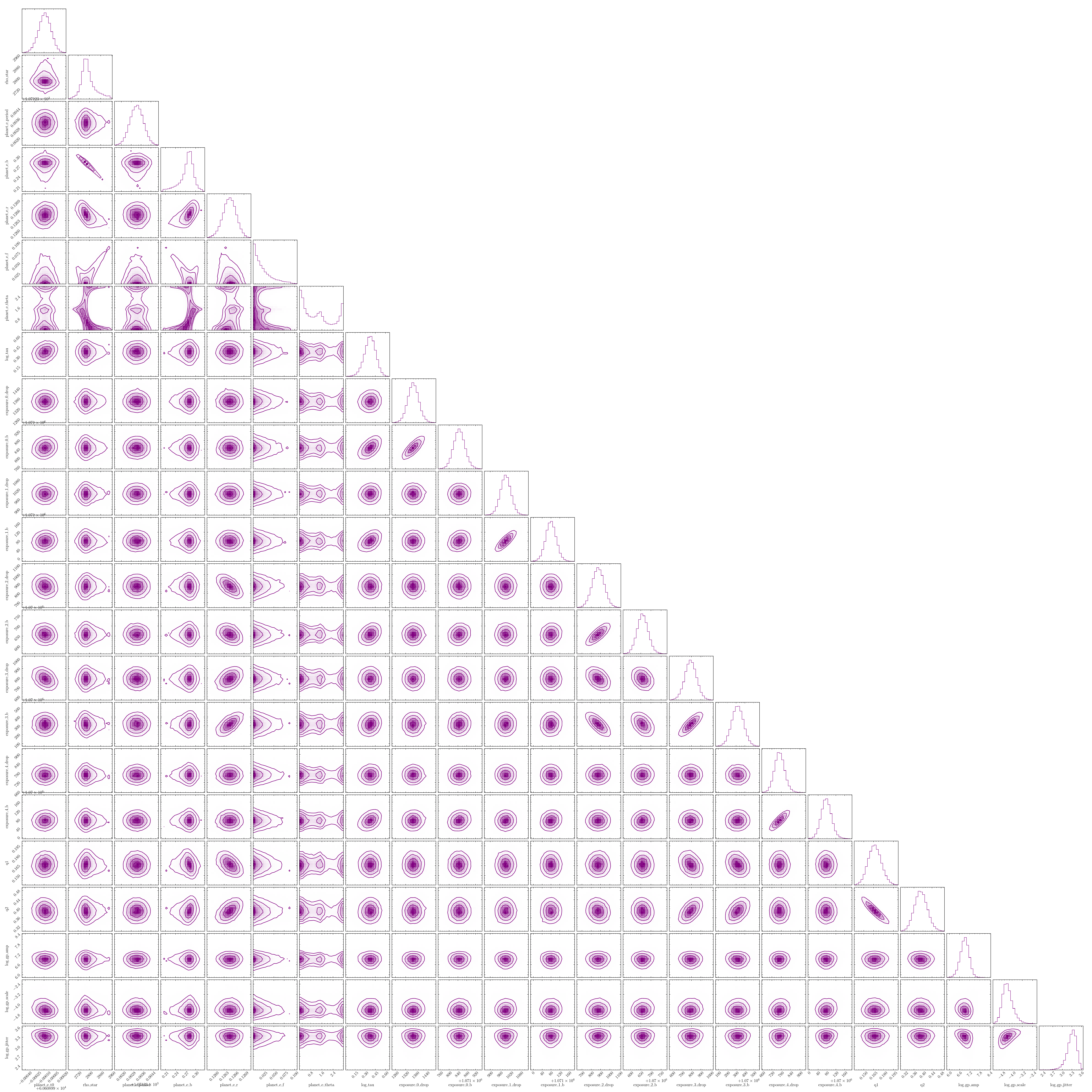}
    \caption{One of the 60 nested sampling fits whose marginal distribution for $f$ is shown in Fig. \ref{fig:nested_f_posts}. This particular run is for katahin\_snr\_25, quadratic limb darkening, and an oblate planet model. The majority of the parameters are for modeling the unique trend of each of the 5 exposures.} 
    \label{fig:rep_corner}
\end{figure*}

\bibliography{references}{}

@article{astropy:2013,
Adsurl = {http://adsabs.harvard.edu/abs/2013A%26A...558A..33A},
Archiveprefix = {arXiv},
Author = {{Astropy Collaboration} and {Robitaille}, T.~P. and {Tollerud}, E.~J. and {Greenfield}, P. and {Droettboom}, M. and {Bray}, E. and {Aldcroft}, T. and {Davis}, M. and {Ginsburg}, A. and {Price-Whelan}, A.~M. and {Kerzendorf}, W.~E. and {Conley}, A. and {Crighton}, N. and {Barbary}, K. and {Muna}, D. and {Ferguson}, H. and {Grollier}, F. and {Parikh}, M.~M. and {Nair}, P.~H. and {Unther}, H.~M. and {Deil}, C. and {Woillez}, J. and {Conseil}, S. and {Kramer}, R. and {Turner}, J.~E.~H. and {Singer}, L. and {Fox}, R. and {Weaver}, B.~A. and {Zabalza}, V. and {Edwards}, Z.~I. and {Azalee Bostroem}, K. and {Burke}, D.~J. and {Casey}, A.~R. and {Crawford}, S.~M. and {Dencheva}, N. and {Ely}, J. and {Jenness}, T. and {Labrie}, K. and {Lim}, P.~L. and {Pierfederici}, F. and {Pontzen}, A. and {Ptak}, A. and {Refsdal}, B. and {Servillat}, M. and {Streicher}, O.},
Doi = {10.1051/0004-6361/201322068},
Eid = {A33},
Eprint = {1307.6212},
Journal = {\aap},
Month = oct,
Pages = {A33},
Primaryclass = {astro-ph.IM},
Title = {{Astropy: A community Python package for astronomy}},
Volume = 558,
Year = 2013}

@ARTICLE{astropy:2018,
       author = {{Astropy Collaboration} and {Price-Whelan}, A.~M. and
         {Sip{\H{o}}cz}, B.~M. and {G{\"u}nther}, H.~M. and {Lim}, P.~L. and
         {Crawford}, S.~M. and {Conseil}, S. and {Shupe}, D.~L. and
         {Craig}, M.~W. and {Dencheva}, N. and {Ginsburg}, A. and {Vand
        erPlas}, J.~T. and {Bradley}, L.~D. and {P{\'e}rez-Su{\'a}rez}, D. and
         {de Val-Borro}, M. and {Aldcroft}, T.~L. and {Cruz}, K.~L. and
         {Robitaille}, T.~P. and {Tollerud}, E.~J. and {Ardelean}, C. and
         {Babej}, T. and {Bach}, Y.~P. and {Bachetti}, M. and {Bakanov}, A.~V. and
         {Bamford}, S.~P. and {Barentsen}, G. and {Barmby}, P. and
         {Baumbach}, A. and {Berry}, K.~L. and {Biscani}, F. and {Boquien}, M. and
         {Bostroem}, K.~A. and {Bouma}, L.~G. and {Brammer}, G.~B. and
         {Bray}, E.~M. and {Breytenbach}, H. and {Buddelmeijer}, H. and
         {Burke}, D.~J. and {Calderone}, G. and {Cano Rodr{\'\i}guez}, J.~L. and
         {Cara}, M. and {Cardoso}, J.~V.~M. and {Cheedella}, S. and {Copin}, Y. and
         {Corrales}, L. and {Crichton}, D. and {D'Avella}, D. and {Deil}, C. and
         {Depagne}, {\'E}. and {Dietrich}, J.~P. and {Donath}, A. and
         {Droettboom}, M. and {Earl}, N. and {Erben}, T. and {Fabbro}, S. and
         {Ferreira}, L.~A. and {Finethy}, T. and {Fox}, R.~T. and
         {Garrison}, L.~H. and {Gibbons}, S.~L.~J. and {Goldstein}, D.~A. and
         {Gommers}, R. and {Greco}, J.~P. and {Greenfield}, P. and
         {Groener}, A.~M. and {Grollier}, F. and {Hagen}, A. and {Hirst}, P. and
         {Homeier}, D. and {Horton}, A.~J. and {Hosseinzadeh}, G. and {Hu}, L. and
         {Hunkeler}, J.~S. and {Ivezi{\'c}}, {\v{Z}}. and {Jain}, A. and
         {Jenness}, T. and {Kanarek}, G. and {Kendrew}, S. and {Kern}, N.~S. and
         {Kerzendorf}, W.~E. and {Khvalko}, A. and {King}, J. and {Kirkby}, D. and
         {Kulkarni}, A.~M. and {Kumar}, A. and {Lee}, A. and {Lenz}, D. and
         {Littlefair}, S.~P. and {Ma}, Z. and {Macleod}, D.~M. and
         {Mastropietro}, M. and {McCully}, C. and {Montagnac}, S. and
         {Morris}, B.~M. and {Mueller}, M. and {Mumford}, S.~J. and {Muna}, D. and
         {Murphy}, N.~A. and {Nelson}, S. and {Nguyen}, G.~H. and
         {Ninan}, J.~P. and {N{\"o}the}, M. and {Ogaz}, S. and {Oh}, S. and
         {Parejko}, J.~K. and {Parley}, N. and {Pascual}, S. and {Patil}, R. and
         {Patil}, A.~A. and {Plunkett}, A.~L. and {Prochaska}, J.~X. and
         {Rastogi}, T. and {Reddy Janga}, V. and {Sabater}, J. and
         {Sakurikar}, P. and {Seifert}, M. and {Sherbert}, L.~E. and
         {Sherwood-Taylor}, H. and {Shih}, A.~Y. and {Sick}, J. and
         {Silbiger}, M.~T. and {Singanamalla}, S. and {Singer}, L.~P. and
         {Sladen}, P.~H. and {Sooley}, K.~A. and {Sornarajah}, S. and
         {Streicher}, O. and {Teuben}, P. and {Thomas}, S.~W. and
         {Tremblay}, G.~R. and {Turner}, J.~E.~H. and {Terr{\'o}n}, V. and
         {van Kerkwijk}, M.~H. and {de la Vega}, A. and {Watkins}, L.~L. and
         {Weaver}, B.~A. and {Whitmore}, J.~B. and {Woillez}, J. and
         {Zabalza}, V. and {Astropy Contributors}},
        title = "{The Astropy Project: Building an Open-science Project and Status of the v2.0 Core Package}",
      journal = {\aj},
         year = 2018,
        month = sep,
       volume = {156},
       number = {3},
          eid = {123},
        pages = {123},
          doi = {10.3847/1538-3881/aabc4f},
archivePrefix = {arXiv},
       eprint = {1801.02634},
 primaryClass = {astro-ph.IM},
       adsurl = {https://ui.adsabs.harvard.edu/abs/2018AJ....156..123A}
}

@ARTICLE{astropy:2022,
       author = {{Astropy Collaboration} and {Price-Whelan}, Adrian M. and {Lim}, Pey Lian and {Earl}, Nicholas and {Starkman}, Nathaniel and {Bradley}, Larry and {Shupe}, David L. and {Patil}, Aarya A. and {Corrales}, Lia and {Brasseur}, C.~E. and {N{"o}the}, Maximilian and {Donath}, Axel and {Tollerud}, Erik and {Morris}, Brett M. and {Ginsburg}, Adam and {Vaher}, Eero and {Weaver}, Benjamin A. and {Tocknell}, James and {Jamieson}, William and {van Kerkwijk}, Marten H. and {Robitaille}, Thomas P. and {Merry}, Bruce and {Bachetti}, Matteo and {G{"u}nther}, H. Moritz and {Aldcroft}, Thomas L. and {Alvarado-Montes}, Jaime A. and {Archibald}, Anne M. and {B{'o}di}, Attila and {Bapat}, Shreyas and {Barentsen}, Geert and {Baz{'a}n}, Juanjo and {Biswas}, Manish and {Boquien}, M{'e}d{'e}ric and {Burke}, D.~J. and {Cara}, Daria and {Cara}, Mihai and {Conroy}, Kyle E. and {Conseil}, Simon and {Craig}, Matthew W. and {Cross}, Robert M. and {Cruz}, Kelle L. and {D'Eugenio}, Francesco and {Dencheva}, Nadia and {Devillepoix}, Hadrien A.~R. and {Dietrich}, J{"o}rg P. and {Eigenbrot}, Arthur Davis and {Erben}, Thomas and {Ferreira}, Leonardo and {Foreman-Mackey}, Daniel and {Fox}, Ryan and {Freij}, Nabil and {Garg}, Suyog and {Geda}, Robel and {Glattly}, Lauren and {Gondhalekar}, Yash and {Gordon}, Karl D. and {Grant}, David and {Greenfield}, Perry and {Groener}, Austen M. and {Guest}, Steve and {Gurovich}, Sebastian and {Handberg}, Rasmus and {Hart}, Akeem and {Hatfield-Dodds}, Zac and {Homeier}, Derek and {Hosseinzadeh}, Griffin and {Jenness}, Tim and {Jones}, Craig K. and {Joseph}, Prajwel and {Kalmbach}, J. Bryce and {Karamehmetoglu}, Emir and {Ka{l}uszy{'n}ski}, Miko{l}aj and {Kelley}, Michael S.~P. and {Kern}, Nicholas and {Kerzendorf}, Wolfgang E. and {Koch}, Eric W. and {Kulumani}, Shankar and {Lee}, Antony and {Ly}, Chun and {Ma}, Zhiyuan and {MacBride}, Conor and {Maljaars}, Jakob M. and {Muna}, Demitri and {Murphy}, N.~A. and {Norman}, Henrik and {O'Steen}, Richard and {Oman}, Kyle A. and {Pacifici}, Camilla and {Pascual}, Sergio and {Pascual-Granado}, J. and {Patil}, Rohit R. and {Perren}, Gabriel I. and {Pickering}, Timothy E. and {Rastogi}, Tanuj and {Roulston}, Benjamin R. and {Ryan}, Daniel F. and {Rykoff}, Eli S. and {Sabater}, Jose and {Sakurikar}, Parikshit and {Salgado}, Jes{'u}s and {Sanghi}, Aniket and {Saunders}, Nicholas and {Savchenko}, Volodymyr and {Schwardt}, Ludwig and {Seifert-Eckert}, Michael and {Shih}, Albert Y. and {Jain}, Anany Shrey and {Shukla}, Gyanendra and {Sick}, Jonathan and {Simpson}, Chris and {Singanamalla}, Sudheesh and {Singer}, Leo P. and {Singhal}, Jaladh and {Sinha}, Manodeep and {Sip{H{o}}cz}, Brigitta M. and {Spitler}, Lee R. and {Stansby}, David and {Streicher}, Ole and {{{S}}umak}, Jani and {Swinbank}, John D. and {Taranu}, Dan S. and {Tewary}, Nikita and {Tremblay}, Grant R. and {Val-Borro}, Miguel de and {Van Kooten}, Samuel J. and {Vasovi{'c}}, Zlatan and {Verma}, Shresth and {de Miranda Cardoso}, Jos{'e} Vin{'i}cius and {Williams}, Peter K.~G. and {Wilson}, Tom J. and {Winkel}, Benjamin and {Wood-Vasey}, W.~M. and {Xue}, Rui and {Yoachim}, Peter and {Zhang}, Chen and {Zonca}, Andrea and {Astropy Project Contributors}},
        title = "{The Astropy Project: Sustaining and Growing a Community-oriented Open-source Project and the Latest Major Release (v5.0) of the Core Package}",
      journal = {\apj},
         year = 2022,
        month = aug,
       volume = {935},
       number = {2},
          eid = {167},
        pages = {167},
          doi = {10.3847/1538-4357/ac7c74},
archivePrefix = {arXiv},
       eprint = {2206.14220},
 primaryClass = {astro-ph.IM},
       adsurl = {https://ui.adsabs.harvard.edu/abs/2022ApJ...935..167A}
}

@ARTICLE{kipping_discovery_2016,
       author = {{Kipping}, D.~M. and {Torres}, G. and {Henze}, C. and {Teachey}, A. and {Isaacson}, H. and {Petigura}, E. and {Marcy}, G.~W. and {Buchhave}, L.~A. and {Chen}, J. and {Bryson}, S.~T. and {Sandford}, E.},
        title = "{A Transiting Jupiter Analog}",
      journal = {\apj},
         year = 2016,
        month = apr,
       volume = {820},
       number = {2},
          eid = {112},
        pages = {112},
          doi = {10.3847/0004-637X/820/2/112},
archivePrefix = {arXiv},
       eprint = {1603.00042},
 primaryClass = {astro-ph.EP},
       adsurl = {https://ui.adsabs.harvard.edu/abs/2016ApJ...820..112K}
}

@ARTICLE{gaia_dr3,
       author = {{Gaia Collaboration} and {Vallenari}, A. and {Brown}, A.~G.~A. and {Prusti}, T. and {de Bruijne}, J.~H.~J. and {Arenou}, F. and {Babusiaux}, C. and {Biermann}, M. and {Creevey}, O.~L. and {Ducourant}, C. and {Evans}, D.~W. and {Eyer}, L. and {Guerra}, R. and {Hutton}, A. and {Jordi}, C. and {Klioner}, S.~A. and {Lammers}, U.~L. and {Lindegren}, L. and {Luri}, X. and {Mignard}, F. and {Panem}, C. and {Pourbaix}, D. and {Randich}, S. and {Sartoretti}, P. and {Soubiran}, C. and {Tanga}, P. and {Walton}, N.~A. and {Bailer-Jones}, C.~A.~L. and {Bastian}, U. and {Drimmel}, R. and {Jansen}, F. and {Katz}, D. and {Lattanzi}, M.~G. and {van Leeuwen}, F. and {Bakker}, J. and {Cacciari}, C. and {Casta{\~n}eda}, J. and {De Angeli}, F. and {Fabricius}, C. and {Fouesneau}, M. and {Fr{\'e}mat}, Y. and {Galluccio}, L. and {Guerrier}, A. and {Heiter}, U. and {Masana}, E. and {Messineo}, R. and {Mowlavi}, N. and {Nicolas}, C. and {Nienartowicz}, K. and {Pailler}, F. and {Panuzzo}, P. and {Riclet}, F. and {Roux}, W. and {Seabroke}, G.~M. and {Sordo}, R. and {Th{\'e}venin}, F. and {Gracia-Abril}, G. and {Portell}, J. and {Teyssier}, D. and {Altmann}, M. and {Andrae}, R. and {Audard}, M. and {Bellas-Velidis}, I. and {Benson}, K. and {Berthier}, J. and {Blomme}, R. and {Burgess}, P.~W. and {Busonero}, D. and {Busso}, G. and {C{\'a}novas}, H. and {Carry}, B. and {Cellino}, A. and {Cheek}, N. and {Clementini}, G. and {Damerdji}, Y. and {Davidson}, M. and {de Teodoro}, P. and {Nu{\~n}ez Campos}, M. and {Delchambre}, L. and {Dell'Oro}, A. and {Esquej}, P. and {Fern{\'a}ndez-Hern{\'a}ndez}, J. and {Fraile}, E. and {Garabato}, D. and {Garc{\'\i}a-Lario}, P. and {Gosset}, E. and {Haigron}, R. and {Halbwachs}, J. -L. and {Hambly}, N.~C. and {Harrison}, D.~L. and {Hern{\'a}ndez}, J. and {Hestroffer}, D. and {Hodgkin}, S.~T. and {Holl}, B. and {Jan{\ss}en}, K. and {Jevardat de Fombelle}, G. and {Jordan}, S. and {Krone-Martins}, A. and {Lanzafame}, A.~C. and {L{\"o}ffler}, W. and {Marchal}, O. and {Marrese}, P.~M. and {Moitinho}, A. and {Muinonen}, K. and {Osborne}, P. and {Pancino}, E. and {Pauwels}, T. and {Recio-Blanco}, A. and {Reyl{\'e}}, C. and {Riello}, M. and {Rimoldini}, L. and {Roegiers}, T. and {Rybizki}, J. and {Sarro}, L.~M. and {Siopis}, C. and {Smith}, M. and {Sozzetti}, A. and {Utrilla}, E. and {van Leeuwen}, M. and {Abbas}, U. and {{\'A}brah{\'a}m}, P. and {Abreu Aramburu}, A. and {Aerts}, C. and {Aguado}, J.~J. and {Ajaj}, M. and {Aldea-Montero}, F. and {Altavilla}, G. and {{\'A}lvarez}, M.~A. and {Alves}, J. and {Anders}, F. and {Anderson}, R.~I. and {Anglada Varela}, E. and {Antoja}, T. and {Baines}, D. and {Baker}, S.~G. and {Balaguer-N{\'u}{\~n}ez}, L. and {Balbinot}, E. and {Balog}, Z. and {Barache}, C. and {Barbato}, D. and {Barros}, M. and {Barstow}, M.~A. and {Bartolom{\'e}}, S. and {Bassilana}, J. -L. and {Bauchet}, N. and {Becciani}, U. and {Bellazzini}, M. and {Berihuete}, A. and {Bernet}, M. and {Bertone}, S. and {Bianchi}, L. and {Binnenfeld}, A. and {Blanco-Cuaresma}, S. and {Blazere}, A. and {Boch}, T. and {Bombrun}, A. and {Bossini}, D. and {Bouquillon}, S. and {Bragaglia}, A. and {Bramante}, L. and {Breedt}, E. and {Bressan}, A. and {Brouillet}, N. and {Brugaletta}, E. and {Bucciarelli}, B. and {Burlacu}, A. and {Butkevich}, A.~G. and {Buzzi}, R. and {Caffau}, E. and {Cancelliere}, R. and {Cantat-Gaudin}, T. and {Carballo}, R. and {Carlucci}, T. and {Carnerero}, M.~I. and {Carrasco}, J.~M. and {Casamiquela}, L. and {Castellani}, M. and {Castro-Ginard}, A. and {Chaoul}, L. and {Charlot}, P. and {Chemin}, L. and {Chiaramida}, V. and {Chiavassa}, A. and {Chornay}, N. and {Comoretto}, G. and {Contursi}, G. and {Cooper}, W.~J. and {Cornez}, T. and {Cowell}, S. and {Crifo}, F. and {Cropper}, M. and {Crosta}, M. and {Crowley}, C. and {Dafonte}, C. and {Dapergolas}, A. and {David}, M. and {David}, P. and {de Laverny}, P. and {De Luise}, F. and {De March}, R.},
        title = "{Gaia Data Release 3. Summary of the content and survey properties}",
      journal = {\aap},
         year = 2023,
        month = jun,
       volume = {674},
          eid = {A1},
        pages = {A1},
          doi = {10.1051/0004-6361/202243940},
archivePrefix = {arXiv},
       eprint = {2208.00211},
 primaryClass = {astro-ph.GA},
       adsurl = {https://ui.adsabs.harvard.edu/abs/2023A&A...674A...1G}
}

@ARTICLE{chachan_rvs_2022,
       author = {{Chachan}, Yayaati and {Dalba}, Paul A. and {Knutson}, Heather A. and {Fulton}, Benjamin J. and {Thorngren}, Daniel and {Beichman}, Charles and {Ciardi}, David R. and {Howard}, Andrew W. and {Van Zandt}, Judah},
        title = "{Kepler-167e as a Probe of the Formation Histories of Cold Giants with Inner Super-Earths}",
      journal = {\apj},
         year = 2022,
        month = feb,
       volume = {926},
       number = {1},
          eid = {62},
        pages = {62},
          doi = {10.3847/1538-4357/ac3ed6},
archivePrefix = {arXiv},
       eprint = {2112.00747},
 primaryClass = {astro-ph.EP},
       adsurl = {https://ui.adsabs.harvard.edu/abs/2022ApJ...926...62C}
}

@misc{simbad_vmag,
       author = {{Zacharias}, N. and {Finch}, C.~T. and {Girard}, T.~M. and {Henden}, A. and {Bartlett}, J.~L. and {Monet}, D.~G. and {Zacharias}, M.~I.},
        title = "{VizieR Online Data Catalog: UCAC4 Catalogue (Zacharias+, 2012)}",
 howpublished = {VizieR On-line Data Catalog: I/322A.  Originally published in: 2013AJ....145...44Z},
         year = 2012,
        month = jul,
          eid = {I/322A},
       adsurl = {https://ui.adsabs.harvard.edu/abs/2012yCat.1322....0Z}
}

@misc{simbad_2mass_phot,
       author = {{Cutri}, R.~M. and {Skrutskie}, M.~F. and {van Dyk}, S. and {Beichman}, C.~A. and {Carpenter}, J.~M. and {Chester}, T. and {Cambresy}, L. and {Evans}, T. and {Fowler}, J. and {Gizis}, J. and {Howard}, E. and {Huchra}, J. and {Jarrett}, T. and {Kopan}, E.~L. and {Kirkpatrick}, J.~D. and {Light}, R.~M. and {Marsh}, K.~A. and {McCallon}, H. and {Schneider}, S. and {Stiening}, R. and {Sykes}, M. and {Weinberg}, M. and {Wheaton}, W.~A. and {Wheelock}, S. and {Zacarias}, N.},
        title = "{VizieR Online Data Catalog: 2MASS All-Sky Catalog of Point Sources (Cutri+ 2003)}",
 howpublished = {VizieR On-line Data Catalog: II/246.  Originally published in: University of Massachusetts and Infrared Processing and Analysis Center, (IPAC/California Institute of Technology) (2003)},
         year = 2003,
        month = jun,
          eid = {II/246},
       adsurl = {https://ui.adsabs.harvard.edu/abs/2003yCat.2246....0C}
}

@ARTICLE{dalba_spitzer_2019,
       author = {{Dalba}, Paul A. and {Tamburo}, Patrick},
        title = "{Spitzer  Detection of the Transiting Jupiter-analog Exoplanet Kepler-167e}",
      journal = {\apjl},
         year = 2019,
        month = mar,
       volume = {873},
       number = {2},
          eid = {L17},
        pages = {L17},
          doi = {10.3847/2041-8213/ab0bb4},
archivePrefix = {arXiv},
       eprint = {1903.01478},
 primaryClass = {astro-ph.EP},
       adsurl = {https://ui.adsabs.harvard.edu/abs/2019ApJ...873L..17D}
}

@ARTICLE{perrocheau_citizen_science_2022,
       author = {{Perrocheau}, Amaury and {Esposito}, Thomas M. and {Dalba}, Paul A. and {Marchis}, Franck and {Avsar}, Arin M. and {Carrera}, Ero and {Douezy}, Michel and {Fukui}, Keiichi and {Gamurot}, Ryan and {Goto}, Tateki and {Guillet}, Bruno and {Kuossari}, Petri and {Laugier}, Jean-Marie and {Lewin}, Pablo and {Loose}, Margaret A. and {Manganese}, Laurent and {Mirwald}, Benjamin and {Mountz}, Hubert and {Mountz}, Marti and {Ostrem}, Cory and {Parker}, Bruce and {Picard}, Patrick and {Primm}, Michael and {Randolph}, Justus and {Runge}, Jay and {Savonnet}, Robert and {Sharon}, Chelsea E. and {Shih}, Jenny and {Shimizu}, Masao and {Silvis}, George and {Simard}, Georges and {Simpson}, Alan and {Sivayogan}, Thusheeta and {Stein}, Meyer and {Trudel}, Denis and {Tsuchiyama}, Hiroaki and {Wagner}, Kevin and {Will}, Stefan},
        title = "{A 16 hr Transit of Kepler-167 e Observed by the Ground-based Unistellar Telescope Network}",
      journal = {\apjl},
         year = 2022,
        month = dec,
       volume = {940},
       number = {2},
          eid = {L39},
        pages = {L39},
          doi = {10.3847/2041-8213/aca073},
archivePrefix = {arXiv},
       eprint = {2211.01532},
 primaryClass = {astro-ph.EP},
       adsurl = {https://ui.adsabs.harvard.edu/abs/2022ApJ...940L..39P}
}

@ARTICLE{nasa_exoplanet_archive,
       author = {{Christiansen}, Jessie L. and {McElroy}, Douglas L. and {Harbut}, Marcy and {Ciardi}, David R. and {Crane}, Megan and {Good}, John and {Hardegree-Ullman}, Kevin K. and {Kesseli}, Aurora Y. and {Lund}, Michael B. and {Lynn}, Meca and {Muthiar}, Ananda and {Nilsson}, Ricky and {Oluyide}, Toba and {Papin}, Michael and {Rivera}, Amalia and {Swain}, Melanie and {Susemiehl}, Nicholas D. and {Tam}, Raymond and {van Eyken}, Julian and {Beichman}, Charles},
        title = "{The NASA Exoplanet Archive and Exoplanet Follow-up Observing Program: Data, Tools, and Usage}",
      journal = {\psj},
         year = 2025,
        month = aug,
       volume = {6},
       number = {8},
          eid = {186},
        pages = {186},
          doi = {10.3847/PSJ/ade3c2},
archivePrefix = {arXiv},
       eprint = {2506.03299},
 primaryClass = {astro-ph.EP},
       adsurl = {https://ui.adsabs.harvard.edu/abs/2025PSJ.....6..186C}
}

@ARTICLE{eureka,
       author = {{Bell}, Taylor and {Ahrer}, Eva-Maria and {Brande}, Jonathan and {Carter}, Aarynn and {Feinstein}, Adina and {Caloca}, Giannina and {Mansfield}, Megan and {Zieba}, Sebastian and {Piaulet}, Caroline and {Benneke}, Bj{\"o}rn and {Filippazzo}, Joseph and {May}, Erin and {Roy}, Pierre-Alexis and {Kreidberg}, Laura and {Stevenson}, Kevin},
        title = "{Eureka!: An End-to-End Pipeline for JWST Time-Series Observations}",
      journal = {The Journal of Open Source Software},
         year = 2022,
        month = nov,
       volume = {7},
       number = {79},
          eid = {4503},
        pages = {4503},
          doi = {10.21105/joss.04503},
archivePrefix = {arXiv},
       eprint = {2207.03585},
 primaryClass = {astro-ph.IM},
       adsurl = {https://ui.adsabs.harvard.edu/abs/2022JOSS....7.4503B}
}

@ARTICLE{transit_ers_2018,
       author = {{Bean}, Jacob L. and {Stevenson}, Kevin B. and {Batalha}, Natalie M. and {Berta-Thompson}, Zachory and {Kreidberg}, Laura and {Crouzet}, Nicolas and {Benneke}, Bj{\"o}rn and {Line}, Michael R. and {Sing}, David K. and {Wakeford}, Hannah R. and {Knutson}, Heather A. and {Kempton}, Eliza M. -R. and {D{\'e}sert}, Jean-Michel and {Crossfield}, Ian and {Batalha}, Natasha E. and {de Wit}, Julien and {Parmentier}, Vivien and {Harrington}, Joseph and {Moses}, Julianne I. and {Lopez-Morales}, Mercedes and {Alam}, Munazza K. and {Blecic}, Jasmina and {Bruno}, Giovanni and {Carter}, Aarynn L. and {Chapman}, John W. and {Decin}, Leen and {Dragomir}, Diana and {Evans}, Thomas M. and {Fortney}, Jonathan J. and {Fraine}, Jonathan D. and {Gao}, Peter and {Garc{\'\i}a Mu{\~n}oz}, Antonio and {Gibson}, Neale P. and {Goyal}, Jayesh M. and {Heng}, Kevin and {Hu}, Renyu and {Kendrew}, Sarah and {Kilpatrick}, Brian M. and {Krick}, Jessica and {Lagage}, Pierre-Olivier and {Lendl}, Monika and {Louden}, Tom and {Madhusudhan}, Nikku and {Mandell}, Avi M. and {Mansfield}, Megan and {May}, Erin M. and {Morello}, Giuseppe and {Morley}, Caroline V. and {Nikolov}, Nikolay and {Redfield}, Seth and {Roberts}, Jessica E. and {Schlawin}, Everett and {Spake}, Jessica J. and {Todorov}, Kamen O. and {Tsiaras}, Angelos and {Venot}, Olivia and {Waalkes}, William C. and {Wheatley}, Peter J. and {Zellem}, Robert T. and {Angerhausen}, Daniel and {Barrado}, David and {Carone}, Ludmila and {Casewell}, Sarah L. and {Cubillos}, Patricio E. and {Damiano}, Mario and {de Val-Borro}, Miguel and {Drummond}, Benjamin and {Edwards}, Billy and {Endl}, Michael and {Espinoza}, Nestor and {France}, Kevin and {Gizis}, John E. and {Greene}, Thomas P. and {Henning}, Thomas K. and {Hong}, Yucian and {Ingalls}, James G. and {Iro}, Nicolas and {Irwin}, Patrick G.~J. and {Kataria}, Tiffany and {Lahuis}, Fred and {Leconte}, J{\'e}r{\'e}my and {Lillo-Box}, Jorge and {Lines}, Stefan and {Lothringer}, Joshua D. and {Mancini}, Luigi and {Marchis}, Franck and {Mayne}, Nathan and {Palle}, Enric and {Rauscher}, Emily and {Roudier}, Ga{\"e}l and {Shkolnik}, Evgenya L. and {Southworth}, John and {Swain}, Mark R. and {Taylor}, Jake and {Teske}, Johanna and {Tinetti}, Giovanna and {Tremblin}, Pascal and {Tucker}, Gregory S. and {van Boekel}, Roy and {Waldmann}, Ingo P. and {Weaver}, Ian C. and {Zingales}, Tiziano},
        title = "{The Transiting Exoplanet Community Early Release Science Program for JWST}",
      journal = {\pasp},
         year = 2018,
        month = nov,
       volume = {130},
       number = {993},
        pages = {114402},
          doi = {10.1088/1538-3873/aadbf3},
archivePrefix = {arXiv},
       eprint = {1803.04985},
 primaryClass = {astro-ph.EP},
       adsurl = {https://ui.adsabs.harvard.edu/abs/2018PASP..130k4402B}
}

@misc{jwst_pipeline,
       author = {{Bushouse}, Howard and {Eisenhamer}, Jonathan and {Dencheva}, Nadia and {Davies}, James and {Greenfield}, Perry and {Morrison}, Jane and {Hodge}, Phil and {Simon}, Bernie and {Grumm}, David and {Droettboom}, Michael and {Slavich}, Edward and {Sosey}, Megan and {Pauly}, Tyler and {Miller}, Todd and {Jedrzejewski}, Robert and {Hack}, Warren and {Davis}, David and {Crawford}, Steven and {Law}, David and {Gordon}, Karl and {Regan}, Michael and {Cara}, Mihai and {MacDonald}, Ken and {Bradley}, Larry and {Shanahan}, Clare and {Jamieson}, William and {Teodoro}, Mairan and {Williams}, Thomas and {Pena-Guerrero}, Maria and {Graham}, Brett and {Molter}, Edward and {Brandt}, Timothy and {Hayes}, Christian and {Cooper}, Rachel and {Clarke}, Melanie and {Filippazzo}, Joseph},
        title = "{JWST Calibration Pipeline}",
         year = 2025,
        month = jul,
          eid = {10.5281/zenodo.6984365},
          doi = {10.5281/zenodo.6984365},
      version = {1.19.1},
    publisher = {Zenodo},
       adsurl = {https://ui.adsabs.harvard.edu/abs/2023zndo...6984365B}
}

@ARTICLE{rauscher_irs2_2017,
       author = {{Rauscher}, Bernard J. and {Arendt}, Richard G. and {Fixsen}, D.~J. and {Greenhouse}, Matthew A. and {Lander}, Matthew and {Lindler}, Don and {Loose}, Markus and {Moseley}, S.~H. and {Mott}, D. Brent and {Wen}, Yiting and {Wilson}, Donna V. and {Xenophontos}, Christos},
        title = "{Improved Reference Sampling and Subtraction: A Technique for Reducing the Read Noise of Near-infrared Detector Systems}",
      journal = {\pasp},
         year = 2017,
        month = oct,
       volume = {129},
       number = {980},
        pages = {105003},
          doi = {10.1088/1538-3873/aa83fd},
archivePrefix = {arXiv},
       eprint = {1707.09387},
 primaryClass = {astro-ph.IM},
       adsurl = {https://ui.adsabs.harvard.edu/abs/2017PASP..129j5003R}
}

@ARTICLE{rustamkulov_lab_timeseries_2022,
       author = {{Rustamkulov}, Zafar and {Sing}, David K. and {Liu}, Rongrong and {Wang}, Ashley},
        title = "{Analysis of a JWST NIRSpec Lab Time Series: Characterizing Systematics, Recovering Exoplanet Transit Spectroscopy, and Constraining a Noise Floor}",
      journal = {\apjl},
         year = 2022,
        month = mar,
       volume = {928},
       number = {1},
          eid = {L7},
        pages = {L7},
          doi = {10.3847/2041-8213/ac5b6f},
archivePrefix = {arXiv},
       eprint = {2203.04173},
 primaryClass = {astro-ph.EP},
       adsurl = {https://ui.adsabs.harvard.edu/abs/2022ApJ...928L...7R}
}

@INPROCEEDINGS{moseley_read_noise_2010,
       author = {{Moseley}, S.~H. and {Arendt}, Richard G. and {Fixsen}, D.~J. and {Lindler}, Don and {Loose}, Markus and {Rauscher}, Bernard J.},
        title = "{Reducing the read noise of H2RG detector arrays: eliminating correlated noise with efficient use of reference signals}",
    booktitle = {High Energy, Optical, and Infrared Detectors for Astronomy IV},
         year = 2010,
       editor = {{Holland}, Andrew D. and {Dorn}, David A.},
       series = {Society of Photo-Optical Instrumentation Engineers (SPIE) Conference Series},
       volume = {7742},
        month = jul,
          eid = {77421B},
        pages = {77421B},
          doi = {10.1117/12.866773},
       adsurl = {https://ui.adsabs.harvard.edu/abs/2010SPIE.7742E..1BM}
}

@ARTICLE{rauscher_nsclean_2024,
       author = {{Rauscher}, Bernard J.},
        title = "{NSClean: An Algorithm for Removing Correlated Noise from JWST NIRSpec Images}",
      journal = {\pasp},
         year = 2024,
        month = jan,
       volume = {136},
       number = {1},
          eid = {015001},
        pages = {015001},
          doi = {10.1088/1538-3873/ad1b36},
archivePrefix = {arXiv},
       eprint = {2306.03250},
 primaryClass = {astro-ph.IM},
       adsurl = {https://ui.adsabs.harvard.edu/abs/2024PASP..136a5001R}
}

@ARTICLE{horne_optimal_extraction_1986,
       author = {{Horne}, K.},
        title = "{An optimal extraction algorithm for CCD spectroscopy.}",
      journal = {\pasp},
         year = 1986,
        month = jun,
       volume = {98},
        pages = {609-617},
          doi = {10.1086/131801},
       adsurl = {https://ui.adsabs.harvard.edu/abs/1986PASP...98..609H}
}

@ARTICLE{alderson_wasp_ers_2023,
       author = {{Alderson}, Lili and {Wakeford}, Hannah R. and {Alam}, Munazza K. and {Batalha}, Natasha E. and {Lothringer}, Joshua D. and {Adams Redai}, Jea and {Barat}, Saugata and {Brande}, Jonathan and {Damiano}, Mario and {Daylan}, Tansu and {Espinoza}, N{\'e}stor and {Flagg}, Laura and {Goyal}, Jayesh M. and {Grant}, David and {Hu}, Renyu and {Inglis}, Julie and {Lee}, Elspeth K.~H. and {Mikal-Evans}, Thomas and {Ramos-Rosado}, Lakeisha and {Roy}, Pierre-Alexis and {Wallack}, Nicole L. and {Batalha}, Natalie M. and {Bean}, Jacob L. and {Benneke}, Bj{\"o}rn and {Berta-Thompson}, Zachory K. and {Carter}, Aarynn L. and {Changeat}, Quentin and {Col{\'o}n}, Knicole D. and {Crossfield}, Ian J.~M. and {D{\'e}sert}, Jean-Michel and {Foreman-Mackey}, Daniel and {Gibson}, Neale P. and {Kreidberg}, Laura and {Line}, Michael R. and {L{\'o}pez-Morales}, Mercedes and {Molaverdikhani}, Karan and {Moran}, Sarah E. and {Morello}, Giuseppe and {Moses}, Julianne I. and {Mukherjee}, Sagnick and {Schlawin}, Everett and {Sing}, David K. and {Stevenson}, Kevin B. and {Taylor}, Jake and {Aggarwal}, Keshav and {Ahrer}, Eva-Maria and {Allen}, Natalie H. and {Barstow}, Joanna K. and {Bell}, Taylor J. and {Blecic}, Jasmina and {Casewell}, Sarah L. and {Chubb}, Katy L. and {Crouzet}, Nicolas and {Cubillos}, Patricio E. and {Decin}, Leen and {Feinstein}, Adina D. and {Fortney}, Joanthan J. and {Harrington}, Joseph and {Heng}, Kevin and {Iro}, Nicolas and {Kempton}, Eliza M. -R. and {Kirk}, James and {Knutson}, Heather A. and {Krick}, Jessica and {Leconte}, J{\'e}r{\'e}my and {Lendl}, Monika and {MacDonald}, Ryan J. and {Mancini}, Luigi and {Mansfield}, Megan and {May}, Erin M. and {Mayne}, Nathan J. and {Miguel}, Yamila and {Nikolov}, Nikolay K. and {Ohno}, Kazumasa and {Palle}, Enric and {Parmentier}, Vivien and {Petit dit de la Roche}, Dominique J.~M. and {Piaulet}, Caroline and {Powell}, Diana and {Rackham}, Benjamin V. and {Redfield}, Seth and {Rogers}, Laura K. and {Rustamkulov}, Zafar and {Tan}, Xianyu and {Tremblin}, P. and {Tsai}, Shang-Min and {Turner}, Jake D. and {de Val-Borro}, Miguel and {Venot}, Olivia and {Welbanks}, Luis and {Wheatley}, Peter J. and {Zhang}, Xi},
        title = "{Early Release Science of the exoplanet WASP-39b with JWST NIRSpec G395H}",
      journal = {\nat},
         year = 2023,
        month = feb,
       volume = {614},
       number = {7949},
        pages = {664-669},
          doi = {10.1038/s41586-022-05591-3},
archivePrefix = {arXiv},
       eprint = {2211.10488},
 primaryClass = {astro-ph.EP},
       adsurl = {https://ui.adsabs.harvard.edu/abs/2023Natur.614..664A}
}

@misc{jax,
  author = {James Bradbury and Roy Frostig and Peter Hawkins and Matthew James Johnson and Chris Leary and Dougal Maclaurin and George Necula and Adam Paszke and Jake Vander{P}las and Skye Wanderman-{M}ilne and Qiao Zhang},
  title = {{JAX}: composable transformations of {P}ython+{N}um{P}y programs},
  url = {http://github.com/jax-ml/jax},
  version = {0.3.13},
  year = {2018},
}

@ARTICLE{kipping_quad_ld_2013,
       author = {{Kipping}, David M.},
        title = "{Efficient, uninformative sampling of limb darkening coefficients for two-parameter laws}",
      journal = {\mnras},
         year = 2013,
        month = nov,
       volume = {435},
       number = {3},
        pages = {2152-2160},
          doi = {10.1093/mnras/stt1435},
archivePrefix = {arXiv},
       eprint = {1308.0009},
 primaryClass = {astro-ph.SR},
       adsurl = {https://ui.adsabs.harvard.edu/abs/2013MNRAS.435.2152K}
}

@ARTICLE{kipping_cube_ld_2016,
       author = {{Kipping}, David M.},
        title = "{Efficient, uninformative sampling of limb-darkening coefficients for a three-parameter law}",
      journal = {\mnras},
         year = 2016,
        month = jan,
       volume = {455},
       number = {2},
        pages = {1680-1690},
          doi = {10.1093/mnras/stv2379},
archivePrefix = {arXiv},
       eprint = {1509.03483},
 primaryClass = {astro-ph.SR},
       adsurl = {https://ui.adsabs.harvard.edu/abs/2016MNRAS.455.1680K}
}

@ARTICLE{lange_nautilus_2023,
       author = {{Lange}, Johannes U.},
        title = "{NAUTILUS: boosting Bayesian importance nested sampling with deep learning}",
      journal = {\mnras},
         year = 2023,
        month = oct,
       volume = {525},
       number = {2},
        pages = {3181-3194},
          doi = {10.1093/mnras/stad2441},
archivePrefix = {arXiv},
       eprint = {2306.16923},
 primaryClass = {astro-ph.IM},
       adsurl = {https://ui.adsabs.harvard.edu/abs/2023MNRAS.525.3181L}
}

@article{numpyro,
  doi = {10.48550/ARXIV.1912.11554},
  url = {https://arxiv.org/abs/1912.11554},
  author = {Phan,  Du and Pradhan,  Neeraj and Jankowiak,  Martin},
  title = {Composable Effects for Flexible and Accelerated Probabilistic Programming in NumPyro},
  publisher = {arXiv},
  journal = {arXiv},
  year = {2019},
  copyright = {arXiv.org perpetual,  non-exclusive license}
}

@article{NUTS,
  doi = {10.48550/ARXIV.1111.4246},
  url = {https://arxiv.org/abs/1111.4246},
  author = {Hoffman,  Matthew D. and Gelman,  Andrew},
  title = {The No-U-Turn Sampler: Adaptively Setting Path Lengths in Hamiltonian Monte Carlo},
  publisher = {arXiv},
  journal = {arXiv},
  year = {2011},
  copyright = {arXiv.org perpetual,  non-exclusive license}
}

@INPROCEEDINGS{skilling_ns_2004,
       author = {{Skilling}, John},
        title = "{Nested Sampling}",
    booktitle = {Bayesian Inference and Maximum Entropy Methods in Science and Engineering: 24th International Workshop on Bayesian Inference and Maximum Entropy Methods in Science and Engineering},
         year = 2004,
       editor = {{Fischer}, Rainer and {Preuss}, Roland and {Toussaint}, Udo Von},
       series = {American Institute of Physics Conference Series},
       volume = {735},
        month = nov,
        pages = {395-405},
          doi = {10.1063/1.1835238},
       adsurl = {https://ui.adsabs.harvard.edu/abs/2004AIPC..735..395S}
}

@ARTICLE{aston_ns_primer_2022,
       author = {{Ashton}, Greg and {Bernstein}, Noam and {Buchner}, Johannes and {Chen}, Xi and {Cs{\'a}nyi}, G{\'a}bor and {Fowlie}, Andrew and {Feroz}, Farhan and {Griffiths}, Matthew and {Handley}, Will and {Habeck}, Michael and {Higson}, Edward and {Hobson}, Michael and {Lasenby}, Anthony and {Parkinson}, David and {P{\'a}rtay}, Livia B. and {Pitkin}, Matthew and {Schneider}, Doris and {Speagle}, Joshua S. and {South}, Leah and {Veitch}, John and {Wacker}, Philipp and {Wales}, David J. and {Yallup}, David},
        title = "{Nested sampling for physical scientists}",
      journal = {Nature Reviews Methods Primers},
         year = 2022,
        month = may,
       volume = {2},
          eid = {39},
        pages = {39},
          doi = {10.1038/s43586-022-00121-x},
archivePrefix = {arXiv},
       eprint = {2205.15570},
 primaryClass = {stat.CO},
       adsurl = {https://ui.adsabs.harvard.edu/abs/2022NRvMP...2...39A}
}

@ARTICLE{multinest_1,
       author = {{Feroz}, F. and {Hobson}, M.~P.},
        title = "{Multimodal nested sampling: an efficient and robust alternative to Markov Chain Monte Carlo methods for astronomical data analyses}",
      journal = {\mnras},
         year = 2008,
        month = feb,
       volume = {384},
       number = {2},
        pages = {449-463},
          doi = {10.1111/j.1365-2966.2007.12353.x},
archivePrefix = {arXiv},
       eprint = {0704.3704},
 primaryClass = {astro-ph},
       adsurl = {https://ui.adsabs.harvard.edu/abs/2008MNRAS.384..449F}
}

@ARTICLE{multinest_2,
       author = {{Feroz}, F. and {Hobson}, M.~P. and {Bridges}, M.},
        title = "{MULTINEST: an efficient and robust Bayesian inference tool for cosmology and particle physics}",
      journal = {\mnras},
         year = 2009,
        month = oct,
       volume = {398},
       number = {4},
        pages = {1601-1614},
          doi = {10.1111/j.1365-2966.2009.14548.x},
archivePrefix = {arXiv},
       eprint = {0809.3437},
 primaryClass = {astro-ph},
       adsurl = {https://ui.adsabs.harvard.edu/abs/2009MNRAS.398.1601F}
}

@ARTICLE{multinest_3,
       author = {{Feroz}, Farhan and {Hobson}, Michael P. and {Cameron}, Ewan and {Pettitt}, Anthony N.},
        title = "{Importance Nested Sampling and the MultiNest Algorithm}",
      journal = {The Open Journal of Astrophysics},
         year = 2019,
        month = nov,
       volume = {2},
       number = {1},
          eid = {10},
        pages = {10},
          doi = {10.21105/astro.1306.2144},
archivePrefix = {arXiv},
       eprint = {1306.2144},
 primaryClass = {astro-ph.IM},
       adsurl = {https://ui.adsabs.harvard.edu/abs/2019OJAp....2E..10F}
}

@ARTICLE{dynesty,
       author = {{Speagle}, Joshua S.},
        title = "{DYNESTY: a dynamic nested sampling package for estimating Bayesian posteriors and evidences}",
      journal = {\mnras},
         year = 2020,
        month = apr,
       volume = {493},
       number = {3},
        pages = {3132-3158},
          doi = {10.1093/mnras/staa278},
archivePrefix = {arXiv},
       eprint = {1904.02180},
 primaryClass = {astro-ph.IM},
       adsurl = {https://ui.adsabs.harvard.edu/abs/2020MNRAS.493.3132S}
}

@ARTICLE{ultranest,
       author = {{Buchner}, Johannes},
        title = "{UltraNest - a robust, general purpose Bayesian inference engine}",
      journal = {The Journal of Open Source Software},
         year = 2021,
        month = apr,
       volume = {6},
       number = {60},
          eid = {3001},
        pages = {3001},
          doi = {10.21105/joss.03001},
archivePrefix = {arXiv},
       eprint = {2101.09604},
 primaryClass = {stat.CO},
       adsurl = {https://ui.adsabs.harvard.edu/abs/2021JOSS....6.3001B}
}

@ARTICLE{buchner_nested_review_2023,
       author = {{Buchner}, Johannes},
        title = "{Nested Sampling Methods}",
      journal = {Statistics Surveys},
         year = 2023,
        month = jan,
       volume = {17},
        pages = {169-215},
          doi = {10.1214/23-SS144},
archivePrefix = {arXiv},
       eprint = {2101.09675},
 primaryClass = {stat.CO},
       adsurl = {https://ui.adsabs.harvard.edu/abs/2023StSur..17..169B}
}

@ARTICLE{exotic_ld,
       author = {{Grant}, David and {Wakeford}, Hannah},
        title = "{ExoTiC-LD: thirty seconds to stellar limb-darkening coefficients}",
      journal = {The Journal of Open Source Software},
         year = 2024,
        month = aug,
       volume = {9},
       number = {100},
          eid = {6816},
        pages = {6816},
          doi = {10.21105/joss.06816},
archivePrefix = {arXiv},
       eprint = {2408.10341},
 primaryClass = {astro-ph.IM},
       adsurl = {https://ui.adsabs.harvard.edu/abs/2024JOSS....9.6816G}
}

@ARTICLE{agol_luger_dfm_2020,
       author = {{Agol}, Eric and {Luger}, Rodrigo and {Foreman-Mackey}, Daniel},
        title = "{Analytic Planetary Transit Light Curves and Derivatives for Stars with Polynomial Limb Darkening}",
      journal = {\aj},
         year = 2020,
        month = mar,
       volume = {159},
       number = {3},
          eid = {123},
        pages = {123},
          doi = {10.3847/1538-3881/ab4fee},
archivePrefix = {arXiv},
       eprint = {1908.03222},
 primaryClass = {astro-ph.EP},
       adsurl = {https://ui.adsabs.harvard.edu/abs/2020AJ....159..123A}
}

@ARTICLE{espinoza_ld_2015,
       author = {{Espinoza}, N{\'e}stor and {Jord{\'a}n}, Andr{\'e}s},
        title = "{Limb darkening and exoplanets: testing stellar model atmospheres and identifying biases in transit parameters}",
      journal = {\mnras},
         year = 2015,
        month = jun,
       volume = {450},
       number = {2},
        pages = {1879-1899},
          doi = {10.1093/mnras/stv744},
archivePrefix = {arXiv},
       eprint = {1503.07020},
 primaryClass = {astro-ph.EP},
       adsurl = {https://ui.adsabs.harvard.edu/abs/2015MNRAS.450.1879E}
}

@ARTICLE{kurucz_grid,
       author = {{Kurucz}, Robert},
        title = "{ATLAS9 Stellar Atmosphere Programs and 2 km/s grid.}",
      journal = {Robert Kurucz CD-ROM},
         year = 1993,
        month = jan,
       volume = {13},
       adsurl = {https://ui.adsabs.harvard.edu/abs/1993KurCD..13.....K}
}

@ARTICLE{phoenix_grid,
       author = {{Husser}, T. -O. and {Wende-von Berg}, S. and {Dreizler}, S. and {Homeier}, D. and {Reiners}, A. and {Barman}, T. and {Hauschildt}, P.~H.},
        title = "{A new extensive library of PHOENIX stellar atmospheres and synthetic spectra}",
      journal = {\aap},
         year = 2013,
        month = may,
       volume = {553},
          eid = {A6},
        pages = {A6},
          doi = {10.1051/0004-6361/201219058},
archivePrefix = {arXiv},
       eprint = {1303.5632},
 primaryClass = {astro-ph.SR},
       adsurl = {https://ui.adsabs.harvard.edu/abs/2013A&A...553A...6H}
}

@ARTICLE{mps_grid_a,
       author = {{Kostogryz}, N. and {Shapiro}, A.~I. and {Witzke}, V. and {Grant}, D. and {Wakeford}, H.~R. and {Stevenson}, K.~B. and {Solanki}, S.~K. and {Gizon}, L.},
        title = "{MPS-ATLAS Library of Stellar Model Atmospheres and Spectra}",
      journal = {Research Notes of the American Astronomical Society},
         year = 2023,
        month = mar,
       volume = {7},
       number = {3},
          eid = {39},
        pages = {39},
          doi = {10.3847/2515-5172/acc180},
archivePrefix = {arXiv},
       eprint = {2303.02685},
 primaryClass = {astro-ph.SR},
       adsurl = {https://ui.adsabs.harvard.edu/abs/2023RNAAS...7...39K}
}

@ARTICLE{mps_grid_b,
       author = {{Kostogryz}, N.~M. and {Witzke}, V. and {Shapiro}, A.~I. and {Solanki}, S.~K. and {Maxted}, P.~F.~L. and {Kurucz}, R.~L. and {Gizon}, L.},
        title = "{Stellar limb darkening. A new MPS-ATLAS library for Kepler, TESS, CHEOPS, and PLATO passbands}",
      journal = {\aap},
         year = 2022,
        month = oct,
       volume = {666},
          eid = {A60},
        pages = {A60},
          doi = {10.1051/0004-6361/202243722},
archivePrefix = {arXiv},
       eprint = {2206.06641},
 primaryClass = {astro-ph.SR},
       adsurl = {https://ui.adsabs.harvard.edu/abs/2022A&A...666A..60K}
}

@ARTICLE{stagger_grid,
       author = {{Magic}, Z. and {Chiavassa}, A. and {Collet}, R. and {Asplund}, M.},
        title = "{The Stagger-grid: A grid of 3D stellar atmosphere models. IV. Limb darkening coefficients}",
      journal = {\aap},
         year = 2015,
        month = jan,
       volume = {573},
          eid = {A90},
        pages = {A90},
          doi = {10.1051/0004-6361/201423804},
archivePrefix = {arXiv},
       eprint = {1403.3487},
 primaryClass = {astro-ph.SR},
       adsurl = {https://ui.adsabs.harvard.edu/abs/2015A&A...573A..90M}
}

@ARTICLE{mandel_agol_2002,
       author = {{Mandel}, Kaisey and {Agol}, Eric},
        title = "{Analytic Light Curves for Planetary Transit Searches}",
      journal = {\apjl},
         year = 2002,
        month = dec,
       volume = {580},
       number = {2},
        pages = {L171-L175},
          doi = {10.1086/345520},
archivePrefix = {arXiv},
       eprint = {astro-ph/0210099},
 primaryClass = {astro-ph},
       adsurl = {https://ui.adsabs.harvard.edu/abs/2002ApJ...580L.171M}
}

@ARTICLE{fowlie_2020,
       author = {{Fowlie}, Andrew and {Handley}, Will and {Su}, Liangliang},
        title = "{Nested sampling cross-checks using order statistics}",
      journal = {\mnras},
         year = 2020,
        month = oct,
       volume = {497},
       number = {4},
        pages = {5256-5263},
          doi = {10.1093/mnras/staa2345},
archivePrefix = {arXiv},
       eprint = {2006.03371},
 primaryClass = {stat.CO},
       adsurl = {https://ui.adsabs.harvard.edu/abs/2020MNRAS.497.5256F}
}

@ARTICLE{berardo_ob_effects_2022,
       author = {{Berardo}, David and {de Wit}, Julien},
        title = "{On the Effects of Planetary Oblateness on Exoplanet Studies}",
      journal = {\apj},
         year = 2022,
        month = aug,
       volume = {935},
       number = {2},
          eid = {178},
        pages = {178},
          doi = {10.3847/1538-4357/ac82b2},
archivePrefix = {arXiv},
       eprint = {2207.07670},
 primaryClass = {astro-ph.EP},
       adsurl = {https://ui.adsabs.harvard.edu/abs/2022ApJ...935..178B}
}

@ARTICLE{claret_2000,
       author = {{Claret}, A.},
        title = "{A new non-linear limb-darkening law for LTE stellar atmosphere models. Calculations for -5.0 <= log[M/H] <= +1, 2000 K <= T$_{eff}$ <= 50000 K at several surface gravities}",
      journal = {\aap},
         year = 2000,
        month = nov,
       volume = {363},
        pages = {1081-1190},
       adsurl = {https://ui.adsabs.harvard.edu/abs/2000A&A...363.1081C}
}

@ARTICLE{alam_compass_2025,
       author = {{Alam}, Munazza K. and {Gao}, Peter and {Adams Redai}, Jea and {Wallack}, Nicole L. and {Wogan}, Nicholas F. and {Aguichine}, Artyom and {Dattilo}, Anne and {Alderson}, Lili and {Batalha}, Natasha E. and {Batalha}, Natalie M. and {Kirk}, James and {L{\'o}pez-Morales}, Mercedes and {Meech}, Annabella and {Moran}, Sarah E. and {Teske}, Johanna and {Wakeford}, Hannah R. and {Wolfgang}, Angie},
        title = "{JWST COMPASS: The First Near- to Mid-infrared Transmission Spectrum of the Hot Super-Earth L 168-9 b}",
      journal = {\aj},
         year = 2025,
        month = jan,
       volume = {169},
       number = {1},
          eid = {15},
        pages = {15},
          doi = {10.3847/1538-3881/ad8eb5},
archivePrefix = {arXiv},
       eprint = {2411.03154},
 primaryClass = {astro-ph.EP},
       adsurl = {https://ui.adsabs.harvard.edu/abs/2025AJ....169...15A}
}

@ARTICLE{squishyplanet,
       author = {{Cassese}, Ben and {Vega}, Justin and {Lu}, Tiger and {Rice}, Malena and {Poddar}, Avishi and {Kipping}, David},
        title = "{squishyplanet: modeling transits of non-spherical exoplanets in JAX}",
      journal = {The Journal of Open Source Software},
         year = 2024,
        month = aug,
       volume = {9},
       number = {100},
          eid = {6972},
        pages = {6972},
          doi = {10.21105/joss.06972},
       adsurl = {https://ui.adsabs.harvard.edu/abs/2024JOSS....9.6972C}
}

@ARTICLE{alam_hst_spectrum_2022,
       author = {{Alam}, Munazza K. and {Kirk}, James and {Dressing}, Courtney D. and {L{\'o}pez-Morales}, Mercedes and {Ohno}, Kazumasa and {Gao}, Peter and {Akinsanmi}, Babatunde and {Santerne}, Alexandre and {Grouffal}, Salom{\'e} and {Adibekyan}, Vardan and {Barros}, Susana C.~C. and {Buchhave}, Lars A. and {Crossfield}, Ian J.~M. and {Dai}, Fei and {Deleuil}, Magali and {Giacalone}, Steven and {Lillo-Box}, Jorge and {Marley}, Mark and {Mayo}, Andrew W. and {Mortier}, Annelies and {Santos}, Nuno C. and {Sousa}, S{\'e}rgio G. and {Turtelboom}, Emma V. and {Wheatley}, Peter J. and {Vanderburg}, Andrew M.},
        title = "{The First Near-infrared Transmission Spectrum of HIP 41378 f, A Low-mass Temperate Jovian World in a Multiplanet System}",
      journal = {\apjl},
         year = 2022,
        month = mar,
       volume = {927},
       number = {1},
          eid = {L5},
        pages = {L5},
          doi = {10.3847/2041-8213/ac559d},
archivePrefix = {arXiv},
       eprint = {2201.02686},
 primaryClass = {astro-ph.EP},
       adsurl = {https://ui.adsabs.harvard.edu/abs/2022ApJ...927L...5A}
}

@ARTICLE{patel_ld_2022,
       author = {{Patel}, Jayshil A. and {Espinoza}, N{\'e}stor},
        title = "{Empirical Limb-darkening Coefficients and Transit Parameters of Known Exoplanets from TESS}",
      journal = {\aj},
         year = 2022,
        month = may,
       volume = {163},
       number = {5},
          eid = {228},
        pages = {228},
          doi = {10.3847/1538-3881/ac5f55},
archivePrefix = {arXiv},
       eprint = {2203.05661},
 primaryClass = {astro-ph.EP},
       adsurl = {https://ui.adsabs.harvard.edu/abs/2022AJ....163..228P}
}

@ARTICLE{wallack_compass_2024,
       author = {{Wallack}, Nicole L. and {Batalha}, Natasha E. and {Alderson}, Lili and {Scarsdale}, Nicholas and {Adams Redai}, Jea I. and {Aguichine}, Artyom and {Alam}, Munazza K. and {Gao}, Peter and {Wolfgang}, Angie and {Batalha}, Natalie M. and {Kirk}, James and {L{\'o}pez-Morales}, Mercedes and {Moran}, Sarah E. and {Teske}, Johanna and {Wakeford}, Hannah R. and {Wogan}, Nicholas F.},
        title = "{JWST COMPASS: A NIRSpec/G395H Transmission Spectrum of the Sub-Neptune TOI-836c}",
      journal = {\aj},
         year = 2024,
        month = aug,
       volume = {168},
       number = {2},
          eid = {77},
        pages = {77},
          doi = {10.3847/1538-3881/ad3917},
archivePrefix = {arXiv},
       eprint = {2404.01264},
 primaryClass = {astro-ph.EP},
       adsurl = {https://ui.adsabs.harvard.edu/abs/2024AJ....168...77W}
}

@ARTICLE{rustamkulov_ers_prism_2023,
       author = {{Rustamkulov}, Zafar and {Sing}, D.~K. and {Mukherjee}, S. and {May}, E.~M. and {Kirk}, J. and {Schlawin}, E. and {Line}, M.~R. and {Piaulet}, C. and {Carter}, A.~L. and {Batalha}, N.~E. and {Goyal}, J.~M. and {L{\'o}pez-Morales}, M. and {Lothringer}, J.~D. and {MacDonald}, R.~J. and {Moran}, S.~E. and {Stevenson}, K.~B. and {Wakeford}, H.~R. and {Espinoza}, N. and {Bean}, J.~L. and {Batalha}, N.~M. and {Benneke}, B. and {Berta-Thompson}, Z.~K. and {Crossfield}, I.~J.~M. and {Gao}, P. and {Kreidberg}, L. and {Powell}, D.~K. and {Cubillos}, P.~E. and {Gibson}, N.~P. and {Leconte}, J. and {Molaverdikhani}, K. and {Nikolov}, N.~K. and {Parmentier}, V. and {Roy}, P. and {Taylor}, J. and {Turner}, J.~D. and {Wheatley}, P.~J. and {Aggarwal}, K. and {Ahrer}, E. and {Alam}, M.~K. and {Alderson}, L. and {Allen}, N.~H. and {Banerjee}, A. and {Barat}, S. and {Barrado}, D. and {Barstow}, J.~K. and {Bell}, T.~J. and {Blecic}, J. and {Brande}, J. and {Casewell}, S. and {Changeat}, Q. and {Chubb}, K.~L. and {Crouzet}, N. and {Daylan}, T. and {Decin}, L. and {D{\'e}sert}, J. and {Mikal-Evans}, T. and {Feinstein}, A.~D. and {Flagg}, L. and {Fortney}, J.~J. and {Harrington}, J. and {Heng}, K. and {Hong}, Y. and {Hu}, R. and {Iro}, N. and {Kataria}, T. and {Kempton}, E.~M. -R. and {Krick}, J. and {Lendl}, M. and {Lillo-Box}, J. and {Louca}, A. and {Lustig-Yaeger}, J. and {Mancini}, L. and {Mansfield}, M. and {Mayne}, N.~J. and {Miguel}, Y. and {Morello}, G. and {Ohno}, K. and {Palle}, E. and {Petit dit de la Roche}, D.~J.~M. and {Rackham}, B.~V. and {Radica}, M. and {Ramos-Rosado}, L. and {Redfield}, S. and {Rogers}, L.~K. and {Shkolnik}, E.~L. and {Southworth}, J. and {Teske}, J. and {Tremblin}, P. and {Tucker}, G.~S. and {Venot}, O. and {Waalkes}, W.~C. and {Welbanks}, L. and {Zhang}, X. and {Zieba}, S.},
        title = "{Early Release Science of the exoplanet WASP-39b with JWST NIRSpec PRISM}",
      journal = {\nat},
         year = 2023,
        month = feb,
       volume = {614},
       number = {7949},
        pages = {659-663},
          doi = {10.1038/s41586-022-05677-y},
archivePrefix = {arXiv},
       eprint = {2211.10487},
 primaryClass = {astro-ph.EP},
       adsurl = {https://ui.adsabs.harvard.edu/abs/2023Natur.614..659R}
}

@ARTICLE{teske_compass_2025,
       author = {{Teske}, Johanna and {Batalha}, Natasha E. and {Wallack}, Nicole L. and {Kirk}, James and {Wogan}, Nicholas F. and {Gordon}, Tyler A. and {Alam}, Munazza K. and {Aguichine}, Artyom and {Wolfgang}, Angie and {Wakeford}, Hannah R. and {Scarsdale}, Nicholas and {Adams Redai}, Jea and {Moran}, Sarah E. and {L{\'o}pez-Morales}, Mercedes and {Meech}, Annabella and {Gao}, Peter and {Batalha}, Natalie M. and {Alderson}, Lili and {Gagnebin}, Anna},
        title = "{JWST COMPASS: NIRSpec/G395H Transmission Observations of TOI-776 c, a 2 R$_{{\ensuremath{\oplus}}}$ M Dwarf Planet}",
      journal = {\aj},
         year = 2025,
        month = may,
       volume = {169},
       number = {5},
          eid = {249},
        pages = {249},
          doi = {10.3847/1538-3881/adb975},
archivePrefix = {arXiv},
       eprint = {2502.20501},
 primaryClass = {astro-ph.EP},
       adsurl = {https://ui.adsabs.harvard.edu/abs/2025AJ....169..249T}
}

@misc{tinygp,
       author = {{Foreman-Mackey}, Daniel},
        title = "{dfm/tinygp: The tiniest of Gaussian Process libraries}",
         year = 2023,
        month = feb,
          eid = {10.5281/zenodo.6389737},
          doi = {10.5281/zenodo.6389737},
      version = {0.3.0},
    publisher = {Zenodo},
       adsurl = {https://ui.adsabs.harvard.edu/abs/2024zndo...6389737F}
}

@article{adam,
  title={Adam: A method for stochastic optimization},
  author={Kingma, Diederik P},
  journal={arXiv preprint arXiv:1412.6980},
  year={2014}
}

@ARTICLE{kipping_binning_sinning_2010,
       author = {{Kipping}, David},
        title = "{Binning is sinning: morphological light-curve distortions due to finite integration time}",
      journal = {\mnras},
         year = 2010,
        month = nov,
       volume = {408},
       number = {3},
        pages = {1758-1769},
          doi = {10.1111/j.1365-2966.2010.17242.x},
archivePrefix = {arXiv},
       eprint = {1004.3741},
 primaryClass = {astro-ph.EP},
       adsurl = {https://ui.adsabs.harvard.edu/abs/2010MNRAS.408.1758K}
}

@ARTICLE{starry,
       author = {{Luger}, Rodrigo and {Agol}, Eric and {Foreman-Mackey}, Daniel and {Fleming}, David P. and {Lustig-Yaeger}, Jacob and {Deitrick}, Russell},
        title = "{starry: Analytic Occultation Light Curves}",
      journal = {\aj},
         year = 2019,
        month = feb,
       volume = {157},
       number = {2},
          eid = {64},
        pages = {64},
          doi = {10.3847/1538-3881/aae8e5},
archivePrefix = {arXiv},
       eprint = {1810.06559},
 primaryClass = {astro-ph.IM},
       adsurl = {https://ui.adsabs.harvard.edu/abs/2019AJ....157...64L}
}

@BOOK{murray_dermott,
       author = {{Murray}, Carl D. and {Dermott}, Stanley F.},
        title = "{Solar System Dynamics}",
         year = 1999,
          doi = {10.1017/CBO9781139174817},
        publisher = "{Cambridge University Press}",
       adsurl = {https://ui.adsabs.harvard.edu/abs/1999ssd..book.....M}
}

@ARTICLE{espinoza_commissioning_2023,
       author = {{Espinoza}, N{\'e}stor and {{\'U}beda}, Leonardo and {Birkmann}, Stephan M. and {Ferruit}, Pierre and {Valenti}, Jeff A. and {Sing}, David K. and {Rustamkulov}, Zafar and {Regan}, Michael and {Kendrew}, Sarah and {Sabbi}, Elena and {Schlawin}, Everett and {Beatty}, Thomas and {Albert}, Lo{\"\i}c and {Greene}, Thomas P. and {Nikolov}, Nikolay and {Karakla}, Diane and {Keyes}, Charles and {Alves de Oliveira}, Catarina and {B{\"o}ker}, Torsten and {Pena-Guerrero}, Maria and {Giardino}, Giovanna and {Kumari}, Nimisha and {Manjavacas}, Elena and {Proffitt}, Charles and {Rawle}, Timothy},
        title = "{Spectroscopic Time-series Performance of JWST/NIRSpec from Commissioning Observations}",
      journal = {\pasp},
         year = 2023,
        month = jan,
       volume = {135},
       number = {1043},
          eid = {018002},
        pages = {018002},
          doi = {10.1088/1538-3873/aca3d3},
archivePrefix = {arXiv},
       eprint = {2211.01459},
 primaryClass = {astro-ph.EP},
       adsurl = {https://ui.adsabs.harvard.edu/abs/2023PASP..135a8002E}
}

@MISC{compass,
       author = {{Batalha}, Natasha and {Teske}, Johanna and {Alam}, Munazza and {Alderson}, Lili and {Batalha}, Natalie and {Gao}, Peter and {Lopez-Morales}, Mercedes and {Marley}, Mark S. and {Shahar}, Anat and {Wakeford}, Hannah and {Wolfgang}, Angie},
        title = "{Seeing the Forest and the Trees: Unveiling Small Planet Atmospheres with a Population-Level Framework}",
 howpublished = {JWST Proposal. Cycle 1, ID. \#2512},
         year = 2021,
        month = mar,
        pages = {2512},
       adsurl = {https://ui.adsabs.harvard.edu/abs/2021jwst.prop.2512B}
}

@MISC{hot_rocks,
       author = {{Diamond-Lowe}, Hannah and {Mendonca}, Joao Manuel and {Akin}, Can Jan and {Allen}, Natalie and {Baungaard}, Mette and {Borsato}, Nicholas and {Buchhave}, Lars A. and {Burgasser}, Adam J. and {Demory}, Brice-Olivier and {Espinoza}, Nestor and {Fisher}, Chloe and {Fortune}, Mark and {Gibson}, Neale and {Gressier}, Amelie and {Guzman Mesa}, Andrea and {Heng}, Kevin and {Hoeijmakers}, Jens and {Hooton}, Matthew and {Jones}, Kathryn and {Kitzmann}, Daniel and {Lueber}, Anna and {Meier Valdes}, Erik Andreas and {Prinoth}, Bibiana and {Rathcke}, Alexander and {Tian}, Meng},
        title = "{The Hot Rocks Survey: Testing 9 Irradiated Terrestrial Exoplanets for Atmospheres}",
 howpublished = {JWST Proposal. Cycle 2, ID. \#3730},
         year = 2023,
        month = may,
        pages = {3730},
       adsurl = {https://ui.adsabs.harvard.edu/abs/2023jwst.prop.3730D}
}

@ARTICLE{hui_seager_2002,
       author = {{Hui}, Lam and {Seager}, Sara},
        title = "{Atmospheric Lensing and Oblateness Effects during an Extrasolar Planetary Transit}",
      journal = {\apj},
         year = 2002,
        month = jun,
       volume = {572},
       number = {1},
        pages = {540-555},
          doi = {10.1086/340017},
archivePrefix = {arXiv},
       eprint = {astro-ph/0103329},
 primaryClass = {astro-ph},
       adsurl = {https://ui.adsabs.harvard.edu/abs/2002ApJ...572..540H}
}

@ARTICLE{gardner_jwst_2006,
       author = {{Gardner}, Jonathan P. and {Mather}, John C. and {Clampin}, Mark and {Doyon}, Rene and {Greenhouse}, Matthew A. and {Hammel}, Heidi B. and {Hutchings}, John B. and {Jakobsen}, Peter and {Lilly}, Simon J. and {Long}, Knox S. and {Lunine}, Jonathan I. and {McCaughrean}, Mark J. and {Mountain}, Matt and {Nella}, John and {Rieke}, George H. and {Rieke}, Marcia J. and {Rix}, Hans-Walter and {Smith}, Eric P. and {Sonneborn}, George and {Stiavelli}, Massimo and {Stockman}, H.~S. and {Windhorst}, Rogier A. and {Wright}, Gillian S.},
        title = "{The James Webb Space Telescope}",
      journal = {\ssr},
         year = 2006,
        month = apr,
       volume = {123},
       number = {4},
        pages = {485-606},
          doi = {10.1007/s11214-006-8315-7},
archivePrefix = {arXiv},
       eprint = {astro-ph/0606175},
 primaryClass = {astro-ph},
       adsurl = {https://ui.adsabs.harvard.edu/abs/2006SSRv..123..485G}
}

@ARTICLE{barnes_fortney,
       author = {{Barnes}, Jason W. and {Fortney}, Jonathan J.},
        title = "{Measuring the Oblateness and Rotation of Transiting Extrasolar Giant Planets}",
      journal = {\apj},
         year = 2003,
        month = may,
       volume = {588},
       number = {1},
        pages = {545-556},
          doi = {10.1086/373893},
archivePrefix = {arXiv},
       eprint = {astro-ph/0301156},
 primaryClass = {astro-ph},
       adsurl = {https://ui.adsabs.harvard.edu/abs/2003ApJ...588..545B}
}

@ARTICLE{leconte_2011,
       author = {{Leconte}, J. and {Lai}, D. and {Chabrier}, G.},
        title = "{Distorted, nonspherical transiting planets: impact on the transit depth and on the radius determination}",
      journal = {\aap},
         year = 2011,
        month = apr,
       volume = {528},
          eid = {A41},
        pages = {A41},
          doi = {10.1051/0004-6361/201015811},
archivePrefix = {arXiv},
       eprint = {1101.2813},
 primaryClass = {astro-ph.EP},
       adsurl = {https://ui.adsabs.harvard.edu/abs/2011A&A...528A..41L}
}

@ARTICLE{akinsanmi_2019,
       author = {{Akinsanmi}, B. and {Barros}, S.~C.~C. and {Santos}, N.~C. and {Correia}, A.~C.~M. and {Maxted}, P.~F.~L. and {Bou{\'e}}, G. and {Laskar}, J.},
        title = "{Detectability of shape deformation in short-period exoplanets}",
      journal = {\aap},
         year = 2019,
        month = jan,
       volume = {621},
          eid = {A117},
        pages = {A117},
          doi = {10.1051/0004-6361/201834215},
archivePrefix = {arXiv},
       eprint = {1812.04538},
 primaryClass = {astro-ph.EP},
       adsurl = {https://ui.adsabs.harvard.edu/abs/2019A&A...621A.117A}
}

@ARTICLE{akinsanmi_2020,
       author = {{Akinsanmi}, B. and {Barros}, S.~C.~C. and {Santos}, N.~C. and {Oshagh}, M. and {Serrano}, L.~M.},
        title = "{Constraining the oblateness of transiting planets with photometry and spectroscopy}",
      journal = {\mnras},
         year = 2020,
        month = sep,
       volume = {497},
       number = {3},
        pages = {3484-3492},
          doi = {10.1093/mnras/staa2164},
archivePrefix = {arXiv},
       eprint = {2007.11221},
 primaryClass = {astro-ph.EP},
       adsurl = {https://ui.adsabs.harvard.edu/abs/2020MNRAS.497.3484A}
}

@ARTICLE{carter_winn_depth_2010,
       author = {{Carter}, Joshua A. and {Winn}, Joshua N.},
        title = "{The Detectability of Transit Depth Variations Due to Exoplanetary Oblateness and Spin Precession}",
      journal = {\apj},
         year = 2010,
        month = jun,
       volume = {716},
       number = {1},
        pages = {850-856},
          doi = {10.1088/0004-637X/716/1/850},
archivePrefix = {arXiv},
       eprint = {1005.1663},
 primaryClass = {astro-ph.EP},
       adsurl = {https://ui.adsabs.harvard.edu/abs/2010ApJ...716..850C}
}

@ARTICLE{barros_2022,
       author = {{Barros}, S.~C.~C. and {Akinsanmi}, B. and {Bou{\'e}}, G. and {Smith}, A.~M.~S. and {Laskar}, J. and {Ulmer-Moll}, S. and {Lillo-Box}, J. and {Queloz}, D. and {Cameron}, A. Collier and {Sousa}, S.~G. and {Ehrenreich}, D. and {Hooton}, M.~J. and {Bruno}, G. and {Demory}, B.-O. and {Correia}, A.~C.~M. and {Demangeon}, O.~D.~S. and {Wilson}, T.~G. and {Bonfanti}, A. and {Hoyer}, S. and {Alibert}, Y. and {Alonso}, R. and {Escud{\'e}}, G. Anglada and {Barbato}, D. and {B{\'a}rczy}, T. and {Barrado}, D. and {Baumjohann}, W. and {Beck}, M. and {Beck}, T. and {Benz}, W. and {Bergomi}, M. and {Billot}, N. and {Bonfils}, X. and {Bouchy}, F. and {Brandeker}, A. and {Broeg}, C. and {Cabrera}, J. and {Cessa}, V. and {Charnoz}, S. and {Damme}, C.~C.~V. and {Davies}, M.~B. and {Deleuil}, M. and {Deline}, A. and {Delrez}, L. and {Erikson}, A. and {Fortier}, A. and {Fossati}, L. and {Fridlund}, M. and {Gandolfi}, D. and {Mu{\~n}oz}, A. Garc{\'\i}a and {Gillon}, M. and {G{\"u}del}, M. and {Isaak}, K.~G. and {Heng}, K. and {Kiss}, L. and {des Etangs}, A. Lecavelier and {Lendl}, M. and {Lovis}, C. and {Magrin}, D. and {Nascimbeni}, V. and {Maxted}, P.~F.~L. and {Olofsson}, G. and {Ottensamer}, R. and {Pagano}, I. and {Pall{\'e}}, E. and {Parviainen}, H. and {Peter}, G. and {Piotto}, G. and {Pollacco}, D. and {Ragazzoni}, R. and {Rando}, N. and {Rauer}, H. and {Ribas}, I. and {Santos}, N.~C. and {Scandariato}, G. and {S{\'e}gransan}, D. and {Simon}, A.~E. and {Steller}, M. and {Szab{\'o}}, Gy. M. and {Thomas}, N. and {Udry}, S. and {Ulmer}, B. and {Van Grootel}, V. and {Walton}, N.~A.},
        title = "{Detection of the tidal deformation of WASP-103b at 3 {\ensuremath{\sigma}} with CHEOPS}",
      journal = {\aap},
         year = 2022,
        month = jan,
       volume = {657},
          eid = {A52},
        pages = {A52},
          doi = {10.1051/0004-6361/202142196},
archivePrefix = {arXiv},
       eprint = {2201.03328},
 primaryClass = {astro-ph.EP},
       adsurl = {https://ui.adsabs.harvard.edu/abs/2022A&A...657A..52B}
}

@ARTICLE{akinsanmi_2024,
       author = {{Akinsanmi}, B. and {Barros}, S.~C.~C. and {Lendl}, M. and {Carone}, L. and {Cubillos}, P.~E. and {Bekkelien}, A. and {Fortier}, A. and {Flor{\'e}n}, H.-G. and {Collier Cameron}, A. and {Bou{\'e}}, G. and {Bruno}, G. and {Demory}, B.-O. and {Brandeker}, A. and {Sousa}, S.~G. and {Wilson}, T.~G. and {Deline}, A. and {Bonfanti}, A. and {Scandariato}, G. and {Hooton}, M.~J. and {Correia}, A.~C.~M. and {Demangeon}, O.~D.~S. and {Smith}, A.~M.~S. and {Singh}, V. and {Alibert}, Y. and {Alonso}, R. and {Asquier}, J. and {B{\'a}rczy}, T. and {Barrado Navascues}, D. and {Baumjohann}, W. and {Beck}, M. and {Beck}, T. and {Benz}, W. and {Billot}, N. and {Bonfils}, X. and {Borsato}, L. and {Broeg}, C. and {Buder}, M. and {Charnoz}, S. and {Csizmadia}, Sz. and {Davies}, M.~B. and {Deleuil}, M. and {Delrez}, L. and {Ehrenreich}, D. and {Erikson}, A. and {Farinato}, J. and {Fossati}, L. and {Fridlund}, M. and {Gandolfi}, D. and {Gillon}, M. and {G{\"u}del}, M. and {G{\"u}nther}, M.~N. and {Heitzmann}, A. and {Helling}, Ch. and {Hoyer}, S. and {Isaak}, K.~G. and {Kiss}, L.~L. and {Lam}, K.~W.~F. and {Laskar}, J. and {Lecavelier des Etangs}, A. and {Magrin}, D. and {Maxted}, P.~F.~L. and {Mecina}, M. and {Mordasini}, C. and {Nascimbeni}, V. and {Olofsson}, G. and {Ottensamer}, R. and {Pagano}, I. and {Pall{\'e}}, E. and {Peter}, G. and {Piazza}, D. and {Piotto}, G. and {Pollacco}, D. and {Queloz}, D. and {Ragazzoni}, R. and {Rando}, N. and {Rauer}, H. and {Ribas}, I. and {Santos}, N.~C. and {S{\'e}gransan}, D. and {Simon}, A.~E. and {Stalport}, M. and {Szab{\'o}}, Gy. M. and {Thomas}, N. and {Udry}, S. and {Van Grootel}, V. and {Venturini}, J. and {Villaver}, E. and {Walton}, N.~A.},
        title = "{The tidal deformation and atmosphere of WASP-12 b from its phase curve★}",
      journal = {\aap},
         year = 2024,
        month = may,
       volume = {685},
          eid = {A63},
        pages = {A63},
          doi = {10.1051/0004-6361/202348502},
archivePrefix = {arXiv},
       eprint = {2402.10486},
 primaryClass = {astro-ph.EP},
       adsurl = {https://ui.adsabs.harvard.edu/abs/2024A&A...685A..63A}
}

@BOOK{hubbard_1984,
       author = {{Hubbard}, W.~B.},
        title = "{Planetary interiors}",
         year = 1984,
    publisher = "{Van Nostrand}",
       adsurl = {https://ui.adsabs.harvard.edu/abs/1984plin.book.....H}
}

@ARTICLE{carter_winn_empirical_2010,
       author = {{Carter}, Joshua A. and {Winn}, Joshua N.},
        title = "{Empirical Constraints on the Oblateness of an Exoplanet}",
      journal = {\apj},
         year = 2010,
        month = feb,
       volume = {709},
       number = {2},
        pages = {1219-1229},
          doi = {10.1088/0004-637X/709/2/1219},
archivePrefix = {arXiv},
       eprint = {0912.1594},
 primaryClass = {astro-ph.EP},
       adsurl = {https://ui.adsabs.harvard.edu/abs/2010ApJ...709.1219C}
}

@ARTICLE{biersteker_2017,
       author = {{Biersteker}, John and {Schlichting}, Hilke},
        title = "{Determining Exoplanetary Oblateness Using Transit Depth Variations}",
      journal = {\aj},
         year = 2017,
        month = oct,
       volume = {154},
       number = {4},
          eid = {164},
        pages = {164},
          doi = {10.3847/1538-3881/aa88c2},
archivePrefix = {arXiv},
       eprint = {1708.08990},
 primaryClass = {astro-ph.EP},
       adsurl = {https://ui.adsabs.harvard.edu/abs/2017AJ....154..164B}
}

@ARTICLE{zhu_2014,
       author = {{Zhu}, Wei and {Huang}, Chelsea X. and {Zhou}, George and {Lin}, D.~N.~C.},
        title = "{Constraining the Oblateness of Kepler Planets}",
      journal = {\apj},
         year = 2014,
        month = nov,
       volume = {796},
       number = {1},
          eid = {67},
        pages = {67},
          doi = {10.1088/0004-637X/796/1/67},
archivePrefix = {arXiv},
       eprint = {1410.0361},
 primaryClass = {astro-ph.EP},
       adsurl = {https://ui.adsabs.harvard.edu/abs/2014ApJ...796...67Z}
}

@ARTICLE{burton_2014,
       author = {{Burton}, J.~R. and {Watson}, C.~A. and {Fitzsimmons}, A. and {Pollacco}, D. and {Moulds}, V. and {Littlefair}, S.~P. and {Wheatley}, P.~J.},
        title = "{Tidally Distorted Exoplanets: Density Corrections for Short-period Hot-Jupiters Based Solely on Observable Parameters}",
      journal = {\apj},
         year = 2014,
        month = jul,
       volume = {789},
       number = {2},
          eid = {113},
        pages = {113},
          doi = {10.1088/0004-637X/789/2/113},
archivePrefix = {arXiv},
       eprint = {1405.1839},
 primaryClass = {astro-ph.EP},
       adsurl = {https://ui.adsabs.harvard.edu/abs/2014ApJ...789..113B}
}

@ARTICLE{batygin_2018,
       author = {{Batygin}, Konstantin},
        title = "{On the Terminal Rotation Rates of Giant Planets}",
      journal = {\aj},
         year = 2018,
        month = apr,
       volume = {155},
       number = {4},
          eid = {178},
        pages = {178},
          doi = {10.3847/1538-3881/aab54e},
archivePrefix = {arXiv},
       eprint = {1803.07106},
 primaryClass = {astro-ph.EP},
       adsurl = {https://ui.adsabs.harvard.edu/abs/2018AJ....155..178B}
}

@ARTICLE{bryan_2018,
       author = {{Bryan}, Marta L. and {Benneke}, Bj{\"o}rn and {Knutson}, Heather A. and {Batygin}, Konstantin and {Bowler}, Brendan P.},
        title = "{Constraints on the spin evolution of young planetary-mass companions}",
      journal = {Nature Astronomy},
         year = 2018,
        month = dec,
       volume = {2},
        pages = {138-144},
          doi = {10.1038/s41550-017-0325-8},
archivePrefix = {arXiv},
       eprint = {1712.00457},
 primaryClass = {astro-ph.EP},
       adsurl = {https://ui.adsabs.harvard.edu/abs/2018NatAs...2..138B}
}

@ARTICLE{berardo_bottleneck_2022,
       author = {{Berardo}, David and {de Wit}, Julien},
        title = "{Tidal Distortions as a Bottleneck on Constraining Exoplanet Compositions}",
      journal = {\apj},
         year = 2022,
        month = dec,
       volume = {941},
       number = {2},
          eid = {155},
        pages = {155},
          doi = {10.3847/1538-4357/aca409},
archivePrefix = {arXiv},
       eprint = {2301.08755},
 primaryClass = {astro-ph.EP},
       adsurl = {https://ui.adsabs.harvard.edu/abs/2022ApJ...941..155B}
}

@ARTICLE{price_2025,
       author = {{Price}, Ellen M. and {Becker}, Juliette and {de Beurs}, Zo{\"e} L. and {Rogers}, Leslie A. and {Vanderburg}, Andrew},
        title = "{A Long Spin Period for a Sub-Neptune-mass Exoplanet}",
      journal = {\apjl},
         year = 2025,
        month = mar,
       volume = {981},
       number = {1},
          eid = {L7},
        pages = {L7},
          doi = {10.3847/2041-8213/adb42b},
archivePrefix = {arXiv},
       eprint = {2410.05408},
 primaryClass = {astro-ph.EP},
       adsurl = {https://ui.adsabs.harvard.edu/abs/2025ApJ...981L...7P}
}

@ARTICLE{dholakia_2025,
       author = {{Dholakia}, Shashank and {Dholakia}, Shishir and {Pope}, Benjamin J.~S.},
        title = "{A General, Differentiable Transit Model for Ellipsoidal Occulters: Derivation, Application, and Forecast of Planetary Oblateness and Obliquity Constraints with JWST}",
      journal = {\apj},
         year = 2025,
        month = jul,
       volume = {987},
       number = {2},
          eid = {150},
        pages = {150},
          doi = {10.3847/1538-4357/addb4e},
archivePrefix = {arXiv},
       eprint = {2410.03449},
 primaryClass = {astro-ph.EP},
       adsurl = {https://ui.adsabs.harvard.edu/abs/2025ApJ...987..150D}
}

@ARTICLE{lammers_2024,
       author = {{Lammers}, Caleb and {Winn}, Joshua N.},
        title = "{Slow Rotation for the Super-puff Planet Kepler-51d}",
      journal = {\apjl},
         year = 2024,
        month = dec,
       volume = {977},
       number = {1},
          eid = {L1},
        pages = {L1},
          doi = {10.3847/2041-8213/ad91ae},
archivePrefix = {arXiv},
       eprint = {2409.06697},
 primaryClass = {astro-ph.EP},
       adsurl = {https://ui.adsabs.harvard.edu/abs/2024ApJ...977L...1L}
}

@ARTICLE{rowe_2014,
       author = {{Rowe}, Jason F. and {Bryson}, Stephen T. and {Marcy}, Geoffrey W. and {Lissauer}, Jack J. and {Jontof-Hutter}, Daniel and {Mullally}, Fergal and {Gilliland}, Ronald L. and {Issacson}, Howard and {Ford}, Eric and {Howell}, Steve B. and {Borucki}, William J. and {Haas}, Michael and {Huber}, Daniel and {Steffen}, Jason H. and {Thompson}, Susan E. and {Quintana}, Elisa and {Barclay}, Thomas and {Still}, Martin and {Fortney}, Jonathan and {Gautier}, III, T.~N. and {Hunter}, Roger and {Caldwell}, Douglas A. and {Ciardi}, David R. and {Devore}, Edna and {Cochran}, William and {Jenkins}, Jon and {Agol}, Eric and {Carter}, Joshua A. and {Geary}, John},
        title = "{Validation of Kepler's Multiple Planet Candidates. III. Light Curve Analysis and Announcement of Hundreds of New Multi-planet Systems}",
      journal = {\apj},
         year = 2014,
        month = mar,
       volume = {784},
       number = {1},
          eid = {45},
        pages = {45},
          doi = {10.1088/0004-637X/784/1/45},
archivePrefix = {arXiv},
       eprint = {1402.6534},
 primaryClass = {astro-ph.EP},
       adsurl = {https://ui.adsabs.harvard.edu/abs/2014ApJ...784...45R}
}

@book{de_Pater_Lissauer_2015, place={Cambridge}, edition={2}, title={Planetary Sciences}, publisher={Cambridge University Press}, author={de Pater, Imke and Lissauer, Jack J.}, year={2015}}

@ARTICLE{hattori_2025,
       author = {{Hattori}, Soichiro and {a}, b and {c}, d},
        title = "",
      journal = {in prep.},
         year = 2025,
       adsurl = {}
}

@article{kass_1995,
author = {Robert E. Kass and Adrian E. Raftery},
title = {Bayes Factors},
journal = {Journal of the American Statistical Association},
volume = {90},
number = {430},
pages = {773--795},
year = {1995},
publisher = {ASA Website},
doi = {10.1080/01621459.1995.10476572},
}

@ARTICLE{mercier_2025,
       author = {{Mercier}, Samson J. and {de Wit}, Julien and {Rackham}, Benjamin V.},
        title = "{What's in Your Transit? Towards Reliably Getting $5\times$ More Science from Exoplanet Transit Data}",
      journal = {arXiv e-prints},
         year = 2025,
        month = sep,
          eid = {arXiv:2510.00124},
        pages = {arXiv:2510.00124},
          doi = {10.48550/arXiv.2510.00124},
archivePrefix = {arXiv},
       eprint = {2510.00124},
 primaryClass = {astro-ph.EP},
       adsurl = {https://ui.adsabs.harvard.edu/abs/2025arXiv251000124M}
}

@ARTICLE{kipping_2025_response,
       author = {{Kipping}, David and {Teachey}, Alex and {Yahalomi}, Daniel A. and {Cassese}, Ben and {Quarles}, Billy and {Bryson}, Steve and {Hansen}, Brad and {Szul{\'a}gyi}, Judit and {Burke}, Chris and {Hardegree-Ullman}, Kevin},
        title = "{Concerning the possible exomoons around Kepler-1625 b and Kepler-1708 b}",
      journal = {Nature Astronomy},
         year = 2025,
        month = jun,
       volume = {9},
        pages = {795-798},
          doi = {10.1038/s41550-025-02547-1},
       adsurl = {https://ui.adsabs.harvard.edu/abs/2025NatAs...9..795K}
}

@ARTICLE{jojo,
       author = {{Liu}, Quanyi and {Zhu}, Wei and {Zhou}, Yifan and {Hu}, Zhecheng and {Lin}, Zitao and {Dai}, Fei and {Masuda}, Kento and {Wang}, Sharon X.},
        title = "{Detecting Planetary Oblateness in the Era of JWST: A Case Study of Kepler-167e}",
      journal = {\aj},
         year = 2025,
        month = feb,
       volume = {169},
       number = {2},
          eid = {79},
        pages = {79},
          doi = {10.3847/1538-3881/ad9b8c},
archivePrefix = {arXiv},
       eprint = {2406.11644},
 primaryClass = {astro-ph.EP},
       adsurl = {https://ui.adsabs.harvard.edu/abs/2025AJ....169...79L}
}

@ARTICLE{liu_kep_51,
       author = {{Liu}, Quanyi and {Zhu}, Wei and {Masuda}, Kento and {Libby-Roberts}, Jessica E. and {Bello-Arufe}, Aaron and {Ca{\~n}as}, Caleb I.},
        title = "{An Extremely Low-density Exoplanet Spins Slow}",
      journal = {\apjl},
         year = 2024,
        month = nov,
       volume = {976},
       number = {1},
          eid = {L14},
        pages = {L14},
          doi = {10.3847/2041-8213/ad8f39},
archivePrefix = {arXiv},
       eprint = {2410.07977},
 primaryClass = {astro-ph.EP},
       adsurl = {https://ui.adsabs.harvard.edu/abs/2024ApJ...976L..14L}
}
\bibliographystyle{aasjournal}

\end{document}